\documentclass[pdflatex,sn-mathphys-num]{sn-jnl}

\usepackage{comment}
\usepackage{graphicx}%
\usepackage{multirow}%
\usepackage{amsmath,amssymb,amsfonts}%
\usepackage{amsthm}%
\usepackage{mathrsfs}%
\usepackage[title]{appendix}%
\usepackage{xcolor}%
\usepackage{textcomp}%
\usepackage{manyfoot}%
\usepackage{booktabs}%
\usepackage{wasysym}
\usepackage{algorithm}%
\usepackage{algorithmicx}%
\usepackage{algpseudocode}%
\usepackage{listings}%
\usepackage{subcaption}%
\usepackage{rotating} 

\theoremstyle{thmstyleone}%
\theoremstyle{thmstyletwo}%

\theoremstyle{thmstylethree}%

\usepackage{xspace}
\newcommand{\degree}{\ensuremath{^\circ}\xspace}

\newcommand{\htp}{\ensuremath{\mathrm{H}_2^+}\xspace}
\newcommand{\dto}{\ensuremath{\mathrm{D}_2\mathrm{O}}\xspace}
\newcommand{\hto}{\ensuremath{\mathrm{H}_2\mathrm{O}}\xspace}

\begin{document}

\title[IsoDAR@Yemilab: PDR - Vol. II]{IsoDAR@Yemilab: Preliminary Design Report - Volume II: Medium Energy Beam Transport, Neutrino Source, and Shielding}

\author[]{\sur{The IsoDAR Collaboration}}
\author*[1]{\fnm{Joshua} \sur{Spitz}}\email{spitzj@umich.edu}
\author[2]{\fnm{Jose R.} \sur{Alonso}}
\author[1]{\fnm{Jon} \sur{Ameel}}
\author[3]{\fnm{Roger} \sur{Barlow}}
\author[4]{\fnm{Larry} \sur{Bartoszek}}
\author[2]{\fnm{Adriana} \sur{Bungau}}
\author[5]{\fnm{Michael H.} \sur{Shaevitz}}
\author[6]{\fnm{Erik A.} \sur{Voirin}}
\author*[2]{\fnm{Daniel} \sur{Winklehner}}\email{winklehn@mit.edu}
\author[2]{\fnm{Janet M.} \sur{Conrad}}
\author[2]{\fnm{Samuel J.} \sur{Engebretson}}
\author[2]{\fnm{Jarrett} \sur{Moon}}
\author[2]{\fnm{Eleanor} \sur{Winkler}}
\author[7]{\fnm{Andreas} \sur{Adelmann}}
\author[8]{\fnm{Spencer N.} \sur{Axani}}
\author[2]{\fnm{William A.} \sur{Barletta}}
\author[9]{\fnm{Luciano} \sur{Calabretta}}
\author[10]{\fnm{Pedro} \sur{Calvo}}
\author[5]{\fnm{Georgia} \sur{Karagiorgi}}
\author[10]{\fnm{Concepti\'on} \sur{Oliver}}
\author[1]{\fnm{Andrew} \sur{Chan}}
\author[1]{\fnm{Emilie} \sur{Lavoie-Ingram}}

\affil[1]{\orgdiv{Physics Department}, \orgname{University of Michigan}, \orgaddress{\street{450 Church St}, \city{Ann Arbor}, \postcode{48109}, \state{MI}, \country{USA}}}

\affil[2]{\orgdiv{Laboratory for Nuclear Science}, \orgname{Massachusetts Institute of Technology}, \orgaddress{\street{77 Massachusetts Ave}, \city{Cambridge}, \postcode{02139}, \state{MA}, \country{USA}}}

\affil[3]{\orgname{University of Huddersfield}, \orgaddress{\city{Huddersfield}, \postcode{HD1 3DH}, \country{UK}}}

\affil[4]{\orgname{Bartoszek Engineering}, \orgaddress{\street{818 W. Downer Place}, \city{Aurora}, \postcode{60506}, \state{IL}, \country{USA}}}

\affil[5]{\orgdiv{Department of Physics}, \orgname{Columbia University}, \orgaddress{\city{New York}, \state{NY} \postcode{10027}, \country{USA}}}

\affil[6]{\orgname{eVoirin Engineering Consulting}, \orgaddress{\city{Batavia}, \state{IL}, \country{USA}}}

\affil[7]{\orgname{Paul Scherrer Institut}, \orgaddress{\city{Villigen PSI}, \postcode{5232}, \country{Switzerland}}}

\affil[8]{\orgdiv{Department of Physics \& Astronomy}, \orgname{University of Delaware}, \orgaddress{\street{210 South College Ave.}, \city{Newark}, \postcode{19716}, \state{DE}, \country{USA}}}

\affil[9]{\orgname{MAGMA srl}, \orgaddress{\street{Via E. A. Pantano 70}, \city{Catania}, \country{Italy}}}

\affil[10]{\orgname{Centro de Investigaciones Energ\'eticas, Medioambientales y Tecnol\'ogicas}, \orgaddress{\city{Madrid}, \country{Spain}}}

\abstract{This Preliminary Design Report (PDR) describes the IsoDAR electron-antineutrino
source in two volumes which are mostly site-independent and describe the cyclotron driver providing a 60 MeV, 10~mA proton beam (Volume I); and the medium energy beam transport line (MEBT) and target (this Volume). The IsoDAR driver and target will produce about \mathversion{normal}$1.15\cdot10^{23}$ electron-antineutrinos over five calendar years. Paired with a kton-scale 
liquid scintillator detector, this will enable a broad particle physics
program including searches for new symmetries, new interactions and new particles. 
Here in Volume II, we describe the medium energy beam transport line, the antineutrino source beam-target and surrounding sleeve, shielding, and plans for monitoring and installation.}





\maketitle

\tableofcontents

\clearpage
\section{Introduction}
This preliminary design report (PDR) is the second of two volumes describing the IsoDAR antineutrino source. The first volume, which also contains a thorough introduction to the particle physics motivation behind the experiment, is focused on the design of the IsoDAR cyclotron-based accelerator system~\cite{IsoDAR:2024rvi}. This second volume covers everything downstream of the proton extraction point at the threshold of the cyclotron system, including the subsequent beam transport line leading to the target, the target, the sleeve surrounding the target (representing the antineutrino source), surrounding shielding (in consideration of safety and physics background), various non-trivial monitoring systems, and installation. In designing these elements, the primary considerations are (1) maximizing sensitivity to well-motivated new physics, (2) timeliness/ease of construction, (3) ease of operation, (4) cost effectiveness, and (5) stability and experimental lifetime. Safety and adhering to local and US engineering codes, laws, and regulations are paramount as well. While the overarching design presented in both volumes is largely applicable to a generic underground location proximal to a large underground detector, we have chosen to add some detail regarding site-specific issues at the Yemilab underground facility~\cite{,Park:2024sio} and pairing with the future Neutrino Experiment at YEmilab detector ($\nu$EYE)\footnote{Formerly known as the Liquid Scintillator Counter (LSC).}~\cite{seo_physics_2023}.

This document is structured as follows:
\begin{itemize}
\item {\bf Chapter 2:}  Describes the design of the Medium Energy Beam Transport (MEBT) system.
\item {\bf Chapter 3:} Describes engineering related to the target design.
\item {\bf Chapter 4:} Describes engineering related to the design of the ``sleeve" surrounding the target.
\item {\bf Chapter 5:} Describes engineering related to the shielding around the IsoDAR target-sleeve antineutrino source.
\item {\bf Chapter 6:} Describes various monitoring systems.
\item {\bf Chapter 7:} Describes the engineering related to installation. 

\end{itemize}

\clearpage
\section{MEBT Design}

\subsection{Introduction}

Though the general desire of this document was to provide generalized designs that are not site specific, it is necessary to present designs in the context of a specific site, in particular for the extended (and therefore cavern- and geometry-dependent) MEBT system described in this chapter.  However, the designs presented, though related specifically to the Yemilab site, can be adapted as necessary to any underground location.  

\begin{figure}
\begin{centering}
    
\includegraphics[width=14 cm]{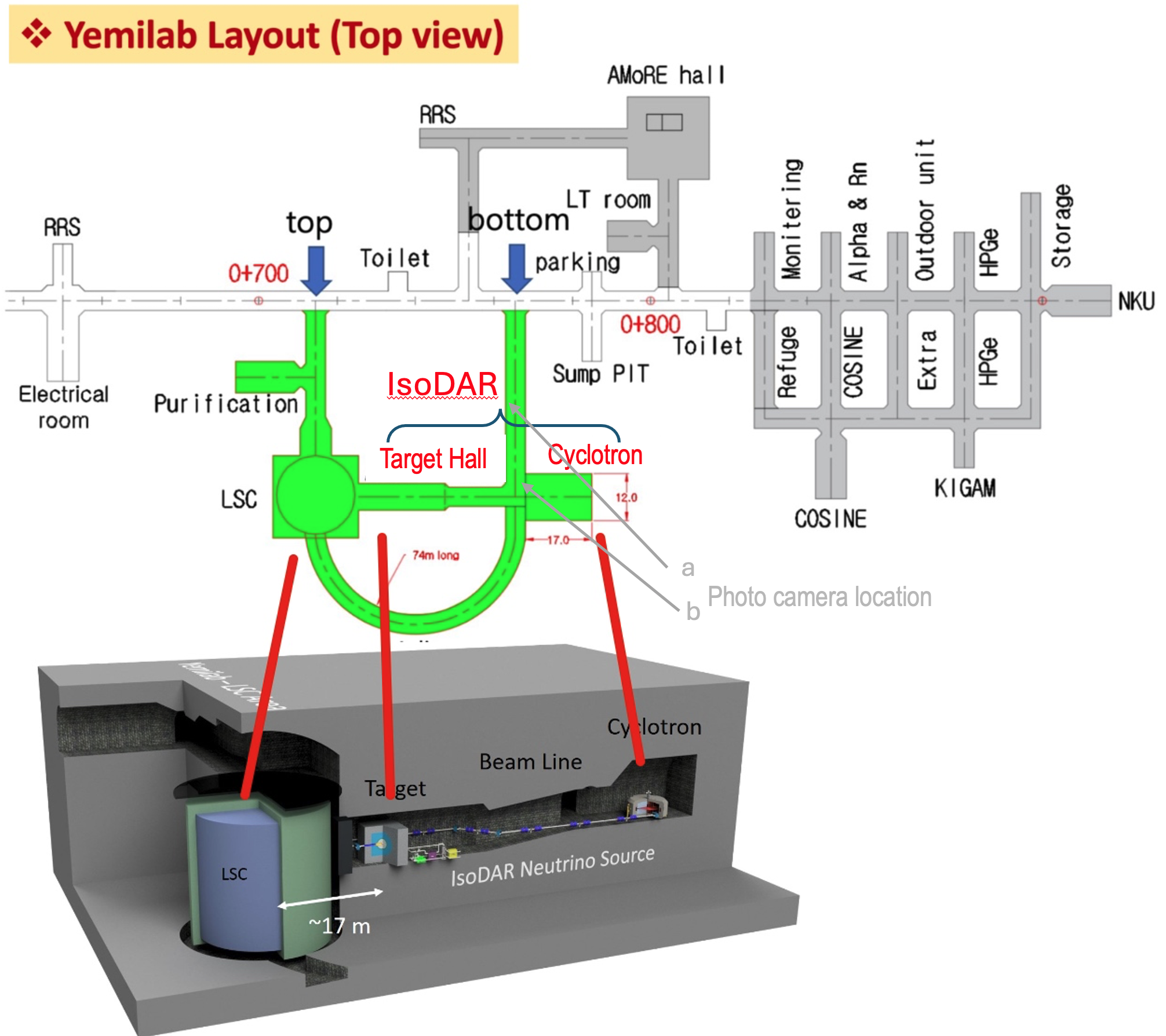}    
\caption{\label{fig:IsoDAR@Yemilab} Deployment of the IsoDAR experiment at Yemilab.  Letters ``a'' and ``b'' denote camera locations for the photos in Fig.~\ref{fig:CavePhotos}.}
\end{centering}
\end{figure}

The Yemilab site for IsoDAR, depicted in Fig.~\ref{fig:IsoDAR@Yemilab}, includes a cavern specifically excavated to the requirements of the cyclotron, and areas, originally used as part of the infrastructure for excavation and rock removal of the large cavern dedicated to a kiloton-scale liquid scintillator detector, referred to in this document as the neutrino Experiment at YEmilab detector ($\nu$EYE)--formerly called the Liquid Scintillator Counter (LSC)--that will house the antineutrino-producing module (target + sleeve) and the transport line that brings the beam from the cyclotron to the target. 

\begin{figure}
\begin{centering}
    
\includegraphics[width=\linewidth]{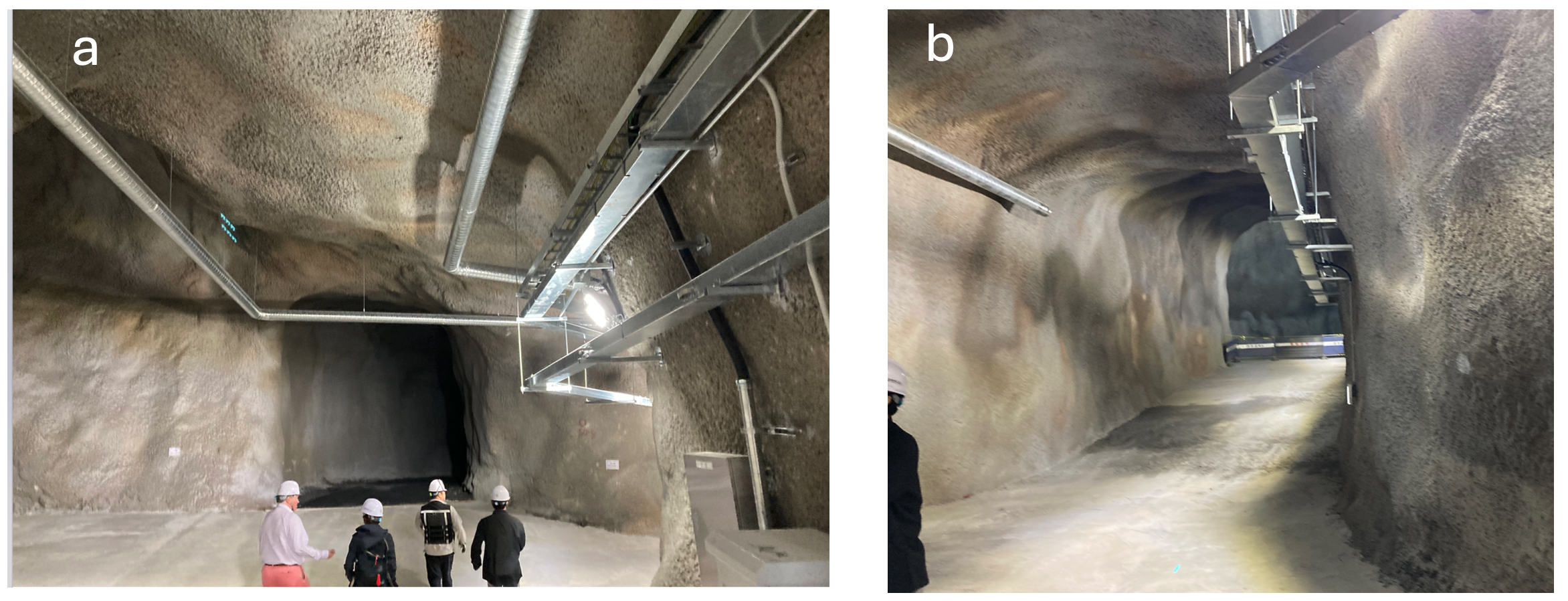}    
\caption{\label{fig:CavePhotos} Photos of the Yemilab site. (a) Looking down the access ramp, the cyclotron cavern is on the left, and the target hall is on the right; (b) Looking down the target hall, the $\nu$EYE cavern can be seen through the end of the target hall.  This opening will be filled with solid iron for fast-neutron shielding.}
\end{centering}
\end{figure}

As depicted in Fig.~\ref{fig:CavePhotos}, the underground infrastructure at Yemilab includes well-finished (with trowelled shotcrete) 5 x 5 meter ramps and drifts, connecting the cyclotron hall and the larger hall adjacent to the detector cavern that will house the target and shielding.
The cyclotron will be installed in its dedicated vault, and the target-sleeve will be located as close to $\nu$EYE as possible, taking into account that a well-designed shield between target and detector is needed to prevent neutrons from reaching the fiducial volume of the detector.  Connecting these two will be the MEBT line and its subsystems which take the 60~MeV/amu beam from the cyclotron to the target. These elements are described in this section.

\subsubsection{Description of the MEBT System}
The beam emerges from the cyclotron as \htp, however it is desired to transport protons to the target rather than \htp.  Not only is the rigidity less so magnets can be smaller, but also uncontrolled beam losses, mainly due to interactions with residual gas, will be substantially reduced if the beam transmitted is protons rather than \htp.  

\begin{figure}
\begin{centering}
\includegraphics[width=\linewidth]{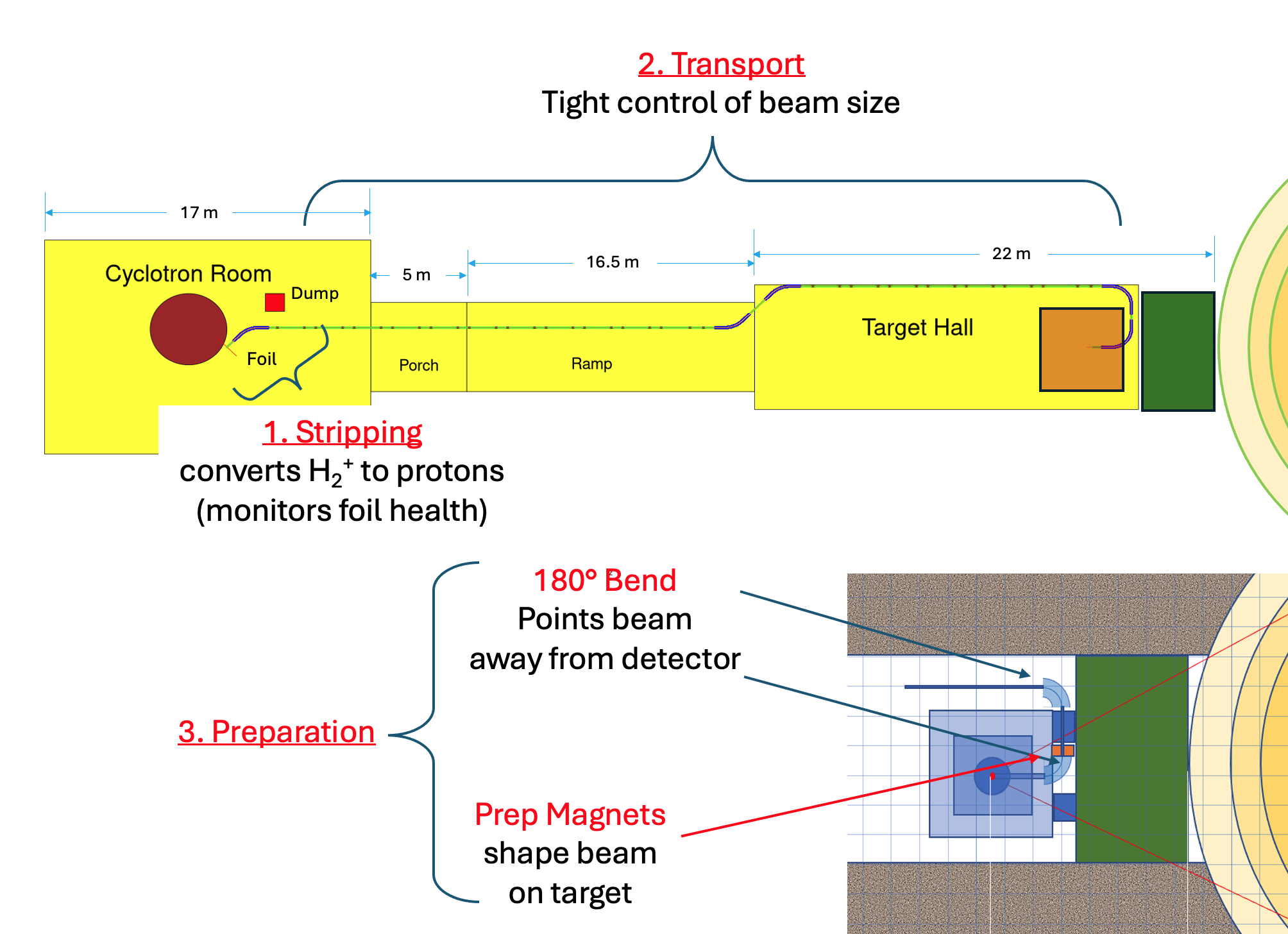}    
\caption{\label{fig:MEBT System} Schematic of the MEBT system with constituent sections identified.}
\end{centering}
\end{figure}

The elements of the MEBT are shown in Fig.~\ref{fig:MEBT System}. The stripping section, located as close to the extraction point of the cyclotron as practical, consists of the stripping foil and an analysis magnet that separates the various elements emerging from the foil, and directs the protons through a collimator into the transport line.  
The transport section takes the beam across the ``porch'' and down the ramp to the target hall. The target hall is about 2 meters lower than the cyclotron room, so some vertical adjustments of the beam height will undoubtedly be necessary, depending upon the height of the cyclotron and target from their respective floors.  
In addition,  horizontal bends will be needed, here shown as two $45^\circ$ dipoles, to move the beam out of direct line with the target. Earlier studies showed the transport line closer to the near wall, opposite from the present far-wall location; the present configuration may be somewhat more convenient in avoiding access conflicts down the main entry ramp, but from a beam transport standpoint either will work.
The preparation section bends the beam through $180^\circ$ (two $90^\circ$ dipoles) and through a vacuum pipe to the target located in the center of the large shielding block (Fig.~\ref{fig:TargetShieldingBlock}).  Figure~\ref{fig:NeutronSpectra} makes clear that neutrons emitted in the backward direction (i.e. going towards the antineutrino detector) are considerably attenuated, both in intensity and energy, justifying the double bend in the beam line.
The last part of the preparation section may include magnets for spreading the beam over the face of the target, to distribute the 600 kW heat load optimally over the target face, though these may not be necessary.

The main points that distinguish the IsoDAR MEBT from the many 
beamlines found in other accelerators 
are the high (10 mA continuous) beam current, which
requires strict conditions on the amount of beam losses that can be tolerated during transport, and the need to spread the beam over the target in a way that maximizes the lifetime of the target in consideration of thermal degradation. 
These two points are discussed in detail in the following subsections.

The study here should be considered as a proof of principle. 
The choices made for various parameters are summarised in Table~\ref{tab:MEBTrequirements}. They are discussed in what follows. Further optimization is certainly possible and 
should be done when resources are available for a final design. Notably as well, as seen in Fig.~\ref{fig:CavePhotos}, the actual excavated cavern is not nearly as smooth and regular as shown in the plan (e.g. the beam line cannot be too close to the wall).

\begin{table}[h]
\begin{centering} 
\caption{\label{tab:MEBTrequirements} Design choices for the MEBT.}
\begin{tabular}{c|c|l}
Parameter& Value & Notes \\
\hline
Loss limit & 1 W/m & usual limit for high power beams  \\
Beam pipe diameter & 10 cm & good vacuum conductance, contains beam, \\ & & but requires larger magnet bores and gaps \\
Stripping foil thickness & 200 $\mu$g/cm$^2$& possibly thinner\\
Field in $45^\circ$ bends & 0.86 T & R=1.3 m\\
Field in $90^\circ$ bends & 1.1 T & R=1 m \\
Quadrupole length & 20 cm & provides reasonable strengths\\
\end{tabular}
\end{centering}
\end{table}

\subsection{The Stripping Section}

\begin{figure}
\begin{centering}

\includegraphics[width=\linewidth]{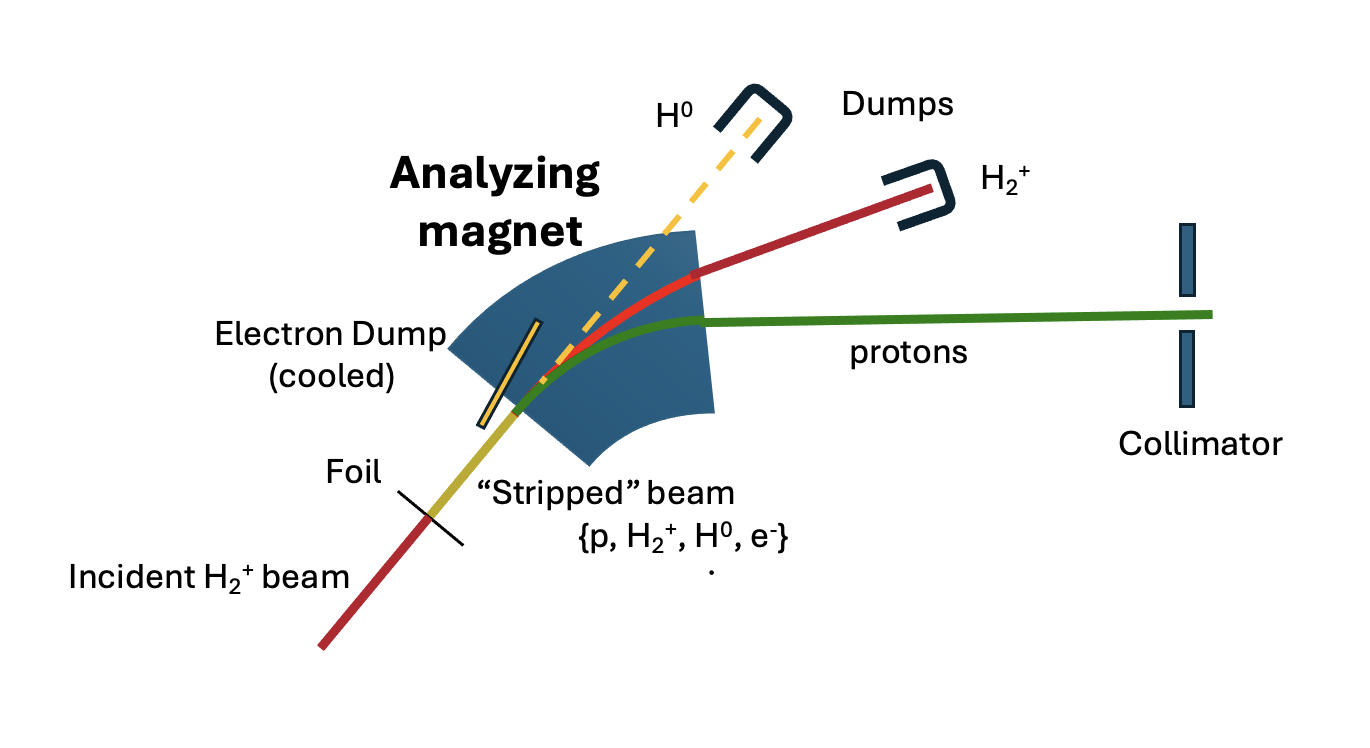}    
\caption{\label{fig:stripper area} Schematic of the stripping section.}
\end{centering}
\end{figure}

Figure~\ref{fig:stripper area} shows a schematic of the stripping section, including the analysis magnet that follows the stripping foil.  

\subsubsection{Analysis Magnet}

Normal constituents of the beam following the stripping foil will be protons and ``convoy'' electrons. The analysis magnet should be optimized for transport of the protons,  the trajectory for these being defined by an instrumented collimator that allows for tuning of the dipole to center the protons into the MEBT channel.
This collimator will undoubtedly scrape edges of the beam, so can be viewed as a ``controlled loss point.''  It must be well cooled, and also should be enclosed in a shielding box to reduce the flux of neutrons -- from stopping protons reacting in the collimator material -- into the cavern. However, a compromised foil in the process of failing, will have holes that can allow unstripped \htp ions to pass through.  
So, there should be an exit port (at double the rigidity) for unstripped \htp ions. A collecting dump after this port should have a sensitive current monitoring capability, to immediately notify of the imminent failure of the foil.

The magnet should also have a port at $0^\circ$ for visual alignment of the magnet in the beamline, as well as for the extremely rare instances where a neutral hydrogen atom emerges from the foil.  This port can also be used for tuning the extracted beam from the cyclotron with no current in the magnet.

The convoy electrons are bent in the opposite direction in the analysis magnet field, and can be dumped on a water-cooled surface inside the vacuum tank. The entrance fringing field is sufficient to bend these electrons into the cooled surface. The maximum heat load from these electrons will be about 150 watts.

\subsubsection{Stripping Foil}

The stripping foil should be located as close as allowed by the extraction optics of the cyclotron.  Ideally it should be at a waist in the beam, to minimize emittance growth due to multiple scattering of the beam in the foil, though even for the thickest foils anticipated ($\approx$200~$\mu$g/cm$^2$) multiple scattering will be at most 0.25 mrad\footnote{Multiple Coulomb scattering of protons  can be expressed by \cite{PDG} 
$\theta_0={\frac{13.6}{\beta c p}} \sqrt{x/X}$, where $X$ is the radiation length of graphite (42.7~g/cm$^2$). This gives, for a 200~$\mu$g/cm$^2$ foil ($x = 2 \cdot 10^{-4}$), a value of 0.25 mrad in both transverse directions.}.
This must be balanced with the need to spread the beam as it traverses the foil to keep the temperature rise in the foil below its failure point.

Tests performed at PSI indicate that foils should have reasonable lifetimes in our MEBT geometry~\cite{dolling_test_2018}.  Foils that are used for extracting beam in H$^-$ cyclotrons have a limit of less than 1 mA, but this is due to the foil being in magnetic fields that bend the convoy electrons (the electrons stripped from the H$^-$ ion) in tight circular orbits so they make many passes through the foil leading to an extremely high heat load.  This is not the case for our stripping foil, which is in a field-free region, so the electrons continue as part of the beam and can be dumped safely on a cooled surface further downstream.
The above-cited test was performed in the transport line at PSI between the 72 MeV ``Injector 2'' cyclotron and their 590 MeV ``Ring Cyclotron'', and showed that a 200 $\mu$g/cm$^2$ carbon foil after a 60-hour exposure to a 1.7 mA proton beam, showed signs of damage but was still functional.

Selecting the parameters for the carbon foil, including its thickness and crystalline composition, is still to be done. There are many foil types: thin diamond films, amorphous CVD, pyrolytic graphite, multilayer graphene, and others.  Each has advantages that need to be explored before selecting the best foil for our application.  Nevertheless, it is certain that foils will eventually fail, and the port designed for \htp ions is required for identifying the need for changing foils, hopefully before the complete failure of the presently running foil.  

Optimizing foil thickness is a tradeoff between stripping efficiency, foil lifetime, and heating.  Thicker foils are mechanically stronger, and provide greater stripping efficiency, but the beam generates more heat because of greater energy loss.  At the very high temperatures at which the foils operate, greater than 2500 K, heat loss is accomplished exclusively by radiation, which has a T$^4$ dependence.  As the radiation cooling is only a function of the surface area, thicker foils will operate at higher temperatures for the same beam spot size.  Stripping efficiency should be excellent with even the thinnest foils, so the tradeoff is really between mechanical strength and temperature levels.  This evaluation will most assuredly be done once high power beams from the cyclotron are available; and as material optimization is an ongoing process, improved performance is likely to be achieved with continued running of the experiment.

Monitoring of foil health is accomplished by looking for the presence of unstripped \htp ions.  A foil that is beginning to fail will develop pinholes or rips that allow a portion of the beam to pass through without intercepting the foil.  As the \htp ion is almost always dissociated within the first few atomic layers of the foil, the presence of unstripped \htp is a clear sign that these ions did not see any stripping material.  Stripping foils are usually mounted on a carousel or conveyor belt (with a capacity typically of about 50 foils) inside the vacuum chamber, so bringing a new foil into the beam can be accomplished very efficiently.

\subsection{The Transport Section}

The line transporting the proton beam to the target can be designed and modeled with well-proven software simulations, as described below.  The tuning and beam monitoring of the proton beam to ensure it conforms to the design, requires instrumentation described in Chapter~\ref{monitoring_chapter}.

We performed studies of possible beamline configurations using MADX~\cite{1289960}  for
preliminary development and the OPAL program~\cite{adelmann_opal_2019}
for detailed simulation. OPAL has been validated through many studies of the accelerators at PSI and elsewhere, for example 
\cite{baumgarten,pogue,snuverink:cyclotrons2019-tha01} and it includes the effect of space charge, which could be important at the high currents and relatively low energies being considered.  The beamline simulations start at the exit of the cyclotron and the stripping foil,
and continue through the analysis magnet, down the ramp and into the target hall and through the final two dipoles, the preparation section, and finish  1 meter before the actual target.

\subsubsection{Dipoles} \label{dipoles}

The distance to the target is large, as shown in Fig.~\ref{fig:MEBT System}, $\sim 60$~m; the path is constrained by drifts and ramps at the Yemilab site. Note that the vertical height of the cyclotron and the target differs by about two meters, requiring vertical adjustment of the beam.
Bends in the beamline are generally undesirable but some will be necessary. 
We consider a situation where the height is adjusted by two $45^\circ$ bending magnets in the beamline, and two $90^\circ$ magnets are used to turn the protons around and back onto the target at the end of the beamline.

For the $45^\circ$ degree bends we propose a bending radius of 1.3 m, which is 1 m end to end,
and a field of 0.86 T. For the $90^\circ$ degree bends 
we take a slightly stronger field, 1.1 T, to give a bending radius of 1 m.

\subsubsection{Quadrupoles and Focussing}

Beam losses in the MEBT must be minimal, to avoid loss of particles on the target and for safety reasons, the latter being a more stringent constraint.   A figure of 1~W/m is generally used as a benchmark for acceptable
losses, and we will work with that.

The IsoDAR 10 mA, 60 MeV beam has a power of 600 kW,  so this gives a requirement that we should not lose more than 1 particle in 600,000 per meter. 
We propose using a beam pipe 10 cm in internal diameter (5 cm radius).
This is a compromise: for a beampipe it is large but not particularly large; a larger diameter would increase the size and costs of the magnets, 
a smaller one would increase losses through the tails of the bunch impacting the walls, and decrease vacuum pumping speed, requiring more pumping infrastructure.   

A radius of 5 cm requires a beam with $\sigma \sim 1$ cm. To maintain such dimensions over a long distance will require continuous focusing and refocusing to keep the beam size under control.

\subsubsection{The Complete Transport Line: Design}\label{TransportLine}

To design the beamline we used simple (Gaussian)
particle bunches with the nominal cyclotron emittance of 18.1~$\pi$-mm-mrad in both directions, manifesting as $\sigma_x=6.03$~mm
and $\sigma_{x'}=3$~mrad.  If we take these as uncorrelated that gives initial Twiss parameters $\alpha=0, \beta=1.42, \gamma=0.70$. 

The actual bunches produced by the cyclotron will not be so simple, and will not be described by simple Gaussians.  The consequences are considered in Section~\ref{sec:MEBTperformance} where 
we take a realistic bunch of particles generated by OPAL
simulations of the IsoDAR cyclotron and
pass it down the beamline we designed based on the simple Gaussian model.

\begin{figure}
\begin{centering}
\includegraphics[width=\linewidth]{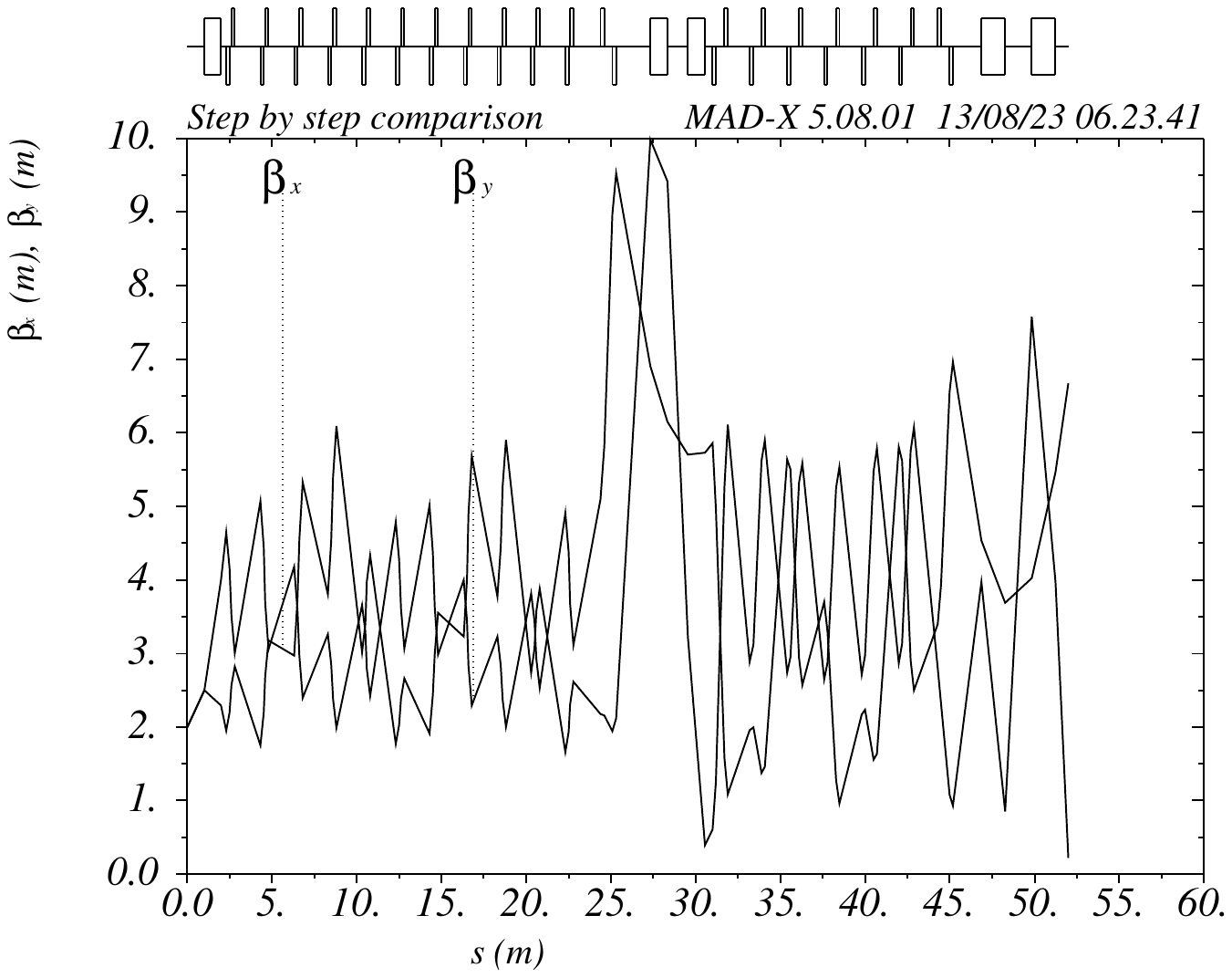}
\includegraphics[width=0.8\linewidth]{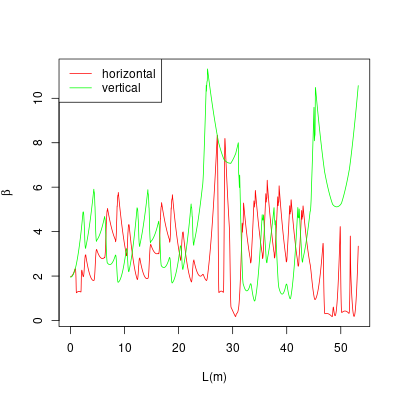}  
\vspace{-0.4cm}
\caption{\label{fig:MEBTbeta1} $\beta$ values from MADX and OPAL.}
\end{centering}
\end{figure}

MADX was used to give a first beamline design, and the $\beta$ values\footnote{For non-accelerator-physicists: $\beta$ is a parameter used to describe the evolution of a bunch along a beamline. The transverse size varies as $\sqrt \beta$. There is a separate horizontal $\beta_x$ 
and vertical $\beta_y.$ The plots show both $\beta$'s changing linearly along drifts as the bunch expands or contracts, with a gradient which decreases or increases sharply at the focussing or defocussing quadrupoles. These are 
displayed as off-axis boxes in the schematic above, with
the dipoles shown as symmetrical. 
} are shown in Fig.~\ref{fig:MEBTbeta1}.
The layout schematic at the top shows the 5 bending magnets and the 38 focusing quadrupoles. 
Numbers and positions were adjusted by hand, and the MADX MATCH feature used to find the quadrupole strengths,
with the aim of keeping the $\beta$ values below about 6 m,
which translates to the desired beam size below 1 cm rms. 
We aim to reduce the large number of magnets in a future optimization study.

This beamline was then simulated using OPAL, which includes edge effects and other details not in MADX. The beta functions from the two simulation programs are similar up to the first bending magnet (note that the display of beta values inside dipoles differs for the two programs): thereafter they are slightly different,
but the beta function is still acceptable. 
So a further optimization using OPAL will need to be done,
once the initial beam parameters are known, but 
the MADX optimized lattice can be used as a base design for the present.
The final configuration will not differ by much.

The quadrupole strengths are shown in Table~\ref{tab:MEBTquads}.  The field gradients are reasonable.  
The magnet length was taken as 20 cm for all quadrupoles: lower fields could be utilized if longer magnets are used (and vice versa).
A full cost-benefit analysis will be done to optimize this choice. 

 \begin{table} [h] 
 \centering 
 \begin{tabular}{ l | l  }  
  \hline Quadrupole & Strength   \\  
 &  (T/m)  \\ \hline  
Odd Q01-Q21:  & -5.912   \\ 
Even Q02-Q22:  & 5.924   \\ 
Q23:  & 1.802   \\ 
Q24:  & -2.692   \\ 
Q25:  & -4.987   \\ 
Q26:  & 7.140   \\ 
Odd Q27-Q35:  & -5.450   \\ 
Even Q28-Q36:  & 4.467   \\ 
Q37:  & 2.178   \\ 
Q38:  & -3.253   \\ 
 \hline \end{tabular} 
 \caption{\label{tab:MEBTquads} Quadrupole strengths.} 
 \end{table}

\begin{figure}
\begin{centering}
    
\includegraphics[width=\linewidth]{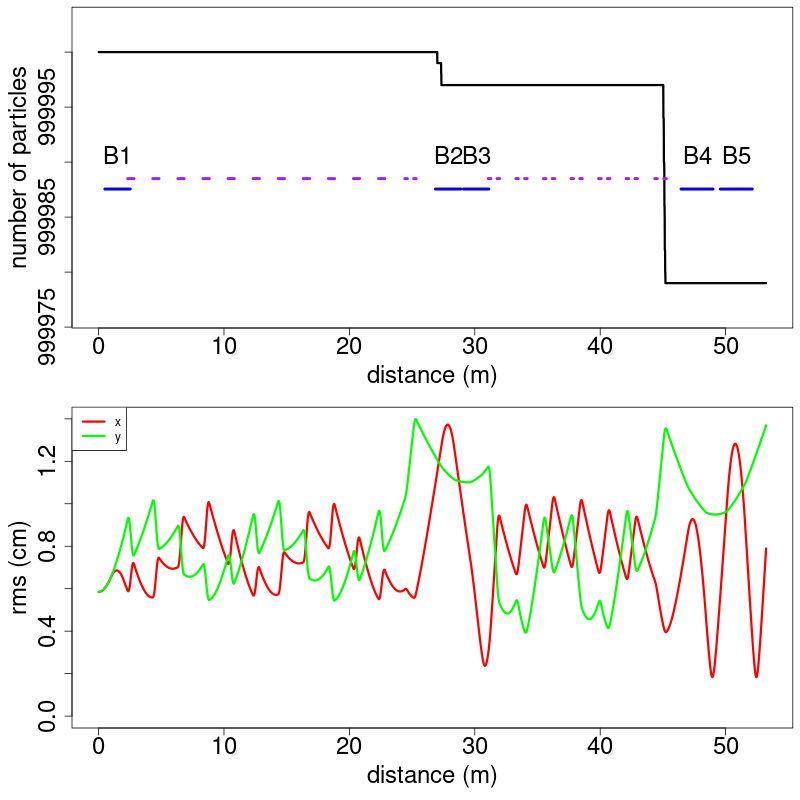}   \caption{\label{fig:MEBTlosses} Particle losses (above) and rms beam size (below) according to OPAL. Note the suppressed zero on the upper plot.}
\end{centering}
\end{figure}

The rms beam size, using an OPAL simulation of 1,000,000 simulated particles, is shown in Fig.~\ref{fig:MEBTlosses}
and, as expected, follows the $\beta$ values. Loosely, one can say that since the rms is well below the 5 cm radius of the beam pipe that losses are small.  More exactly (as the bunch will not be exactly Gaussian) the simulation gives the actual losses, also shown in the figure: only 21 of the 1,000,000 protons are lost, mostly at the start of the final two bending magnets, which is acceptable.   

\begin{figure}
\begin{centering}
    
\includegraphics[width=0.49\textwidth]{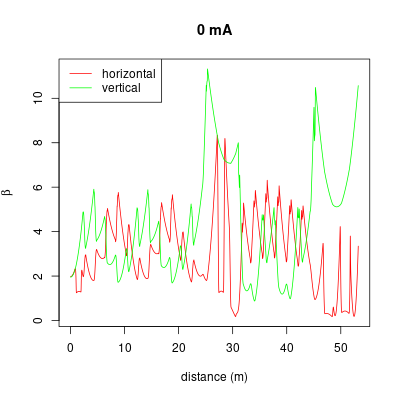}   
\includegraphics[width=0.49\textwidth]{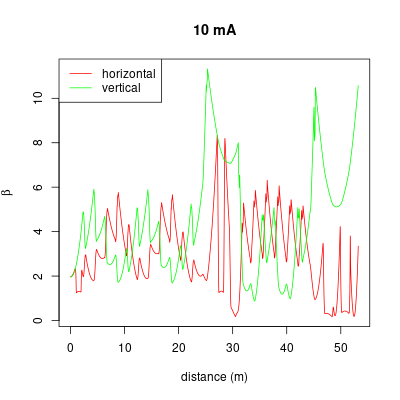}   
\includegraphics[width=0.49\textwidth]{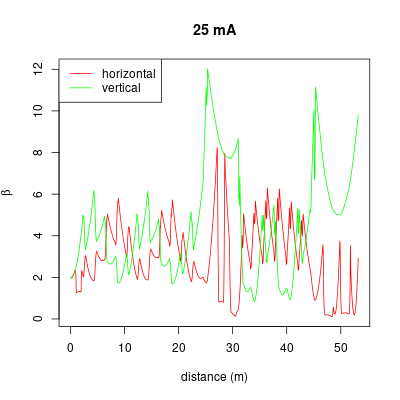}   
\includegraphics[width=0.49\textwidth]{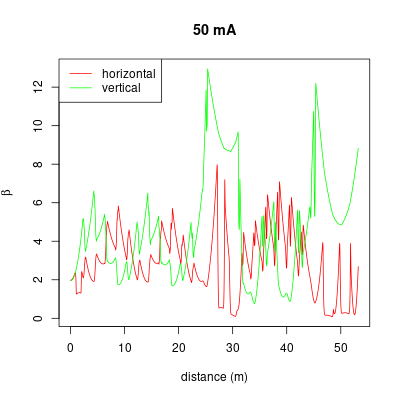}   

\caption{\label{fig:MEBTspacecharge} Effect of space charge on the  beta functions, according to OPAL. Plots show nominal currents of 0, 10, 25 and 50~mA.}
\end{centering}
\end{figure}

The effects of space charge are confirmed to be small at 60~MeV. 
Furthermore, the short cell size with repeated focussing/defocussing not only stops the beam size
in either direction from getting large but also stops it from getting small, which is where space charge effects would be enhanced. 
Figure~\ref{fig:MEBTspacecharge} shows the beta functions for different currents and there is no discernible effect at the nominal 10 mA beam current. Increasing the current to an unrealistic 50~mA starts to produce an effect, but even then this is small.

\begin{figure}
\begin{centering}
    
\includegraphics[width=\linewidth]{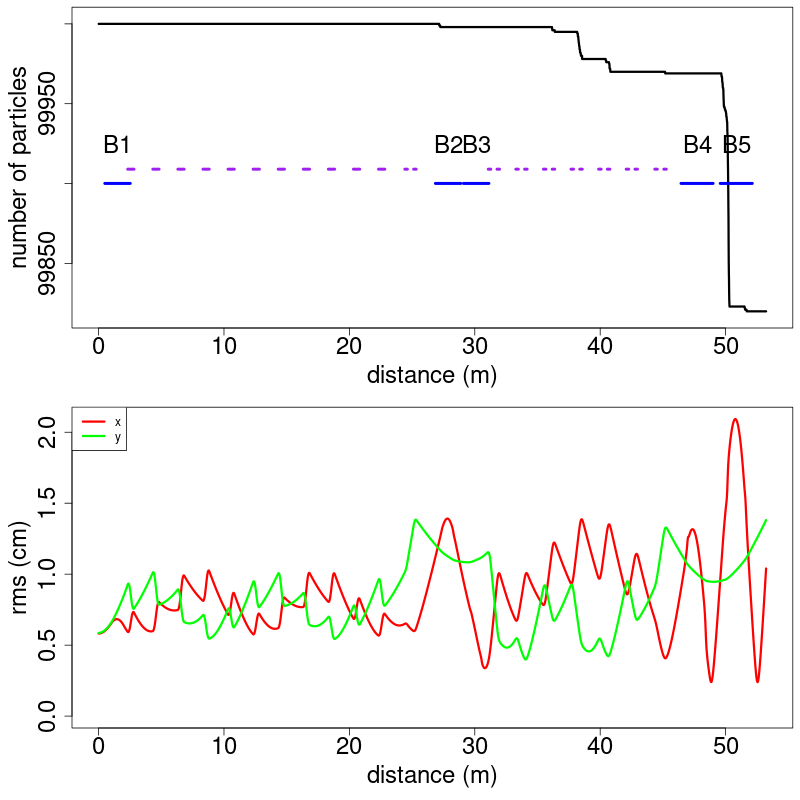}   \caption{\label{fig:MEBTEspreadlosses} Particle losses (above) and rms beam size (below) according to OPAL with a beam energy spread of 0.17 MeV.}
\end{centering}
\end{figure}

A more serious consideration comes from the energy spread of the beam: in the simulations above all particles have the nominal beam energy.
If they are given the quoted energy spread of 0.17 MeV
then the losses and rms are as shown in Fig.~\ref{fig:MEBTEspreadlosses}.
This suggests that around 0.2\% of the beam will be lost 
at the entrance to the final dipole. This should be studied further, when more details of the beam are known; if necessary it can be accommodated by increasing the bore of the final magnet.

\subsubsection{The Complete Transport Line:  Performance}

\label{sec:MEBTperformance}

\begin{figure}
\begin{centering}

\includegraphics[width=1.0\textwidth]{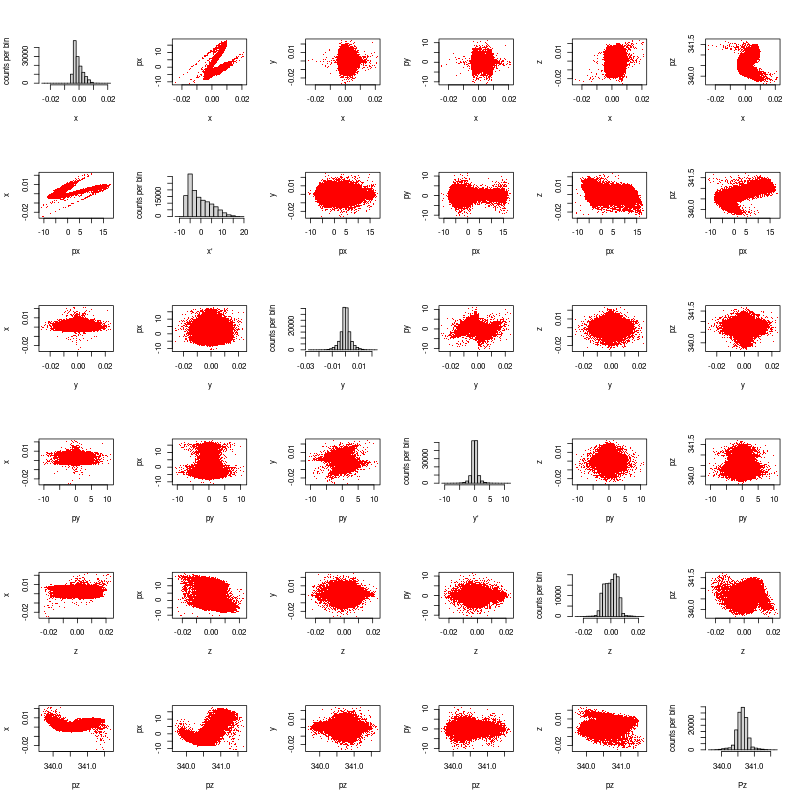}   \caption{\label{fig:MEBTinput} Distributions of particle properties $x,x',y,y',z,P_z$ at the entrance to the MEBT.
Lengths ($x,y,z$) are in metres, angles ($x',y'$) are in mrad, and $P_z$ is in MeV/c.}

\end{centering}
\end{figure}

The beam produced by the cyclotron is, unsurprisingly, not a simple Gaussian.  
Figure~\ref{fig:MEBTinput} displays the properties of a beam, according to an OPAL simulation of the cyclotron~\cite{Engebretson}.

\begin{figure}
\begin{centering}

\includegraphics[width=1.0\textwidth]{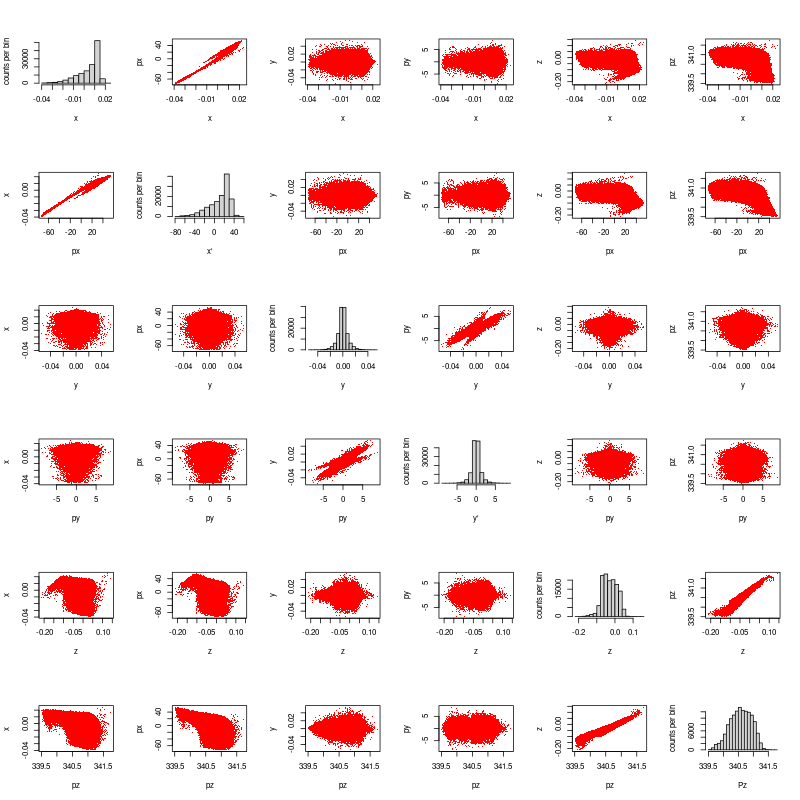}   \caption{\label{fig:MEBToutput} Distributions of particle properties $x,x',y,y',z,P_z$ at the end of the MEBT.
Lengths ($x,y,z$) are in meters, angles ($x',y'$) are in mrad, and $P_z$ is in MeV/c.}

\end{centering}
\end{figure}

\begin{figure}
\begin{centering}
    
\includegraphics[width=1.0\textwidth]{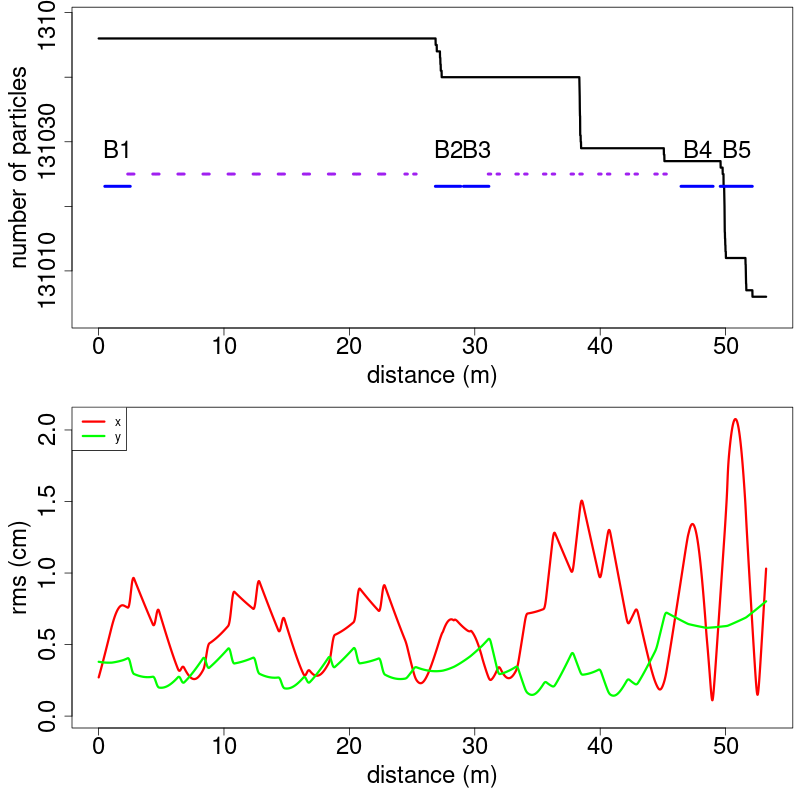}   \caption{\label{fig:MEBTfulllosses} Particle losses (above) and rms beam size (below) according to OPAL using the beam as provided by the cyclotron simulation.}
\end{centering}
\end{figure}

After tracking this beam through the MEBT (as designed for the simple Gaussian beam, above), the distributions are as shown in  Fig.~\ref{fig:MEBToutput}.  A dynamic plot showing the evolution of the bunch down the beamline, in 1 ns steps, can be seen at~Ref.~\cite{roger_website}. The losses for this simulated beam are as shown in Fig.~\ref{fig:MEBTfulllosses} - compare with Figs.~\ref{fig:MEBTlosses} and \ref{fig:MEBTEspreadlosses}.
They are still acceptable, with 40 particles lost out of 131046 - a rate of 0.03\%.
It is noteworthy that in this scenario the 
transverse spread is significantly greater than the vertical, which suggests that the beamline 
could be optimized further with a smaller beampipe or fewer quadrupoles,  which would lower the cost without increasing the losses above their present small values.

\subsubsection{The Beam on the Target}

\begin{figure}
\begin{centering}
\includegraphics[height=0.3\textheight] {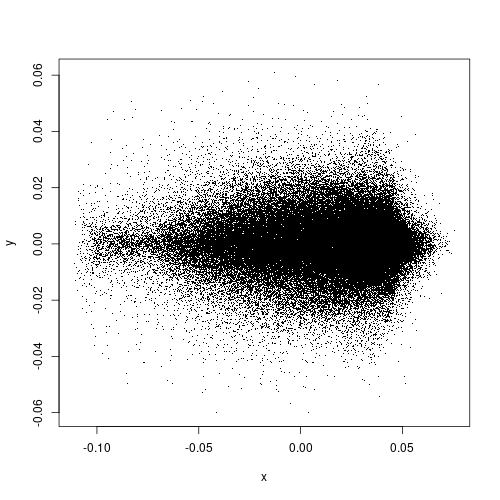}
\includegraphics[height=0.3\textheight] {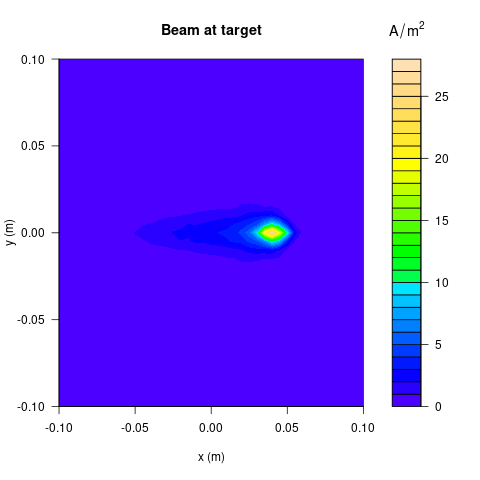}
\includegraphics[height=0.3\textheight] {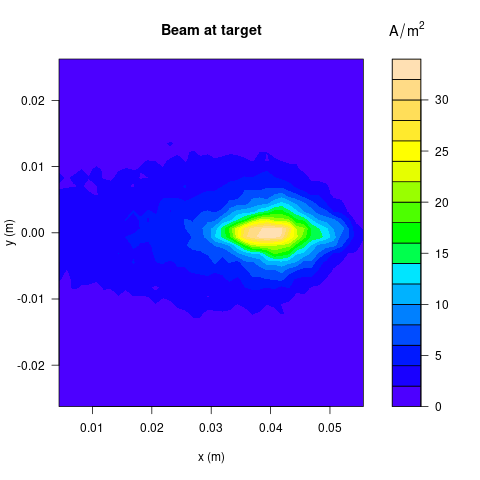}
\includegraphics[height=0.3\textheight] {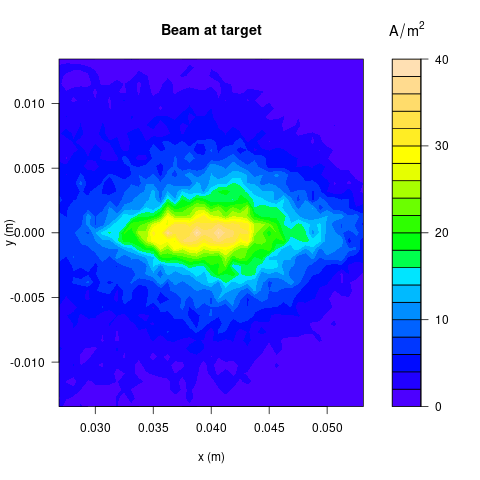}
\caption{\label{fig:MEBTtarget} The beam at the target location, showing a scatter plot and three contour plots at different magnifications. Plots are in the (x,y) plane, perpendicular to the beam-propagation direction.}
\end{centering}
\end{figure}

The beamline simulation ends 1 m before the actual target position, but Fig.~\ref{fig:MEBTtarget} shows the beam shapes extended to the location of the target. These are all (x,y) plots, the top left is the scatter plot output by the OPAL simulation, while the other three show rough contour maps of the beam at different magnifications. 
The scale of the top two figures are roughly equal.
The top right scale encloses the outer diameter of the target, 20 cm, while the lower two give more details of the falloff of the beam density.  
Note, the details of this distribution are dependent the phase space parameters at the beginning of the transport line (at the cyclotron exit).  In addition, no attempt has been done in this simulation to spread the beam over the full target face, to optimize power deposition.  Spreading the beam can be done with adjustments of the last quadrupoles in the beamline, possibly necessitating increasing the bore in the last bending magnet.  Undoubtedly as well, collimation will be needed at the outer radii of the target to collect the wings of the beam distribution.

The beam spot is centered vertically, but is some 4 cm off-centre horizontally. This can readily be 
adjusted by changing the field strength of the final magnet.
This can be included as part of the beam setup, and 
should be monitored and adjusted as necessary during running.

\subsection {Transport summary}

The MEBT design is straightforward, as expected and low ($<1~$W/m) losses are achievable.
We will use MADX to optimise and OPAL simulations for detailed verification.
The beam on target will be a Gaussian ellipse with dimensions around 1 cm, so it will need to be defocussed to avoid target damage.

The current lattice is an existence proof for the final design,
which will need another round of optimisation to reduce $\beta$ values 
upstream of both dipole magnet pairs, and the  horizontal spread in the latter part of the beamline, 
 to accommodate $\sigma_E$ (or an increase in the  beam pipe size in the final few metres).
We will also explore  lattice designs using fewer quadrupoles.

\subsection{Presentation of Beam to the Target}

Once beam is brought to the vicinity of the target, it must be shaped to maximize the effectiveness of the experiment.  The main concerns are keeping fast-neutron background in the $\nu$EYE detector to the lowest reasonable level, and enabling the longest possible lifetime of the target primarily in consideration of thermal stress and heat dissipation. 

\subsubsection{Beam Orientation with Respect to the Antineutrino Detector}\label{BeamOnTarget}

The orientation of the beam on target can significantly affect the fast neutron flux directed towards the $\nu$EYE detector at Yemilab.  We see in Fig.~\ref{fig:NeutronSpectra} that the flux of neutrons in the backward direction is significantly reduced, both in maximum energy as well as total flux.  It is desirable then to bend the beam through $180^\circ$ so that when it strikes the target it is heading away from the detector (cf. Fig.~\ref{fig:MEBT System}, bottom).  This will reduce the amount of shielding required. Specifications for the two $90^\circ$ dipoles, required to orient the beam in the backward direction, have been given above in Section~\ref{dipoles}.  

Notably, there will be a small loss of the antineutrino signal in this configuration.  Antineutrino production is isotropic, so the signal is related to the solid angle subtended by the detector from the target area.  This goes roughly as $1/r^2$, so the placement of the two $90^\circ$ magnets requires that the source is moved back by about 1 meter. 
However, the harder forward-going neutron spectrum would require at least an extra meter of steel to achieve the same degree of shielding.

There is a tradeoff, however, in that there is now a hole in the target shielding: the vacuum beam pipe, that points directly towards the detector.  
This should not be a significant issue, again due to the vastly reduced fast neutron flux emitted in the backward direction.  
An additional advantage of this orientation is that servicing the target torpedo is greatly simplified, it can just be pulled out into the hall from the back of the target shielding block (see Fig.~\ref{fig:target-change}.)

\subsubsection{Beam Distribution on the Target Face}

The total power of the IsoDAR beam is 600 kW.  This must be managed carefully to allow survival of the target.  Specifically, the shape of the beam, and how it is distributed across the target, needs to be well specified.

\begin{figure}
\begin{centering}

\includegraphics[width=\textwidth]{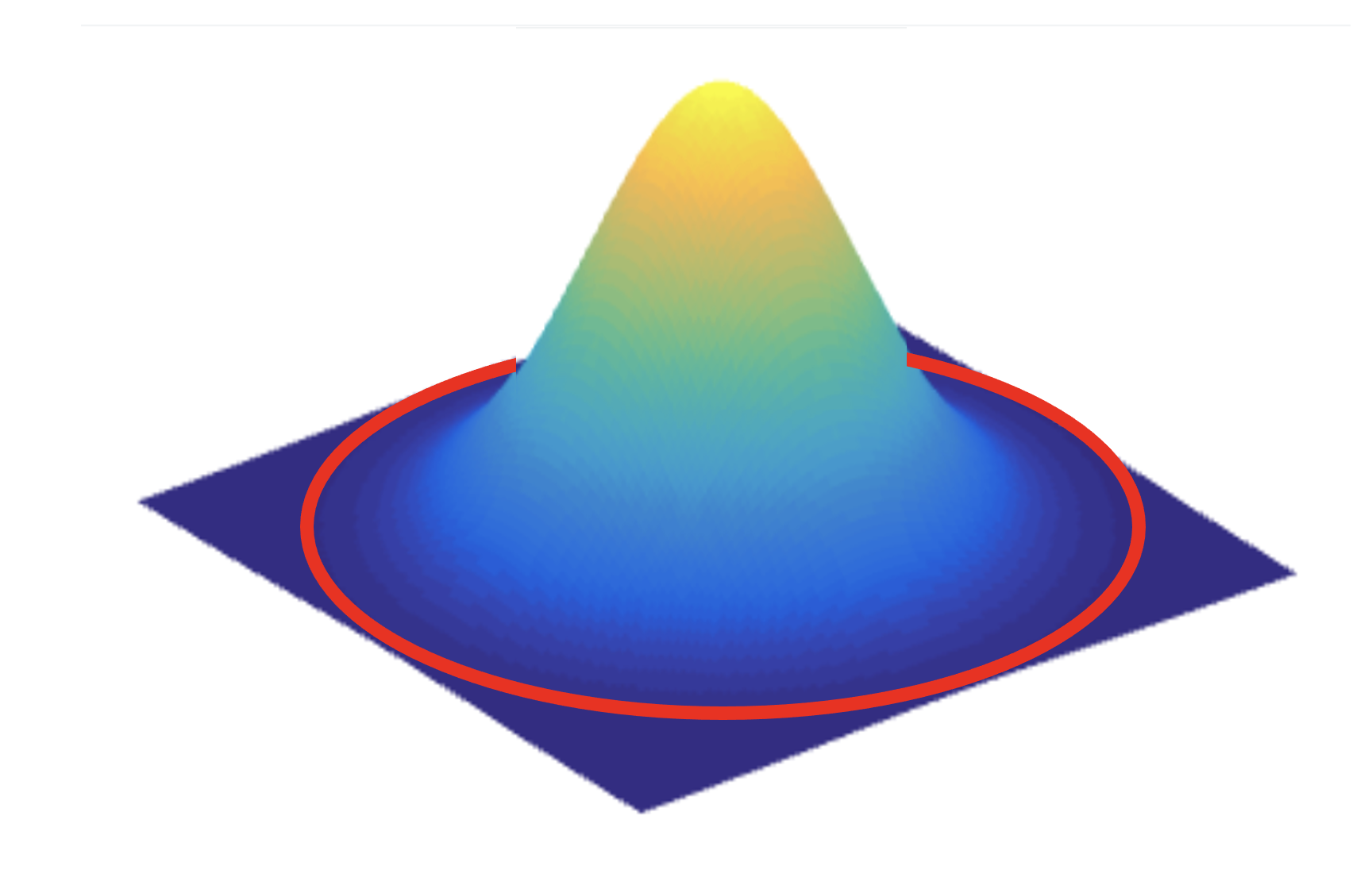}    
\caption{\label{fig:GaussianBeam} This figure shows an ideal gaussian beam distribution on target face, with a sigma of 3.3 cm.  The red circle represents the 20 cm outer diameter of the torpedo. The actual beam will not look like this, but the response of the target to the maximum power density and falloff are important in establishing acceptable parameters for the actual beam shape.
}
\end{centering}
\end{figure}

Thermal analyses presented in the following chapter (See Fig.~\ref{fig:Temp-1P-GaussB}) indicate that a Gaussian beam with a sigma of 3.3 cm (with negligible tails beyond 8.5 cm radius) can be adequately handled with a coolant flow of 750 gpm (50 lps).  This beam profile, shown in Fig.~\ref{fig:GaussianBeam} has a peak power deposition of about 9 W/mm$^2$ at the center of the outer shell, and generates a temperature of about 500 K on the outer surface of this first hemisphere.  

Thermal gradients in the beryllium shell material, related to beam profile gradients as well as heat conduction to cooled surfaces, can have severe consequences for the structural integrity and lifetime of the beryllium shells. 
For beryllium, the maximum allowed stress is 36 ksi (thousand pounds per square inches) or 250 MPa (mega-pascals).  For the case of the gaussian beam, see Fig.~\ref{fig:Tstress-boil-GaussB}, the maximum stress is 30 ksi (200 MPa), within the acceptable limit.

In addition, we see from 
Fig.~\ref{fig:Tstress-flatB}, that beam must not be allowed to hit the lateral edges of the hemispherical target, as water flow in this area is irregular, providing little cooling in these regions.

The key points from these analyses regarding the shape of the beam are:

\begin{itemize}
    \item A peak power deposition of 9 W/mm$^2$ can be handled with adequate cooling flow.
    \item Thermal gradients in the beryllium shell material related to beam profile gradients and heat flow to cooling surfaces must be carefully calculated, to ensure the shape of the beam hitting the target does not produce stresses exceeding the 250 MPa value at any point.
    \item Beam should not extend beyond 8.5 cm from target center.
\end{itemize}

The beam profile shown in Fig.~\ref{fig:MEBTtarget}
is very different, in both shape and size, from that of the ideal of Fig.~\ref{fig:GaussianBeam}, and the design must be modified accordingly. There are three possibilities for achieving this, in order of preference:
\begin{itemize}
    \item Modifying the final quadrupoles (Q37 and Q38). This is a simple solution, if it can be achieved, though the size of the bunch in the preparation section will increase, which may require an increased beam pipe diameter in this final section.
    \item Inserting one or two new quads between the two dipoles, or after the final dipole. As the distance to the target is short, they would have to be quite strong.
    \item Using a magnetic wobbler to spread the beam spot across the target. This has the advantage that any desired distribution can be produced,  but
    adds significant complexity, both in the need for fast sweeping magnets, but also in the time variation of the heat pattern swept across the target face.
\end{itemize}

The distribution of Fig.~\ref{fig:MEBTtarget} has a large tail and a small core, and expanding the core will expand the tail. So inevitably, for all three options, a fraction of the beam will miss the target.  
A well-cooled collimator, probably a circular disk with an inner diameter of 18 cm, should be mounted just upstream of the target, to protect the edges of the target.  Segmenting this ring into four sectors wired as current monitors would allow for steering the beam onto the horizontal and vertical center of the target.

An important requirement will be having appropriate instrumentation with adequate spatial resolution for monitoring the target temperature.  This will be critical for tuning the last beam-optical elements as well as for preventing excessive temperatures from destroying the target. As an example, the FRIB target-monitoring optical IR system is a good model~\cite{FRIB3}.

\subsection{Conclusions}

While straightforward, transporting the beam from the cyclotron to the target still presents challenges, and we have identified in the text above the areas needing special attention. 

In the stripping area, the foil itself requires careful optimization, and though lifetimes and efficiencies will be acceptable with our baseline specifications, there will be undoubtedly room for lifetime improvements with future advances in stripping foil technology.

The collimator defining the entrance to the main MEBT transport section must be carefully designed, as some beam loss can be expected on this collimator. Adequate cooling and radiation shielding will be provided.  

The transport section is straightforward, and beam loss is not anticipated, though confirmation is needed that proper matching of the extracted beam from the cyclotron with the first elements of the MEBT has been achieved.  
The need for halo scrapers, and shielding enclosures along the beamline has not been seen, but again further confirmatory studies will be conducted on this question.

The final shaping of the beam on the target requires careful attention to ensure maximum allowed power density on any spot of the target is not exceeded.  We have determined that the target can handle the 600 kW of beam power provided the shape of the beam is reasonably gaussian, with a 3.3 cm sigma, and any beam shape that does not exceed the highest power density or gradient falloff of this gaussian will be acceptable.  
Careful thermal monitoring of the target face must be performed to ensure expected target longevity.

\subsection{Risks and Mitigation}

\textbf {\noindent Risk: Stripping foil lifetime is shorter than anticipated.} We are exploring new intensity levels with our \htp beams, and it is possible that the survival of stripping foils may be less than anticipated.

\textit{Mitigation:  Foil type, thickness, mounting technique, and most importantly beam size going through the foil will affect foil lifetime.  The most straightforward mitigation is to not focus the beam tightly, but this will cause emittance growth of the beam.  Adequate aperture in the transport line should be provided to accommodate for any increase in beam size needed to ensure foil lifetime is suitably long (greater than a few days).  Reducing foil thickness to an optimal level will also contribute to prolonging foil lifetime, by decreasing the heat deposition in the foil.}

\bigskip 
\noindent \textbf {Risk: High beam losses in transport line.}  The beam carries a large amount of power and small losses could have strong effects. These could be caused by varying conditions in the cyclotron and thus differences in the beam structure.

\textit{Mitigation: Provide many beam loss monitors (they are not expensive) and monitor continuously. Adjust quad currents and be prepared to install collimators if required.}

\bigskip
\noindent \textbf {Risk: Beam damages target.}  If the beam is too focussed it may damage the target.

\textit{Mitigation:  
Further studies of beam defocussing in the final few meters. Monitor the temperature distribution of the target and be prepared to switch off the accelerator if necessary.}

\bigskip 
\noindent \textbf {Risk: Components and vacuum pipe lifetime shortened by radiation damage.}  
Beam losses - especially unplanned trips - will cause radiation damage to essential components such as magnet cooling pipes.

\textit{Mitigation:  
Through understanding of all system components, including the basic low-technology ones, their possibilities for 
damage, and a rolling replacement program.}

\clearpage
\section{Target Design}

\subsection{Introduction to the Two Main Target Assemblies}
The IsoDAR target is divided into two main components.  The central structure inside the beam vacuum enclosure is a water-cooled beryllium ($^9$Be) target which absorbs the proton beam from the cyclotron and emits neutrons.  The neutrons enter a surrounding pressure vessel filled with a mixture of natural beryllium ($^9$Be) and 99.99\% isotopically-pure $^7$Li, where the $^7$Li is then transmuted to $^8$Li via neutron capture, which  beta-decays in less than one second to $^8$Be and emits the desired electron antineutrino. 
Extensive studies with Geant4 have established the concept and basic parameters for the target and sleeve~\cite{bungau_optimisation_2013,bungau_neutron_2015,bungau_optimizing_2019}. However, sophisticated engineering work, described in these next two chapters, has been required to ensure the target can handle the extremely high beam power with the most efficient use of the expensive beryllium and highly-enriched $^7$Li inventories.
This chapter describes the structure of the water-cooled target (also referred to as the ``torpedo"), the following chapter provides the details of the sleeve surrounding the target. 

\subsection{Water-cooled Beryllium Target Mechanical Design}
\begin{figure}
\begin{centering}
\includegraphics[width=\textwidth]{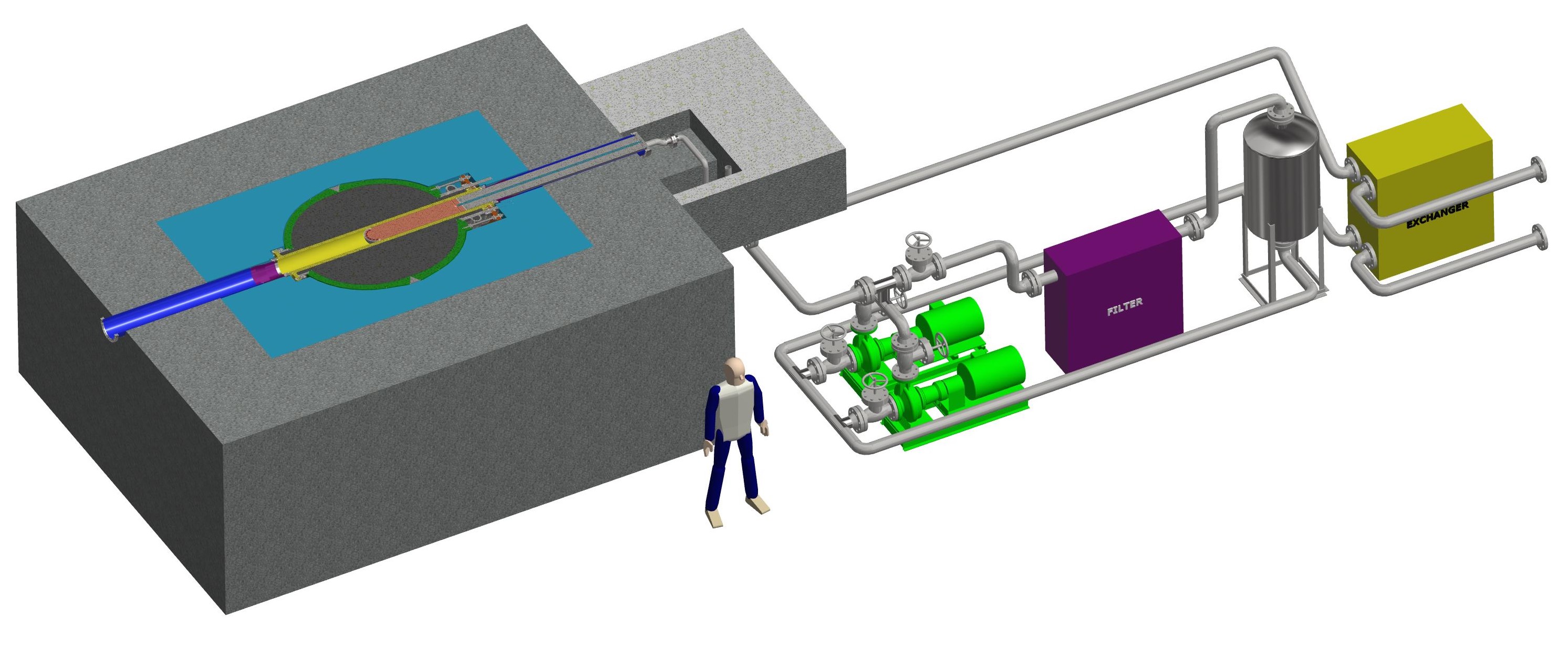}    \caption{\label{fig:ISO21-01} Horizontal section view showing a preliminary design for the water-cooled target recirculation system. The recirculation system cools and filters the water passing through the water-cooled target.}
\end{centering}
\end{figure}

\begin{figure}
\begin{centering}
\includegraphics[width=\textwidth]{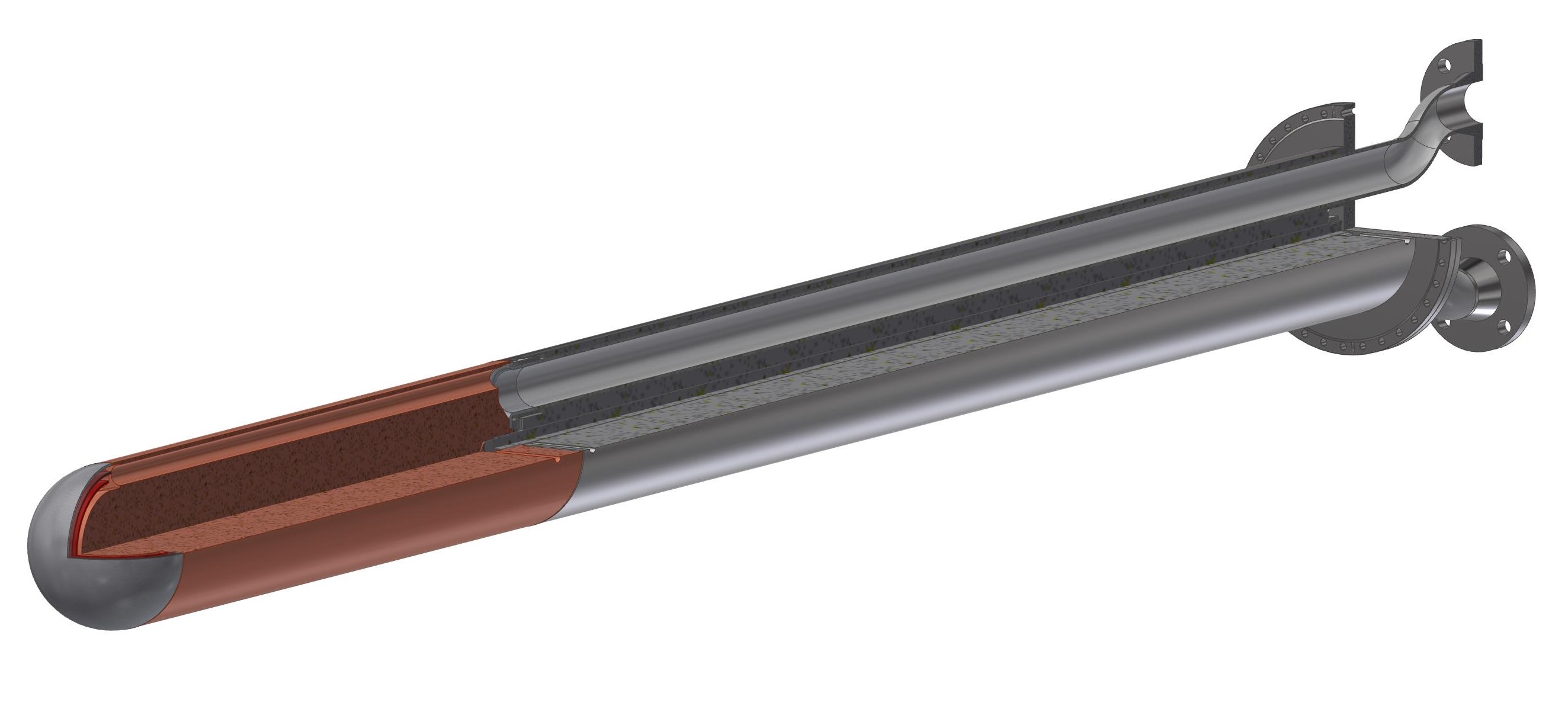}    \caption{\label{fig:WT1-2} Three-quarter section view of the target assembled with steel shielding and water pipes. The diameter of the target is 8 inches (203.2 mm) and the overall length is 127.53 inches (3.24 m).}
\end{centering}
\end{figure}

\begin{figure}
\begin{centering}
\includegraphics[width=\textwidth]{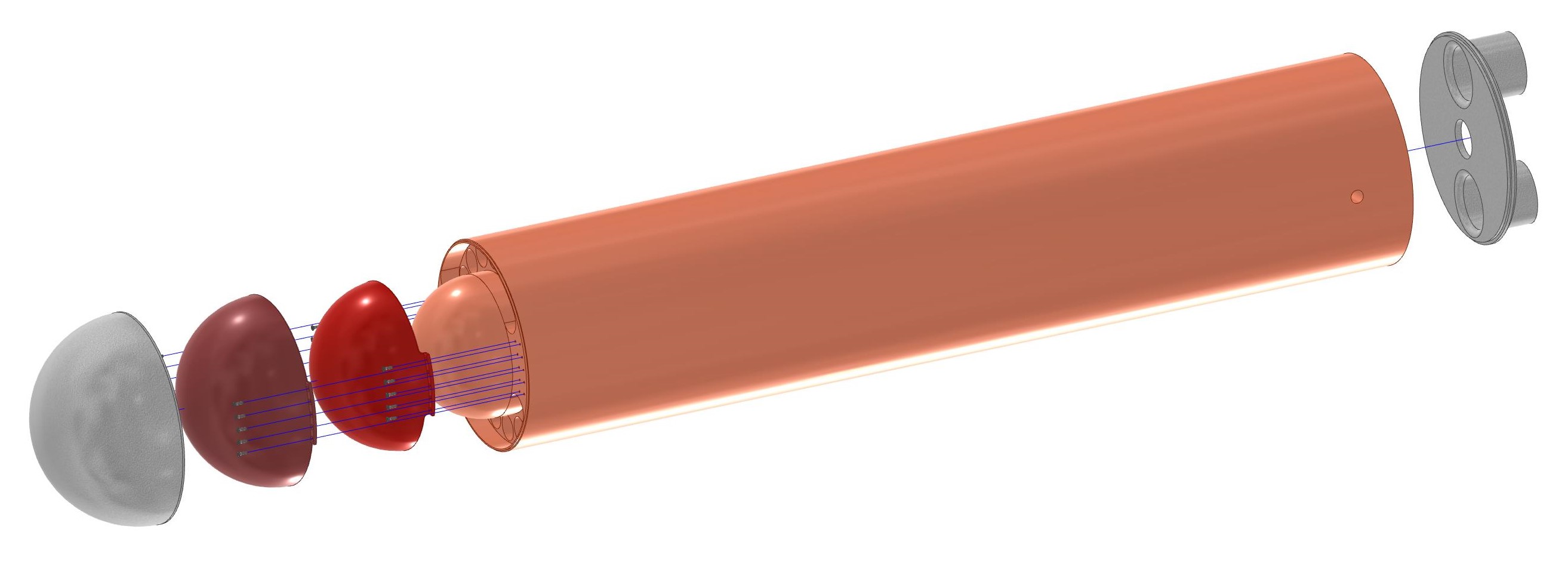}    \caption{\label{fig:WT6-4} Exploded view of the beryllium parts of the target.}
\end{centering}
\end{figure}

The basic design principle of the water-cooled target is to absorb the 600 kW proton beam 
power while ensuring the temperature and thermal stresses in the beryllium remain within acceptable limits.
To accomplish this, the target features three shells of beryllium separated by heavy water that flows between them.  The shells are each 0.120 inches (3.05 mm) thick with 0.274 inches (6.95 mm) space between them for water flow. The outermost shell is the interface between the beam vacuum space and the internal water space.  This shell is critical in that it cannot fail. Figure~\ref{fig:ISO21-01} shows a conceptual design of the components needed to recirculate between 500 and 750 gallons per minute of heavy water through the target. Figure~\ref{fig:WT1-2} shows the overall layout of the target with a cutaway section showing the water channel.   Figure~\ref{fig:WT6-4} shows a close-up exploded view of all the beryllium components of the target.  The inner two hemispherical shells of beryllium are attached to the body of the target with bolts.  The outer shell is brazed or e-beam welded to the salmon-colored body.

\begin{figure}
\begin{centering}
\includegraphics[width=\textwidth]{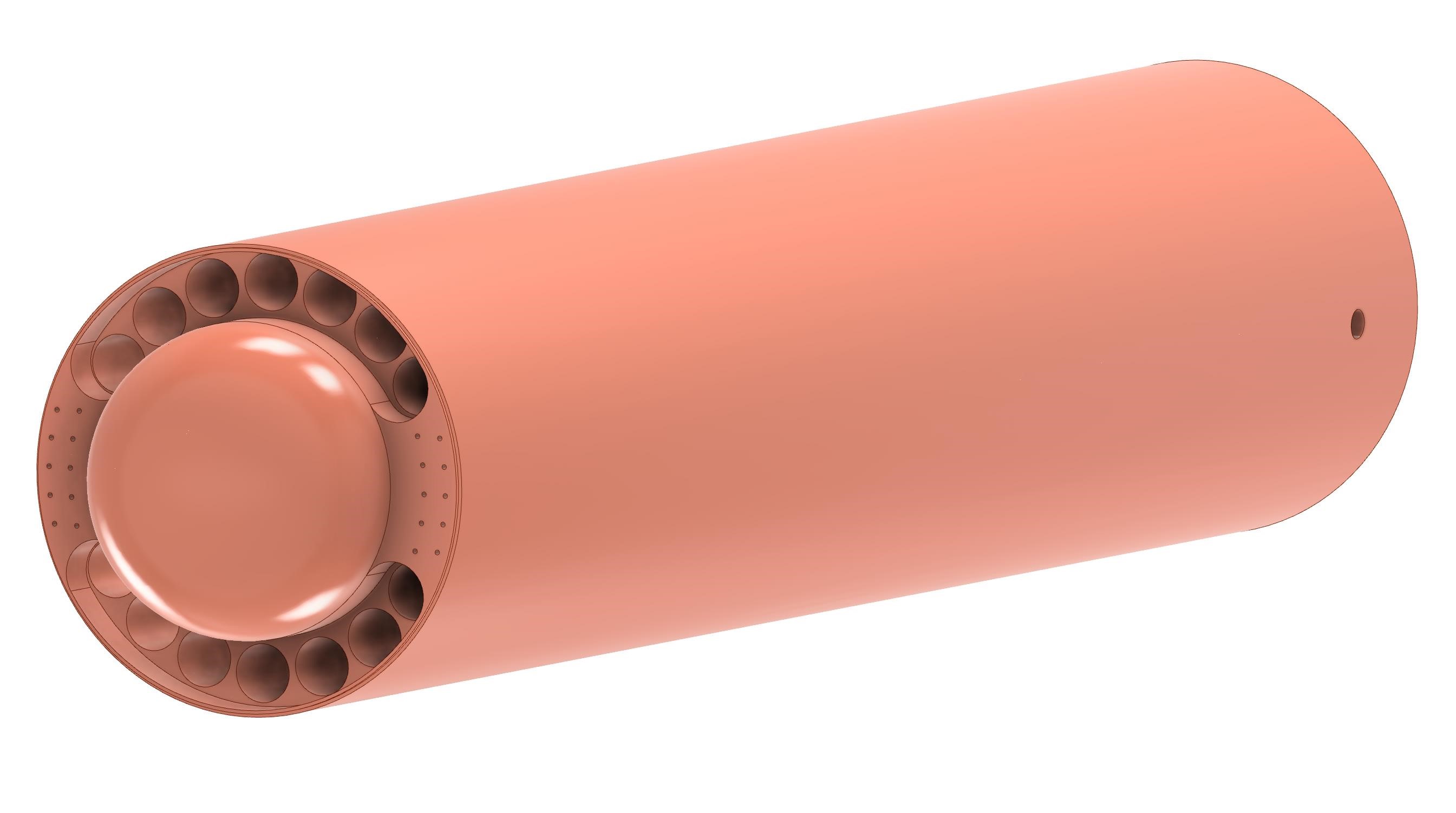}    \caption{\label{fig:GDB-1} View of the gun drilled beryllium body of the target showing the water channels.}
\end{centering}
\end{figure}

\begin{figure}
\begin{centering}
\includegraphics[width=\textwidth]{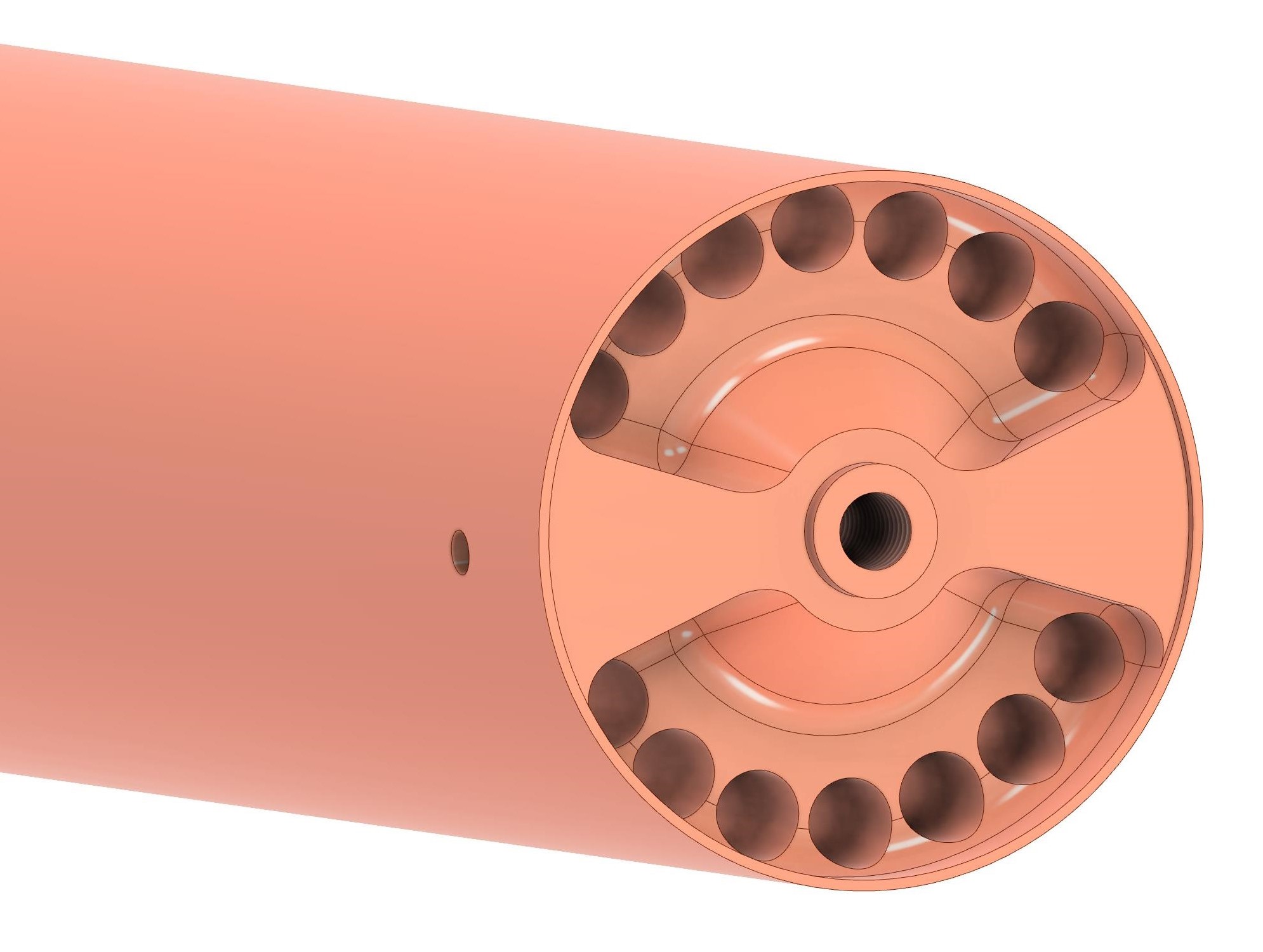}    \caption{\label{fig:GDB-2} View of the downstream end of the gun drilled body of the target.}
\end{centering}
\end{figure}

\begin{figure}
\begin{centering}
\includegraphics[width=\textwidth]{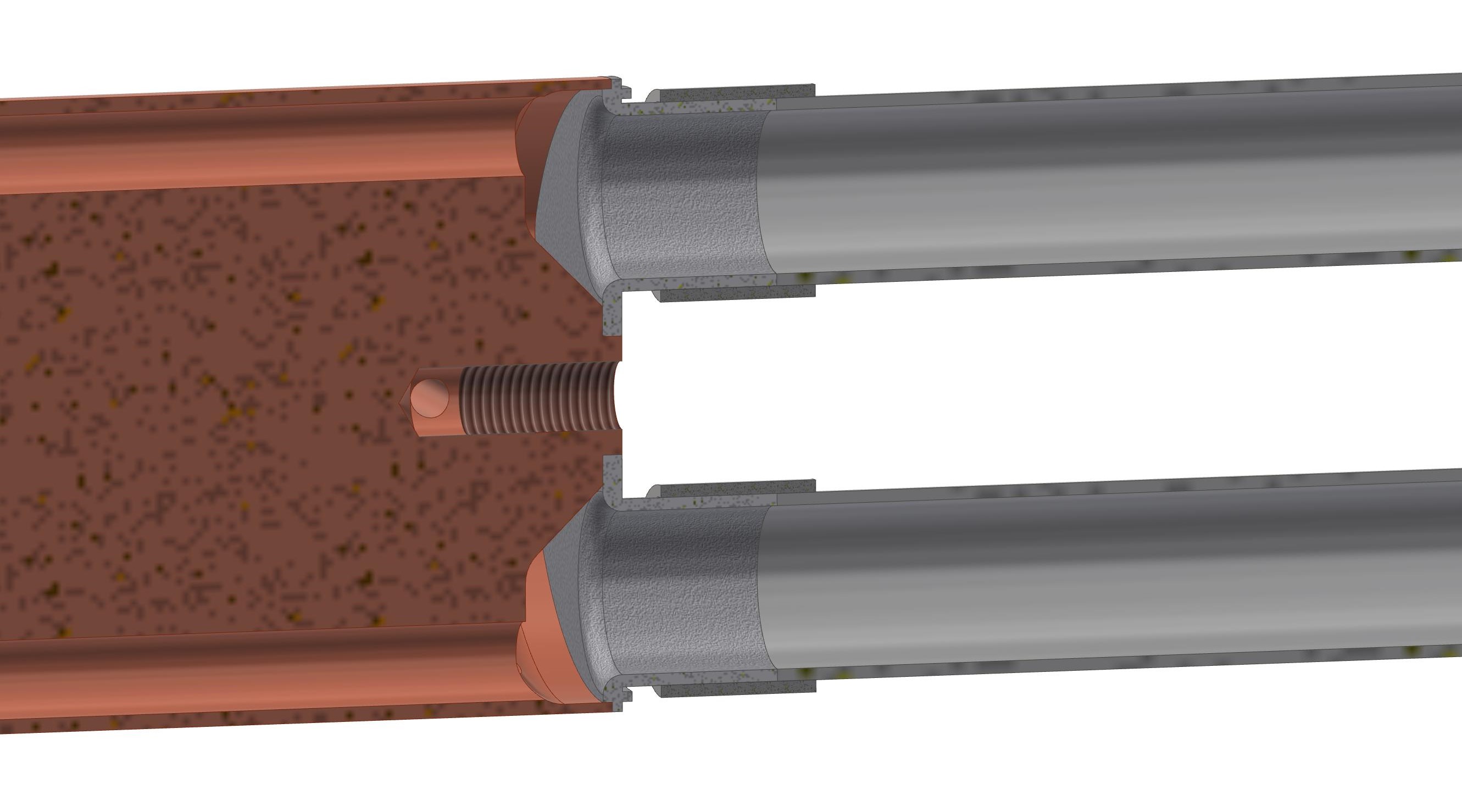}    \caption{\label{fig:WT6-3} Section view close-up of the brazed connections between stainless water pipes and the beryllium tube stubs on the downstream end of the target.}
\end{centering}
\end{figure}

Much of the design of the water-cooled target was determined by the manufacturing capabilities of Materion~\cite{materion}, the only US company to supply beryllium in manufactured shapes.  Figures~\ref{fig:GDB-1} and~\ref{fig:GDB-2} show details of the beryllium body with holes gun drilled through its length to supply water to the hemispherical shells.  The design has been iterated several times with Materion to get to this point.  Figure~\ref{fig:WT6-3} shows a close-up section view of the connection between the stainless steel pipes brazed to the beryllium end plate.  The exact manufacturing details of this connection remain to be worked out with Materion to avoid galvanic corrosion between the stainless and beryllium.

\begin{figure}
\begin{centering}
\includegraphics[width=\textwidth]{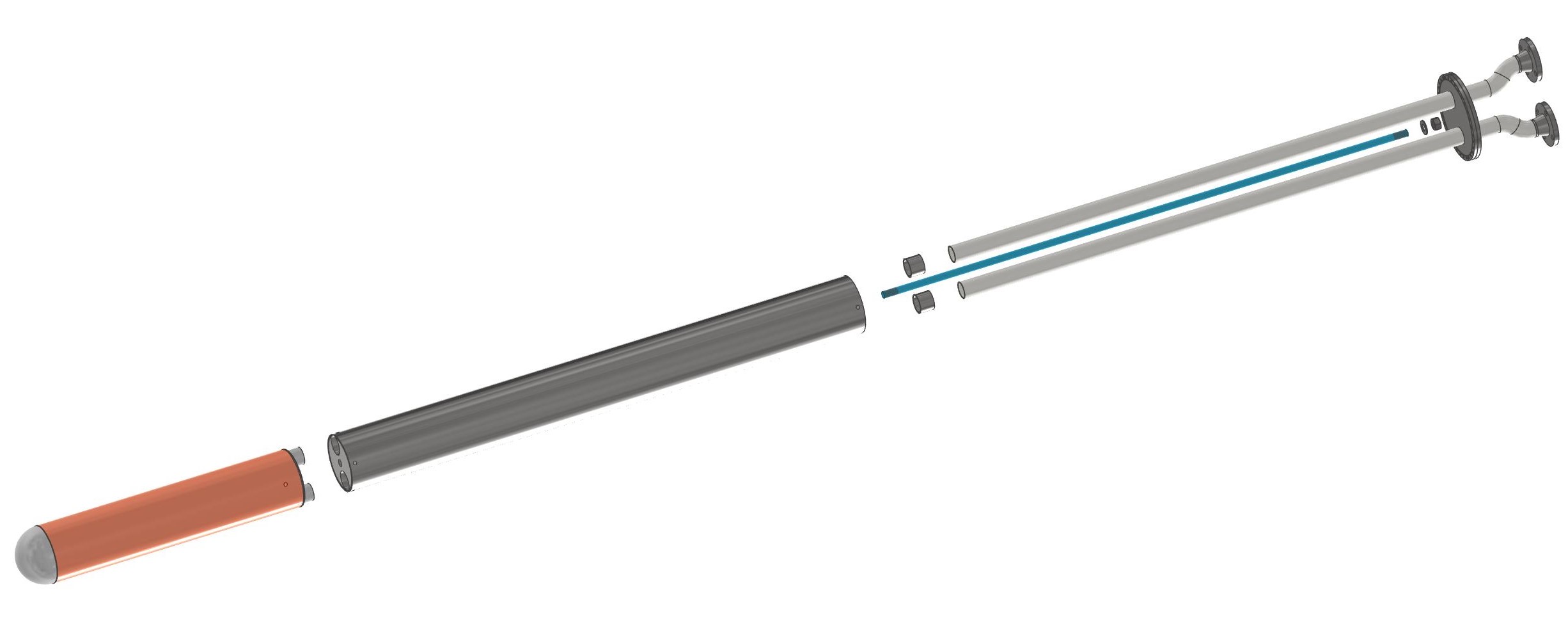}    \caption{\label{fig:WT6-6} Exploded view of the major components of the water target showing the beryllium section on the left, the downstream steel shielding, the water pipes, and vacuum flange.}
\end{centering}
\end{figure}

\begin{figure}
\begin{centering}
\includegraphics[width=\textwidth]{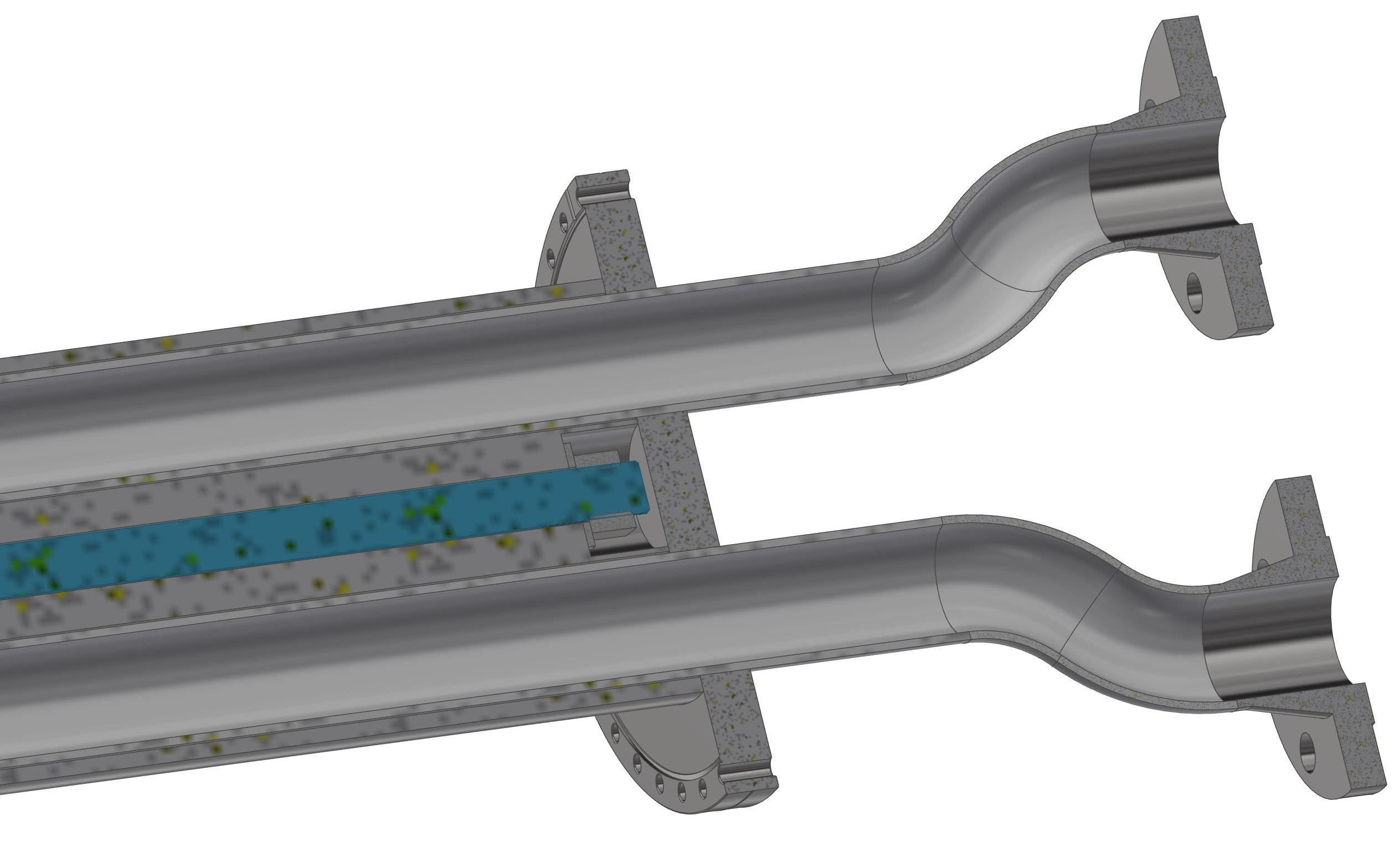}    \caption{\label{fig:WT6-5} Section view of the downstream end of the water target.  The turquoise part is the long threaded rod that draws the steel shielding against the beryllium section of the target.  The stainless pipes are vacuum-tight welded to the 12 inch ConFlat flange.  Structural welding is done between the steel shielding and the ConFlat flange.}
\end{centering}
\end{figure}

The gray material immediately downstream of the beryllium body of the target is a cylindrical mass of shielding steel to prevent excessive neutron flux out of the downstream end of the target where the water recirculation components are.  This steel shielding is pulled tight against the beryllium body by a long threaded rod, shown in turquoise in Figs.~\ref{fig:WT6-6} and~\ref{fig:WT6-5}.  Figure~\ref{fig:WT6-3} shows the tapped hole in the beryllium body that the long threaded rod is assembled to. 

\subsection{CFD Analysis of the Water-cooled Target}

\begin{figure}
\begin{centering}
\includegraphics[width=\textwidth]{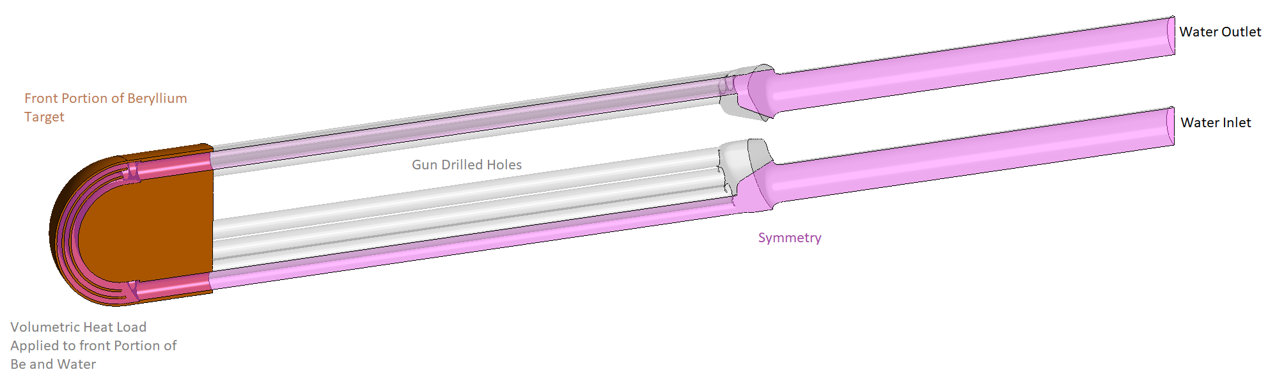}    \caption{\label{fig:CFD01} View of the half-symmetric model of the water cooling channel in the target for CFD analysis.}
\end{centering}
\end{figure}

\begin{figure}
\begin{centering}
\includegraphics[width=\textwidth]{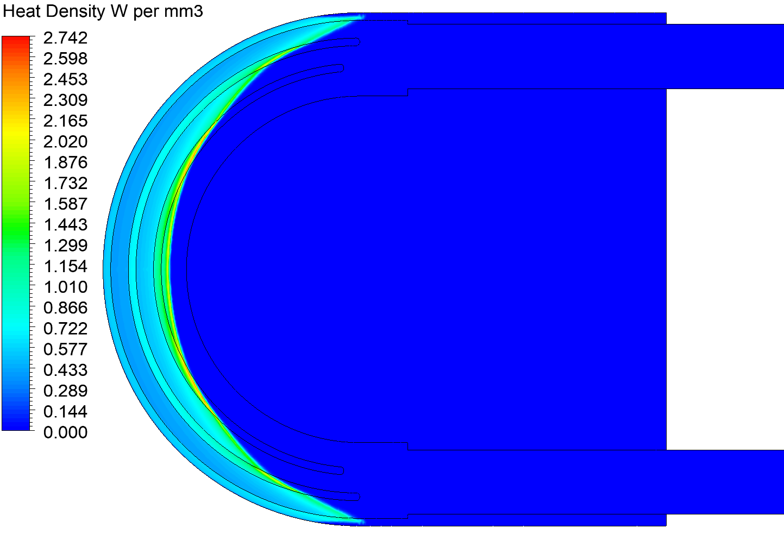}    \caption{\label{fig:FlatB} Heat deposition from a flat beam whose diameter is almost the same as the target.}
\end{centering}
\end{figure}

Figure~\ref{fig:CFD01} shows the half-symmetric model of the water flow channels through the target and the upstream beryllium hemispheres.  Several different Computational Fluid Dynamics (CFD) simulations were studied with different beam profile depositions onto the target, including Gaussian with $\sigma = 3.3$~cm and FWHM (full-width at half maximum) $= 7.8$~cm,
 and uniform.  Figure~\ref{fig:FlatB} shows energy being deposited from the beam with a uniform power distribution by radius.  The beam in this simulation was almost 8 inches (200 mm) in diameter. 

\begin{figure}
\begin{centering}
\includegraphics[width=\textwidth]{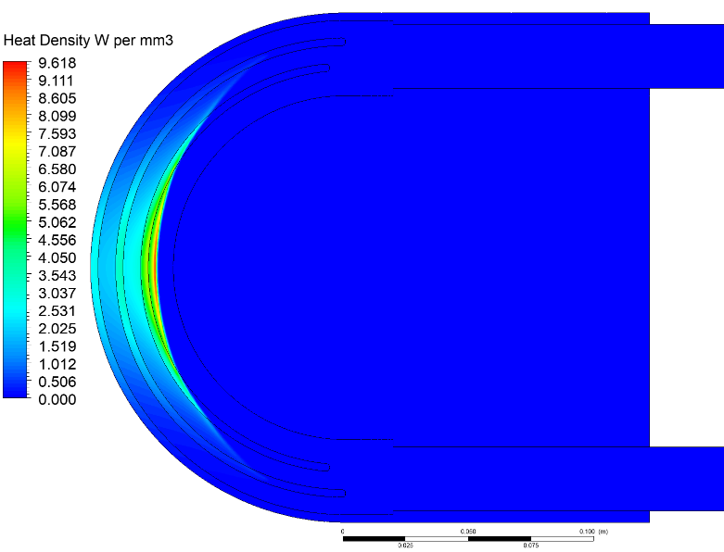}    \caption{\label{fig:GaussB} Heat deposition from a Gaussian beam collimated to less than seven inches in diameter.}
\end{centering}
\end{figure}

Figure~\ref{fig:GaussB} shows the contours of the heat power deposited in the water-cooled target from a Gaussian beam ($\sigma = 3.3$~cm) that is truncated to be less than 7 inches (178~mm) in diameter.

The CFD analyses performed on the water-target all used temperature dependent materials properties for the beryllium and heavy water (\dto).  The inlet temperature of the water was set to 60$^{\circ}$~F (15.56$^{\circ}$~C).  The inlet flow rate for the single-phase \dto was initially 500 gallons per minute (GPM), but was increased to 750~GPM to suppress boiling.  A two-phase analysis was attempted, but the program could not converge the two-phase equation of state for \dto, so \hto was used instead at 550~GPM, (to get an equivalent mass flow and Reynolds number as 500~GPM \dto.)  The two-phase simulation also suffered from inconsistencies in handling the properties of the vapor so the results for the two-phase analysis are not reliable.

An important figure of merit in these studies, Critical Heat Flux is the condition in which a continuous layer of vapor separates the surface being cooled from any liquid.  In this condition, the convection heat transfer coefficient goes down sharply and the temperature of the body being cooled goes up rapidly.  Destruction of the component often follows. The Critical Heat Flux from correlations in Ref.~\cite{liu2021critical} is higher than the maximum heat flux, a very important and promising result.  The Critical Heat Flux was calculated to be $14.6~\mathrm{MW/m}^2$, while the maximum heat flux is about $9~\mathrm{MW/m}^2$ in the Gaussian beam distribution model.  

\begin{figure}
\begin{centering}
\includegraphics[width=\textwidth]{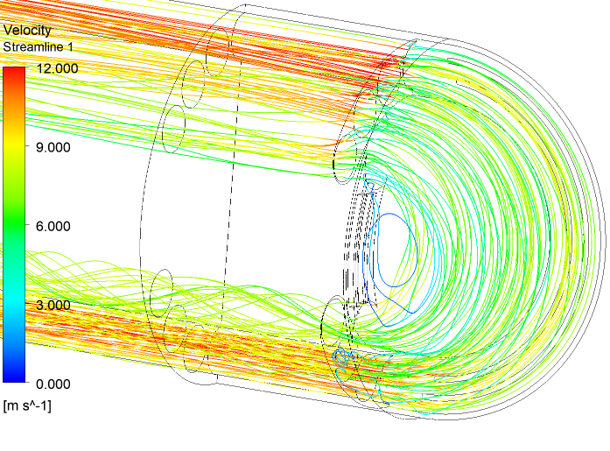}
\caption{\label{fig:flowV-flatB} Fluid velocity for 500 GPM flow in the flat beam profile simulation.}
\end{centering}
\end{figure}

\begin{figure}
\begin{centering}
\includegraphics[width=\textwidth]{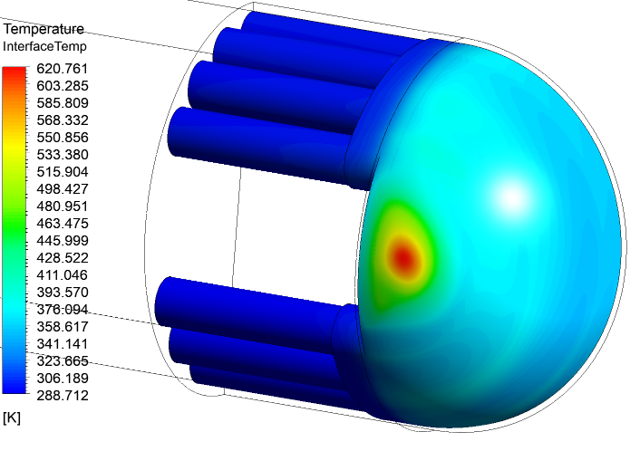}
\caption{\label{fig:Intemp-flatB} Interface temperature for 500 GPM flow in the flat beam profile single-phase simulation.  The interface temperature goes up to 621 K (347.8$^\circ$C), well above the boiling point, because of low flow in eddies at the sides of the hemispheres.}
\end{centering}
\end{figure}

\begin{figure}
\begin{centering}
\includegraphics[width=\textwidth]{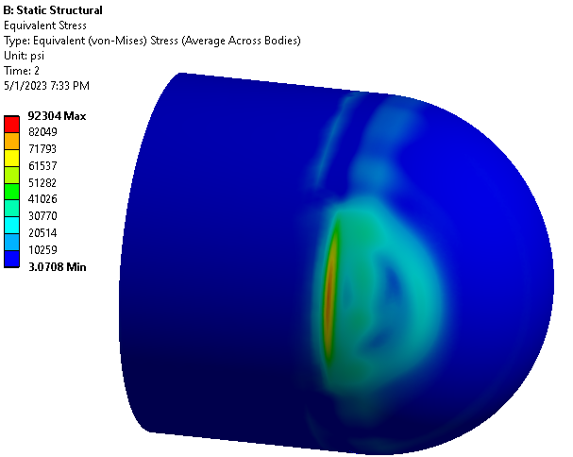}
\caption{\label{fig:Tstress-flatB} The thermal stress in the beryllium for the flat beam profile two-phase simulation.  The themal stress exceeds the allowed stress.}
\end{centering}
\end{figure}

The flat beam power profile gave the important result that a proton beam that is too large in diameter deposits energy in areas of the hemispheres that are only cooled by low velocity eddies of water.  Figure~\ref{fig:flowV-flatB} shows the velocity profile of the \dto, while Fig.~\ref{fig:Intemp-flatB} shows the interface temperature between the water and beryllium.  The interface temperature goes up to 621 K (347.8$^\circ$~C), well above the boiling point.  The simulation was changed to two-phase flow with water, but the thermal stress in the beryllium was still too high from the hot spot.  Figure~\ref{fig:Tstress-flatB} shows the thermal stress in the beryllium.  The allowable thermal stress is 36~ksi, but the maximum stress found in this analysis was 92.3~ksi. The conclusion is that a beam with a flat energy profile needs to be smaller in diameter to work.

\begin{figure}
\begin{centering}
\includegraphics[width=\textwidth]{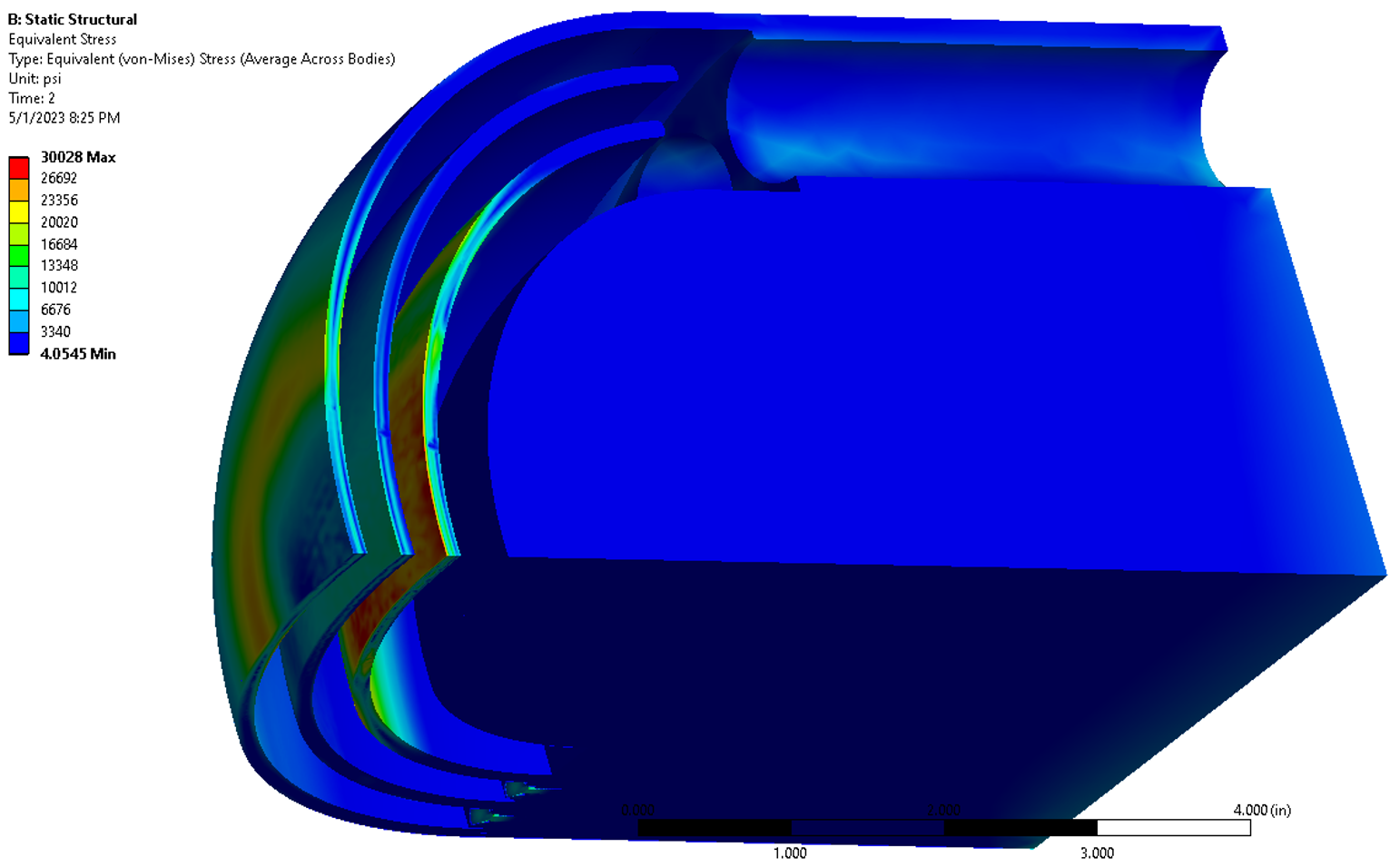}
\caption{\label{fig:Tstress-boil-GaussB} The stress profile in the beryllium for the Gaussian beam energy deposition, two-phase simulation, 550 GPM flow. Maximum stress is 30 ksi.}
\end{centering}
\end{figure}

\begin{figure}
\begin{centering}
\includegraphics[width=\textwidth]{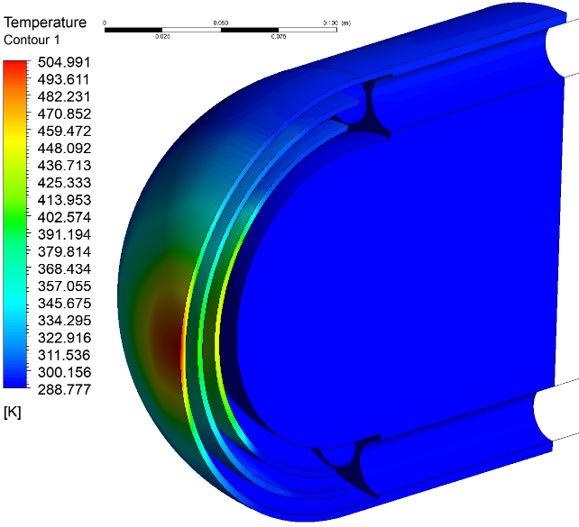}
\caption{\label{fig:Temp-1P-GaussB} The temperature in the beryllium for a single phase simulation with 750 GPM of heavy water, and a gaussian beam with sigma of 3.3 cm.}
\end{centering}
\end{figure}

The smaller diameter Gaussian beam in a two-phase simulation does not cause the thermal stress to exceed the allowable stress in the beryllium, even though the power deposition in the center of the hemisphere is higher than in the flat beam distribution.  Figure~\ref{fig:Tstress-boil-GaussB} shows the stress contours in the beryllium for 550~GPM flow, using a two-phase analysis. The maximum water temperature in the two phase simulation was high enough to run the simulation again as a single-phase analysis at 750 GPM of \dto to suppress boiling. Figure~\ref{fig:Temp-1P-GaussB} shows the temperature profile of the beryllium for the Gaussian beam with 750~GPM of heavy water cooling it.

\begin{figure}
\begin{centering}
\includegraphics[width=\textwidth]{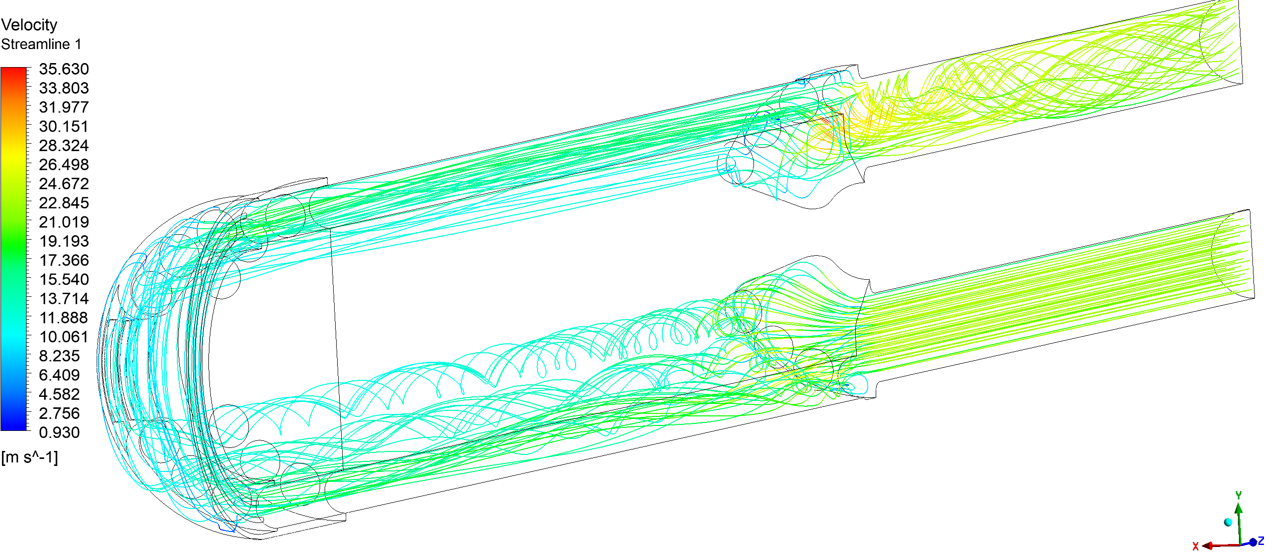}
\caption{\label{fig:StreamlV-GaussB} Velocity streamlines for 750 GPM flow through the water-cooled target.}
\end{centering}
\end{figure}

\begin{figure}
\begin{centering}
\includegraphics[width=\textwidth]{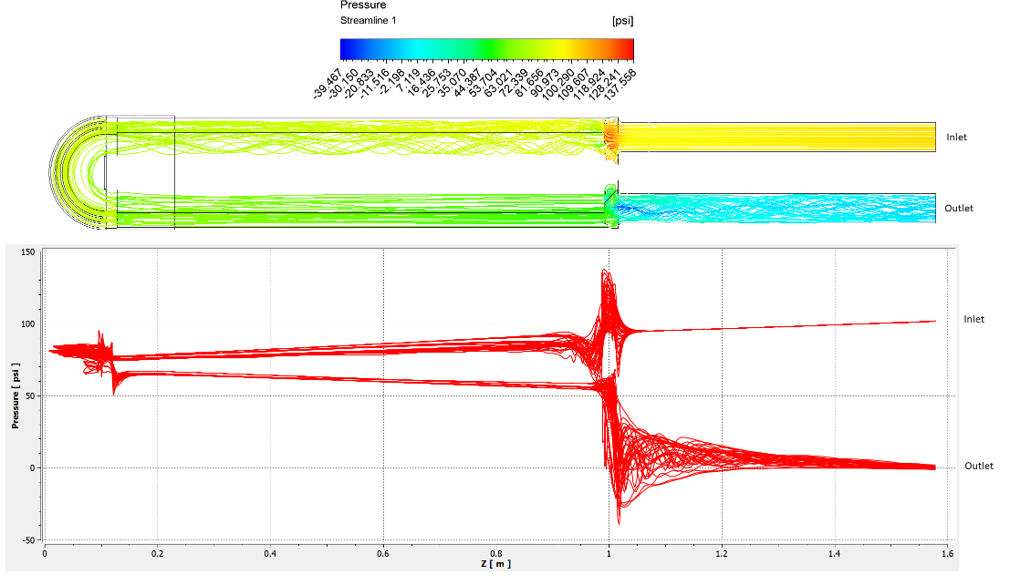}
\caption{\label{fig:Pdrop-750} Pressure drop across the target for 750 GPM flow. The transition from the stainless pipes to the beryllium body sees the highest pressure drop. Note that the inlet and outlet are switched vertically in the picture above to be consistent with the graph.}
\end{centering}
\end{figure}

Figure~\ref{fig:StreamlV-GaussB} shows the velocity streamlines of the water going through the cooling channels in the target.  Figure~\ref{fig:Pdrop-750} shows the pressure drop going around the loop of the cooling channel.  The transition between the stainless pipes bringing water to and from the target, and the beryllium body is where the pressure drop is the highest.  The total pressure drop from inlet to outlet is about 100~psi.  The negative pressure relative to the outlet means that there is a risk of cavitation at the outlet unless the system pressure at the outlet is greater than 69.1~psia.

\begin{figure}
\begin{centering}
\includegraphics[width=\textwidth]{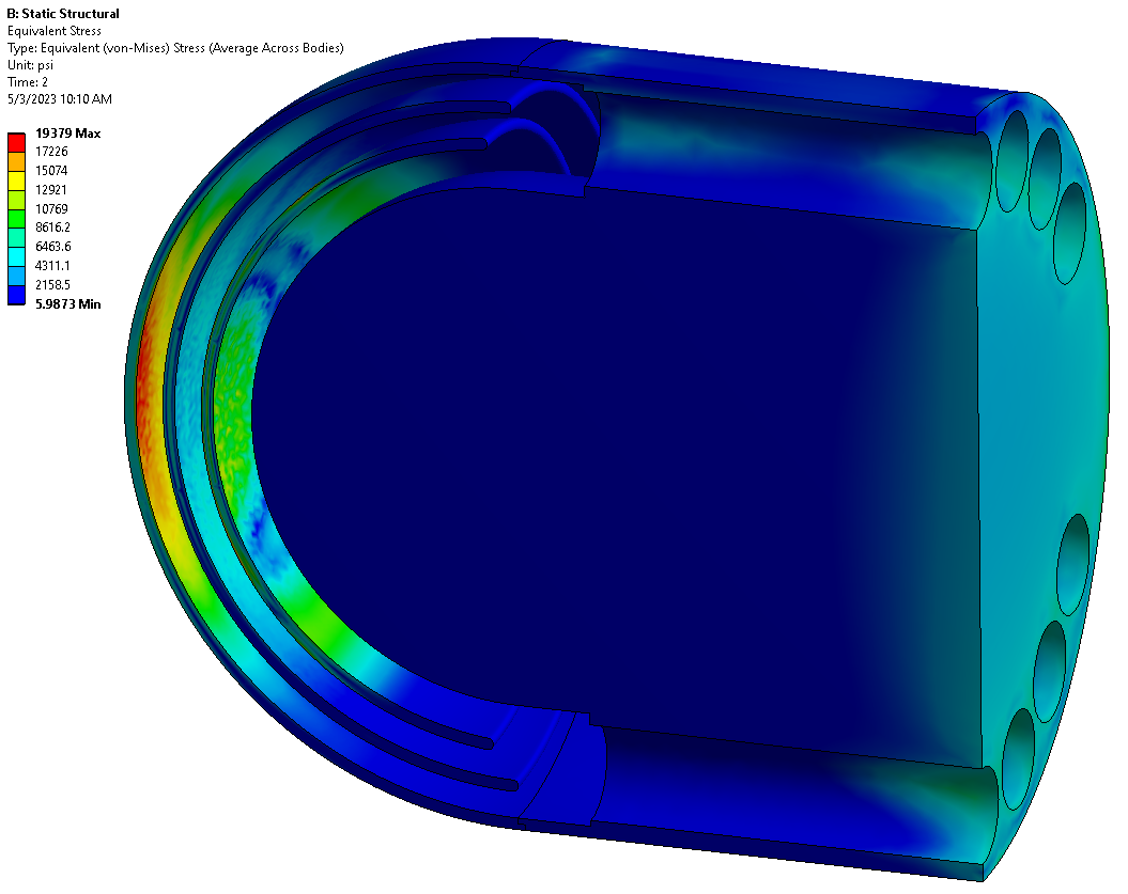}
\caption{\label{fig:Tstress-1P-750} Thermal stress in the single phase analysis at 750 GPM flow of D$_2$O. The maximum thermal stress is 19.4 ksi, which is less than the allowable of 36 ksi.}
\end{centering}
\end{figure}

The target passed a variety of checks based on the ASME Boiler and Pressure Vessel Code~\cite{ASME_VIII_1_2021,ASME_BPVC_VIII_2}.  The Maximum Allowable Working Pressure (MAWP) of the pressure vessel is greater than 600 psi. Figure~\ref{fig:Tstress-1P-750} shows the thermal stress contours from the single-phase analysis of the target running 750~GPM of \dto through the cooling channels.

A structural conclusion from the analysis is that the dynamic flow pressure is low, so the shell mounting hardware is adequate.  Thermal-fluid conclusions are that the flat beam must be collimated more due to the hot spot in the turnaround section at the mid-plane of the target. If this flat beam is collimated to not extend laterally more than 17 cm in diameter,
then 500 GPM water flow should be sufficient for adequate cooling.

The Gaussian beam has intense heat in the central region of the target, nearly 9 MW/m$^2$ of maximum heat flux at the fluid/solid interface. Boiling will occur at 500 GPM. The two-phase CFD model employs the RPI (Rensselaer Polytechnic Institute) wall boiling model~\cite{Kurul1990}.  While this model has been extensively validated, especially in applications involving subcooled boiling of water, it does have limitations, and obtaining accurate results in extreme conditions can be very challenging. The model relies on a set of empirical correlations to determine parameters like bubble departure diameter, lift-off diameter, and the influence area of nucleation sites.  These correlations are often developed based on specific experimental data and may not be universally applicable to different geometries or operating conditions. This reliance on empirical data can lead to reduced predictive accuracy when the model is applied to situations outside the range of the experimental data used to develop the correlations. Due to the limitations mentioned above, the RPI model often requires calibration and tuning on a case-by-case basis to achieve acceptable accuracy, even with limited experimental data.  The combination of intense heat flux, high degree of subcooling ($>$100 K), high wall superheating ($>$100 K), very high fluid interface thermal gradients (2000 K/mm), unconventional geometry, and water mass flux above the validated range make accurate two-phase simulation very challenging.  Results of our model showed the area fraction of bubble influence rise to the maximum user input value of 50\% vapor even at areas near the outer radius of the beam with relatively low wall heat flux. Clipping at this maximum 50\% value was seen through all of the inner radius regions of interest with higher interface heat flux values, indicating RPI model limits may have been exceeded, or are not applicable to this extreme case without modification.  Due to these factors as well as others regarding the wall boiling model, we currently do not have enough confidence in the two-phase simulation results to recommend proceeding with a two-phase cooling design if single phase cooling can be accomplished.

In addition to the challenge of obtaining meaningful results for the two-phase boiling model, there is also the possibility of damaging cavitation at the boiling wall interfaces due to the extremely high heat fluxes and great degree of bulk liquid subcooling, known as subcooled boiling.  Literature sources \cite{Damage1968}, \cite{hammitt1967cavitation}, and \cite{mayinger2006heat}, indicate highly subcooled boiling, such as what could occur in our case, can cause cavitation damage to walls as the vapor bubbles collapse in the highly turbulent, much colder liquid flow.  Bubble collapse in liquids that have a temperature below the saturation temperature of the vapor in the bubble can be controlled by two mechanisms: heat transfer (boiling) or inertia forces (cavitation). The transition from heat transfer-controlled to inertia-controlled condensation can be described with the Jakob number:

\begin{equation}J_a=\frac{\rho_{Liq}}{\rho_{Vap}}\frac{C_{pLiq}(T_{sat}-T_{bulk})}{h_{fg}}\end{equation}

This number represents the ratio of sensible heat of the liquid phase per unit volume to the latent heat of the gas phase during phase change. Transition from heat transfer-controlled to inertia-controlled (cavitation) bubble collapse starts at $ J_a = 30 $ and at $ J_a = 70 $ cavitation is fully developed. With the bulk water temperature being quite constant and very close to our inlet temperature of 60\degree~F, the Jakob number is then a function of pressure only.  Using deuterium oxide fluid properties to compute our Jakob number vs pressure, we can see in the graph below that cavitation would be expected at our operating conditions, or at least the possibility cannot be dismissed.

\begin{figure}[h]
        \centering
        \includegraphics[width=0.8\linewidth]{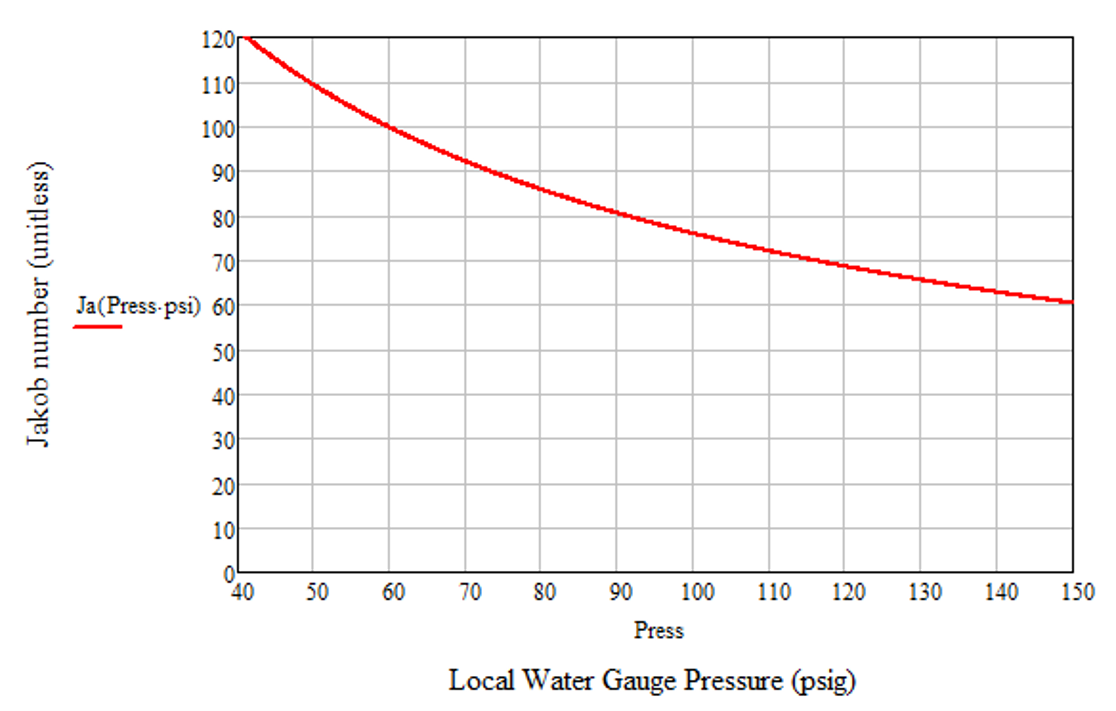}
        \caption{Jakob number as a function of local water gauge pressure.}
        \label{fig:Jakobnumber}
    \end{figure}

The higher flow rate of 750 GPM results in convective cooling coefficients high enough to suppress boiling, depending on the system pressure, resulting in single-phase flow. Lower flow rates with higher system pressures and lower wall heat flux values can also suppress boiling. While it is certainly possible a two-phase boiling design would work without incident, the associated highly complex fluid physics results in some uncertainty of success.  Without extensive experimental testing, a single-phase cooling design is preferred at this time.

\begin{figure}
\begin{centering}
\includegraphics[width=\textwidth]{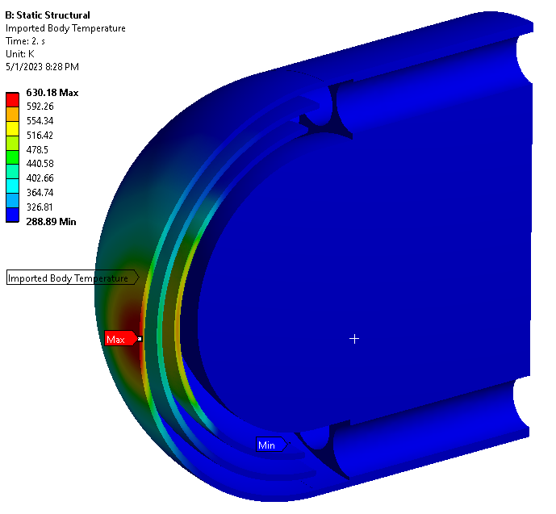}
\caption{\label{fig:Temp-boil-GaussB} The temperature profile in the beryllium for the Gaussian beam energy deposition, two-phase simulation, 550 GPM flow.}
\end{centering}
\end{figure}

\subsection{Risks and Mitigation}

\noindent \textbf{Risk:  Shortened target lifetime due to beryllium embrittlement.}  Solubility of hydrogen gas in beryllium is very low~\cite{kizu1995hydrogen}.  This has led to target blistering in applications such as BNCT where high-current proton beams have been allowed to stop in the beryllium layers of the target. Referring to Fig.~\ref{fig:FlatB}, one can see that there are a few places where the protons range out (stop) inside the beryllium hemispheres of our target. The small stopping distance and low solubility over time could lead  to buildup of hydrogen gas in the areas where beam stops, with pressures sufficient to blister the surface of the hemisphere~\cite{Rinckel}.  In our target such blistering is likely to create turbulence and possibly cavitation in the cooling water flow, which could lead to premature failure of the target.

\textit{Mitigation:  First, careful analysis of hydrogen buildup in our target to ascertain if this is a significant risk, folding in the planned spatial distribution of the beam over the face of the target.  If so, a small redesign of the middle hemisphere -- increasing its thickness, and collimating the beam to 18 cm outer diameter, which must be done anyway to keep beam away from areas where water flow is stagnant (see, for example, Figs.~\ref{fig:flowV-flatB} and ~\ref{fig:Intemp-flatB}), can have the beam stopping entirely in water.}

\bigskip
\noindent  \textbf {Risk: Failure of outer shell leading to vacuum breach.}

\textit{Mitigation: Fast acting valve far enough upstream to close in time to avoid compromising systems, including a system for efficient recovery of any heavy water introduced into the vacuum pipe.}  

\bigskip
\noindent \textbf {Risk:  Failure of an inner shell of the target, leading to blockage of coolant flow.} 

\textit{Mitigation:  Diagnostics in water circuits to carefully monitor water flow and detect deviation in flow rate or pressure that might indicate such problems. In addition, systems to quickly shut down affected systems, including beam-delivery to shut off the source of heat, are required.}   

\bigskip
\noindent \textbf {Risk:  Inadequate monitoring of target.} Preventing target failure requires thorough monitoring of all parameters that can lead to such failure.  The primary monitor will be an infrared optical camera that measures the temperature of the target.  However, thermal stresses have been identified as a possible failure mode, and at present there are no available concepts for a monitor suitable for direct measurement of stress levels in our target.

\textit{Mitigation: While direct stress measurements may not be feasible, use of an infrared camera may provide an indirect method for this measurement.  Thermal stresses arise from temperature gradients, and different expansion rates for different temperatures.  The gradients and stresses for each point on each hemisphere can be calculated from the beam profiles and the beam intensity at each point, so a measurement of  beam intensity gradients can serve as a monitoring technique.  These can be obtained from a careful analysis of the infrared camera viewing the entire target surface. 
The thermal and heat-transfer calculations can translate these surface measurements to the internal status at each point of the target.
The design of this camera and its specifications should provide adequate sensitivity to perform this function.}

\bigskip
\noindent \textbf {Risk:  Leak of heavy water cooling system leading to loss of valuable \boldmath{\dto}.}

\textit{Mitigation:  Careful engineering design of all cooling circuits handling heavy water to minimize chances for leakage.
Preventive maintenance procedures will be established to ensure continued leak-tight status of the cooling system.} 

\clearpage
\section{Sleeve Design}

\subsection{Overall Structure of the Lithium-beryllium Filled Sleeve}

\begin{figure}
\begin{centering}
\includegraphics[width=\textwidth]{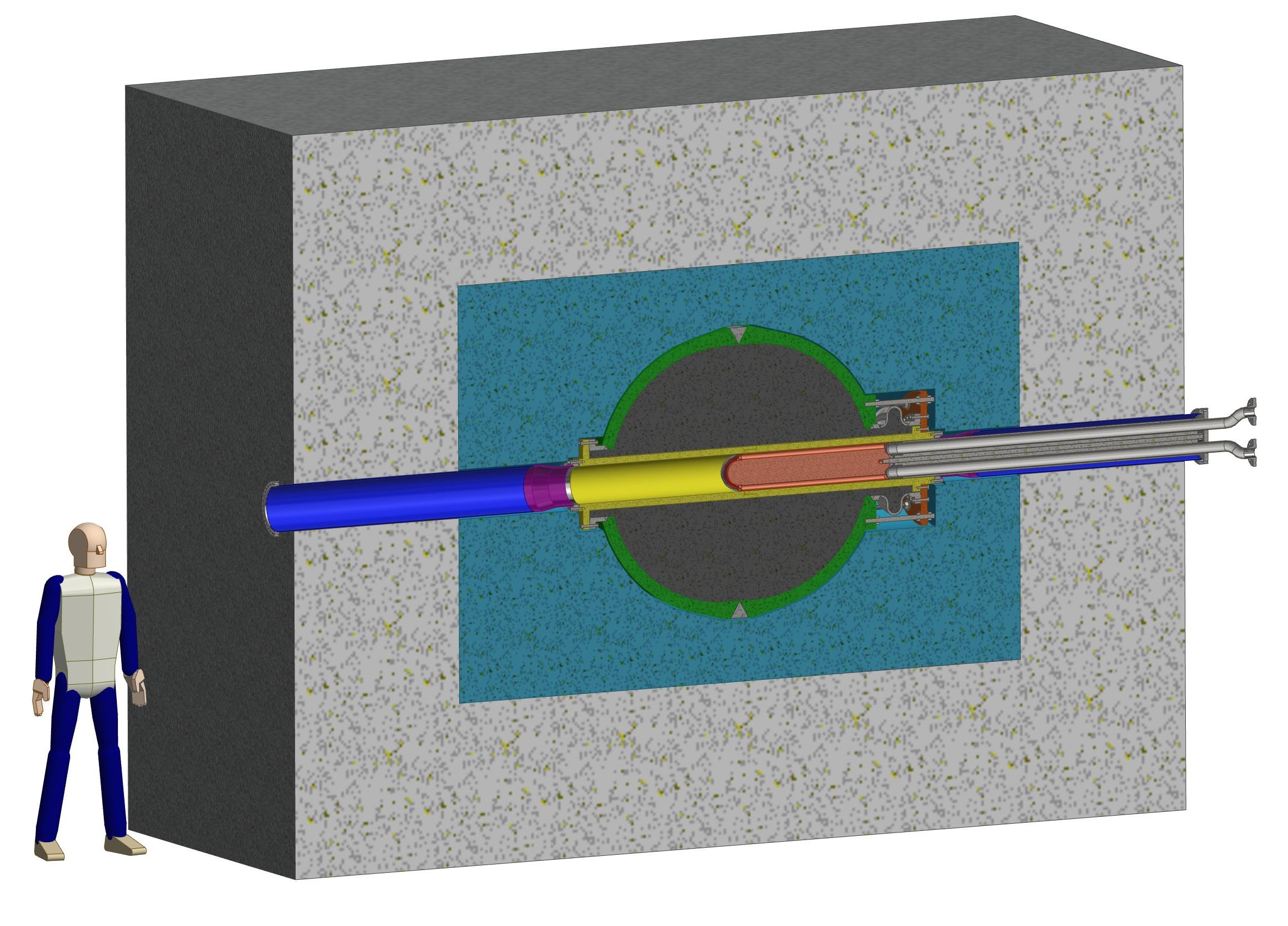}    \caption{\label{fig:ID-22-yemilab} Overall section view of the IsoDAR target and shielding, showing the water-cooled target and the sleeve.}
\end{centering}
\end{figure}

\begin{figure}
\begin{centering}
\includegraphics[width=\textwidth]{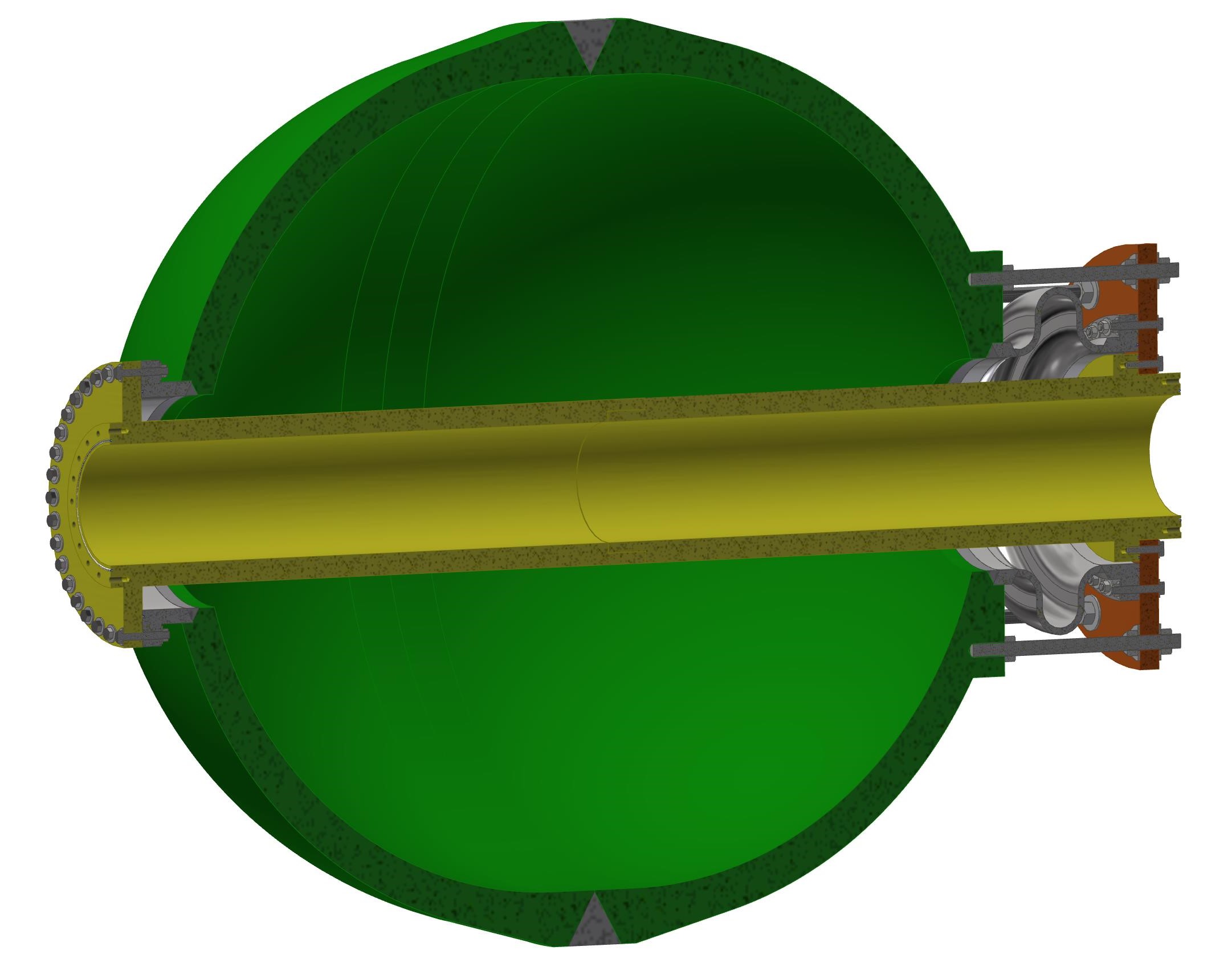}    \caption{\label{fig:li-be-5-001} View of the empty pressure vessel of the sleeve.}
\end{centering}
\end{figure}

\begin{figure}
\begin{centering}
\includegraphics[width=\textwidth]{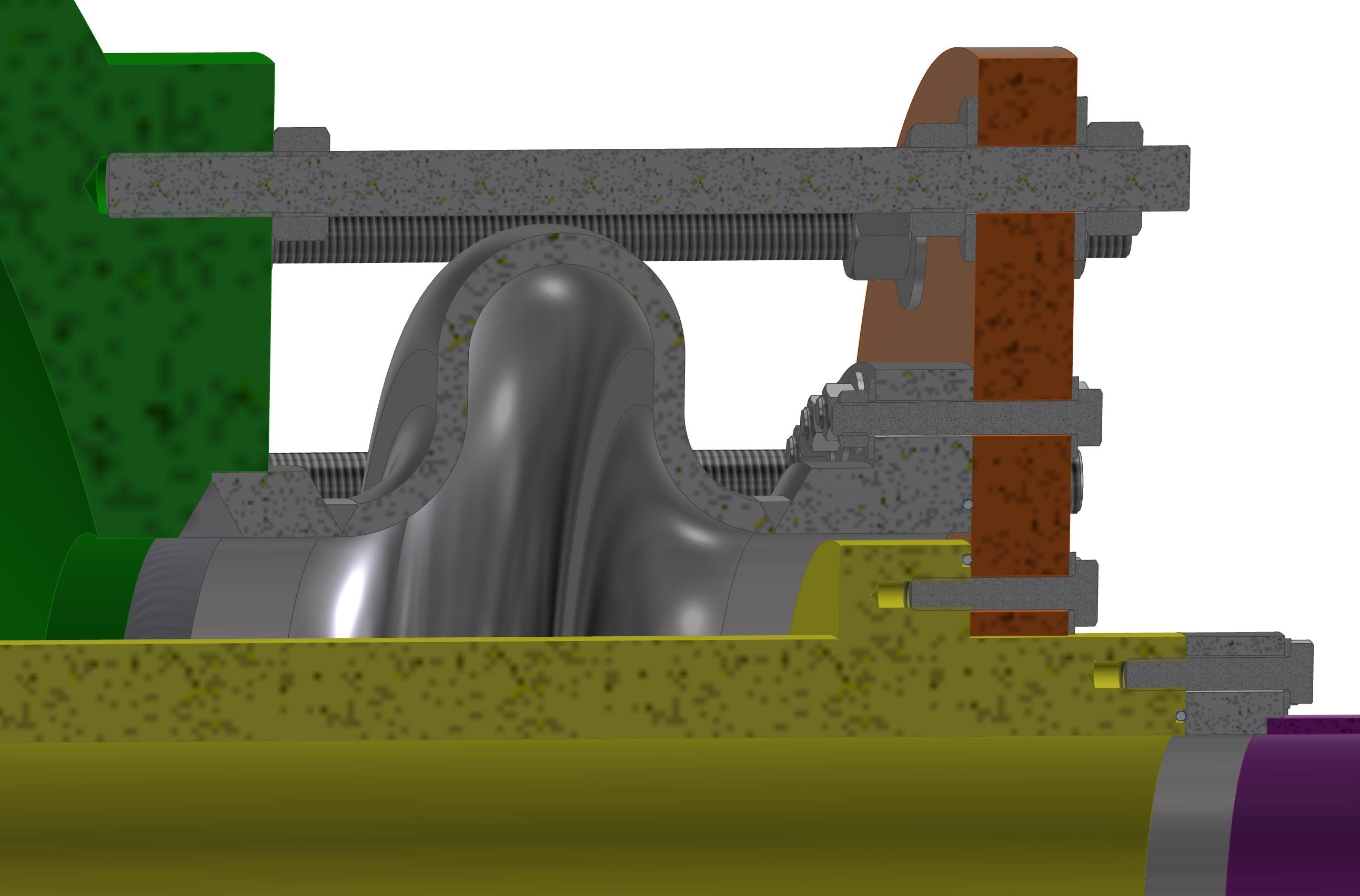}    \caption{\label{fig:li-be-5-seal} Close-up section view of the expansion joint and the Helicoflex seals on the orange interface plate.}
\end{centering}
\end{figure}

\begin{figure}
\begin{centering}
\includegraphics[width=\textwidth]{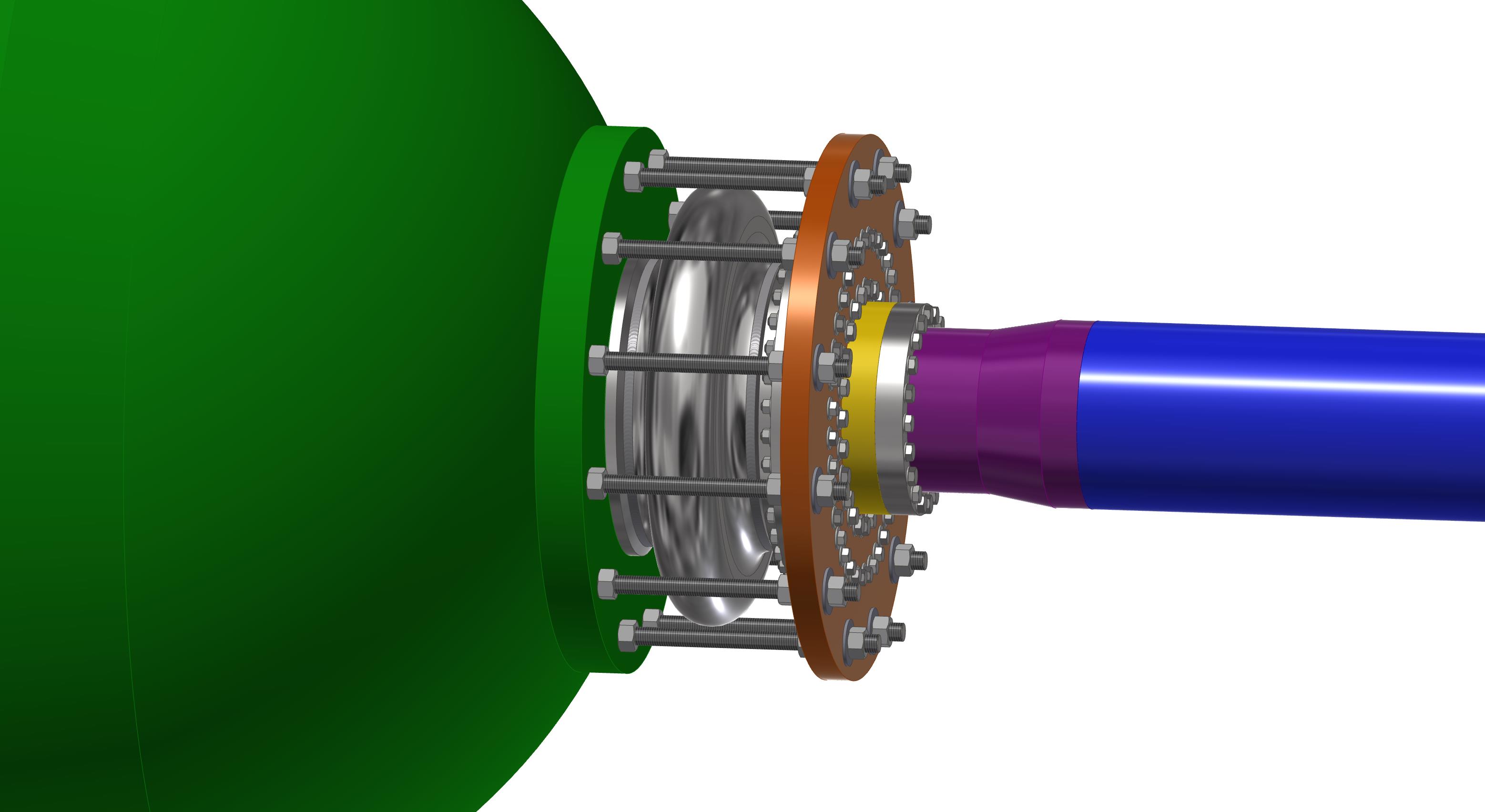}    \caption{\label{fig:li-be-5-001a} Close-up of the downstream end of the sleeve showing the expansion joint and the connection to the downstream vacuum pipe.}
\end{centering}
\end{figure}

\begin{figure}
\begin{centering}
\includegraphics[width=\textwidth]{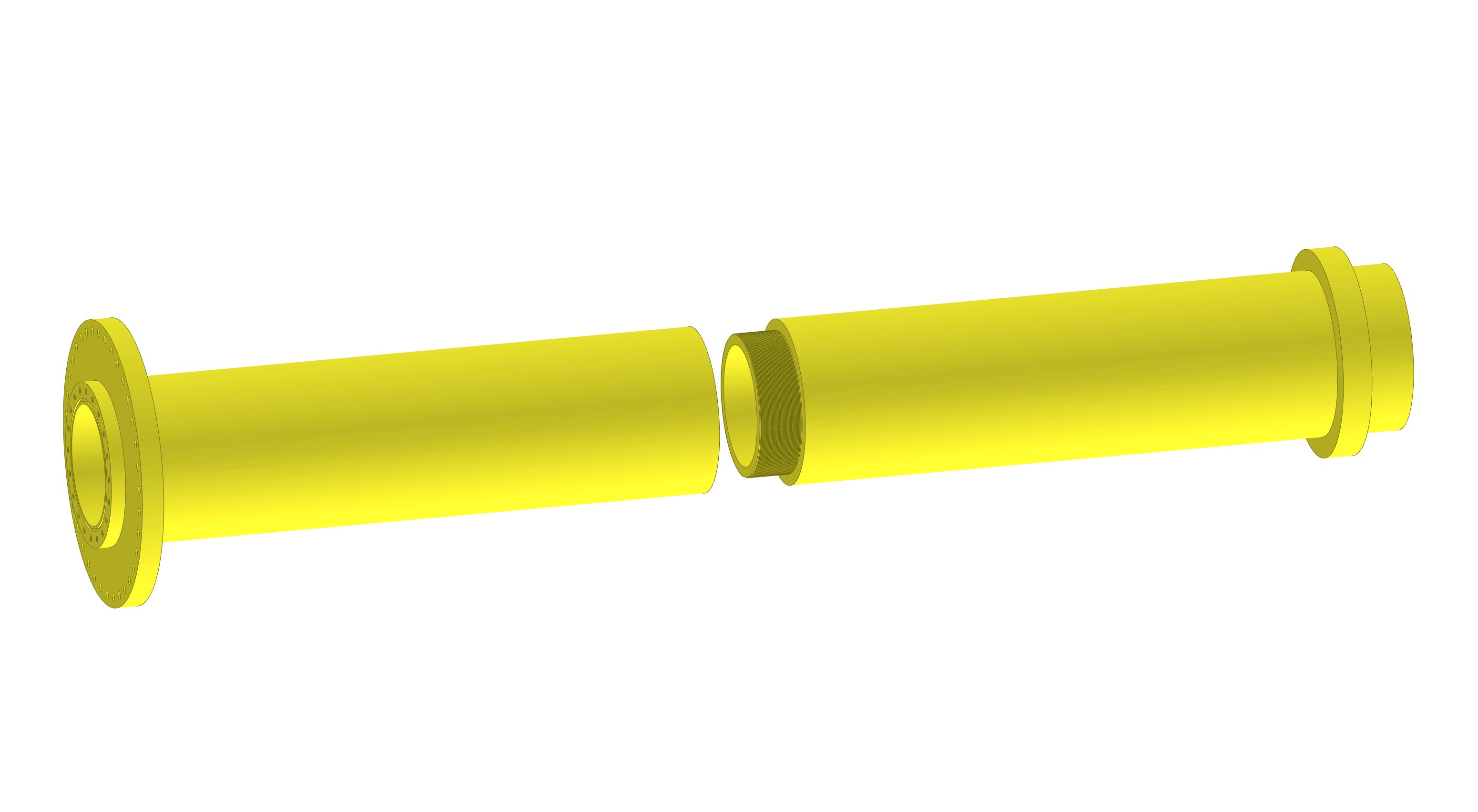}    \caption{\label{fig:Becoretubeexp} View of the beryllium core tube of the sleeve.}
\end{centering}
\end{figure}

As detailed in the following section, the IsoDAR sleeve surrounding the target will contain a mixture of approximately 75\% $^9$Be, 25\% $^7$Li by mass.  This mixture produces an optimized level of antineutrinos when exposed to the neutron flux from the water-cooled target, according to a detailed analysis presented in Ref.~\cite{bungau_optimizing_2019}.  
Lithium and beryllium do not form alloys at any reasonable temperature and pressure, so we will create a solid mixture of the two metals by injecting molten $^7$Li into beryllium powder of the appropriate bulk density.  The total amount of beryllium powder in the sleeve is 1,437 kg and the amount of $^7$Li is 479~kg. 
The injection is designed to take place at up to $300^\circ$C and 2,500 psig.  Towards constructing the sleeve, it is not possible to simulate the pressure drop in the beryllium powder inside the sleeve.  Therefore, we have designed an injection test, described below, which will demonstrate whether 2,500 psig is sufficient to push the molten $^7$Li to the bottom of the vessel of beryllium and form the required mixture. In addition, casting at high pressure is required in order to disallow void creation in the mixture. Such voids could lead to unwanted thermal and mechanical gradients in the sleeve during beam-on and, in general, reduce antineutrino production. This test, utilizing the designed injector for the full-sleeve, will create a small-scale sleeve with the correct mixture ratio.
 
 The spherical container for the mixture must be an ASME Boiler and Pressure Vessel Code stamped vessel due to the high pressure inside. Figure~\ref{fig:ID-22-yemilab} shows the spherical pressure vessel in green, made from two halves welded together at the mid-plane.  The vessel is made from K13049 steel.  The center of the vessel has a beryllium pipe through it to create the volume for the water-cooled target torpedo.  Figure~\ref{fig:li-be-5-001} is a close-up of the empty pressure vessel and the internal beryllium pipe.  The steel and beryllium have different coefficients of thermal expansion, and there will be manufacturing tolerances on the lengths of both that must be bridged to make a seal.  They must be connected together to form a leak-tight volume for the molten $^7$Li, so the downstream connection is made with a custom expansion joint that allows the beryllium pipe and steel vessel to be different lengths, yet still connect through a flexible interface.  The restraining rods around the expansion joint are there to prevent the internal pressure of the molten $^7$Li from overextending the expansion joint, a typical design of any bellows connection in a vacuum system.  The original design showed brazes or e-beam welds between the beryllium core tube and the steel pressure vessel, but Materion cannot make those joints.  All of the connections between the beryllium core tube and steel pressure vessel were changed to bolted joints.  Helicoflex seals are used at bolted joints to contain the molten $^7$Li.

Figure~\ref{fig:li-be-5-001a} shows a close-up view of the expansion joint between the steel vessel and beryllium core pipe.  The expansion joint is made of steel, so it can be welded to the spherical pressure vessel.  Two bolted joint seals need to be made at the dowsntream end on the orange interface flange.  Figure~\ref{fig:li-be-5-seal} shows a cross-section through the expansion joint that shows where the Helicoflex seals are.  The downstream vacuum pipe is bolted to the downstream beryllium flange of the core pipe and also sealed with a Helicoflex seal.  

Figure~\ref{fig:Becoretubeexp} shows another example of a design iteration with Materion on the beryllium core tube.  Materion does not have large enough hot isostatic presses to make the core tube from one piece, so they asked that the core tube be made from two pieces that thread together.  Once the parts are threaded together, there must be an electron beam weld or braze at the interface to make a vacuum and molten-lithium-tight seal.

\subsection{Sleeve Composition}

Geant4 studies were performed including the geometry of the set-up with the corresponding material properties as well as the characteristics of the incoming proton beam. Appropriate physics packages were selected for the interactions of protons with beryllium and for neutron interactions. These included the latest data-driven models for both protons and neutrons, using the ENDF experimental database for both incident particles~\cite{ENDF-VII}. In cases where data for the target elements is not yet available, the code uses the TALYS code database~\cite{KONING20122841}. All these models have been validated for the energies and materials relevant to our studies in previous publications~\cite{bungau_shielding_2020,bungau_optimizing_2019}.

\begin{figure}
\begin{centering}
\includegraphics[width=1.0\textwidth, height=0.29\textheight]{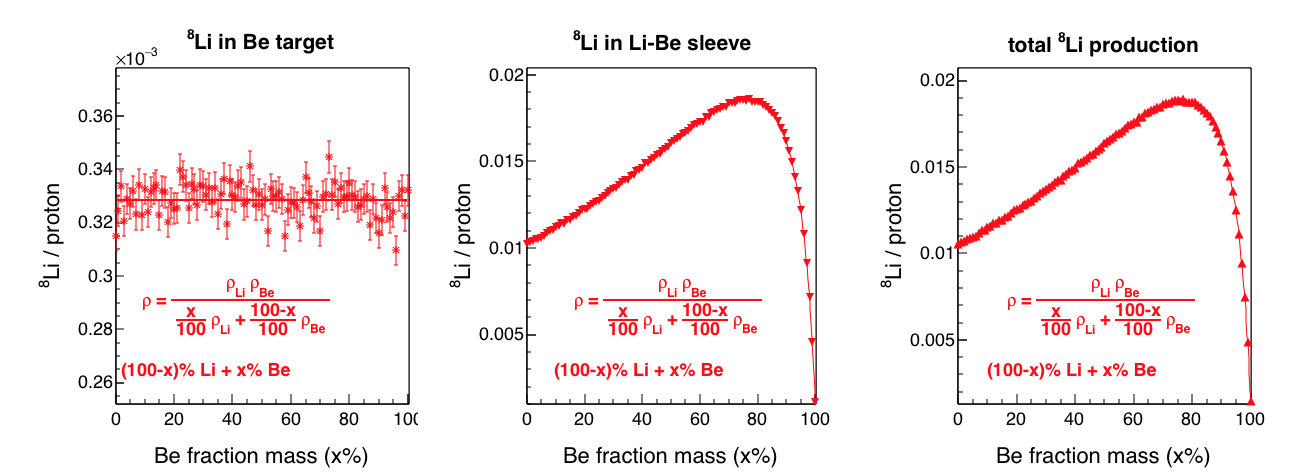} 
\vspace{0.3cm}
\caption{\label{fig:Be-fraction-mass} The overall $^8$Li production in the target and sleeve for various concentrations of lithium and beryllium shows a maximum for 75\% beryllium weight fraction mass~\cite{bungau_optimizing_2019}.}
\end{centering}
\end{figure}

\begin{figure}
\begin{centering}
\includegraphics[width=0.8\textwidth ]{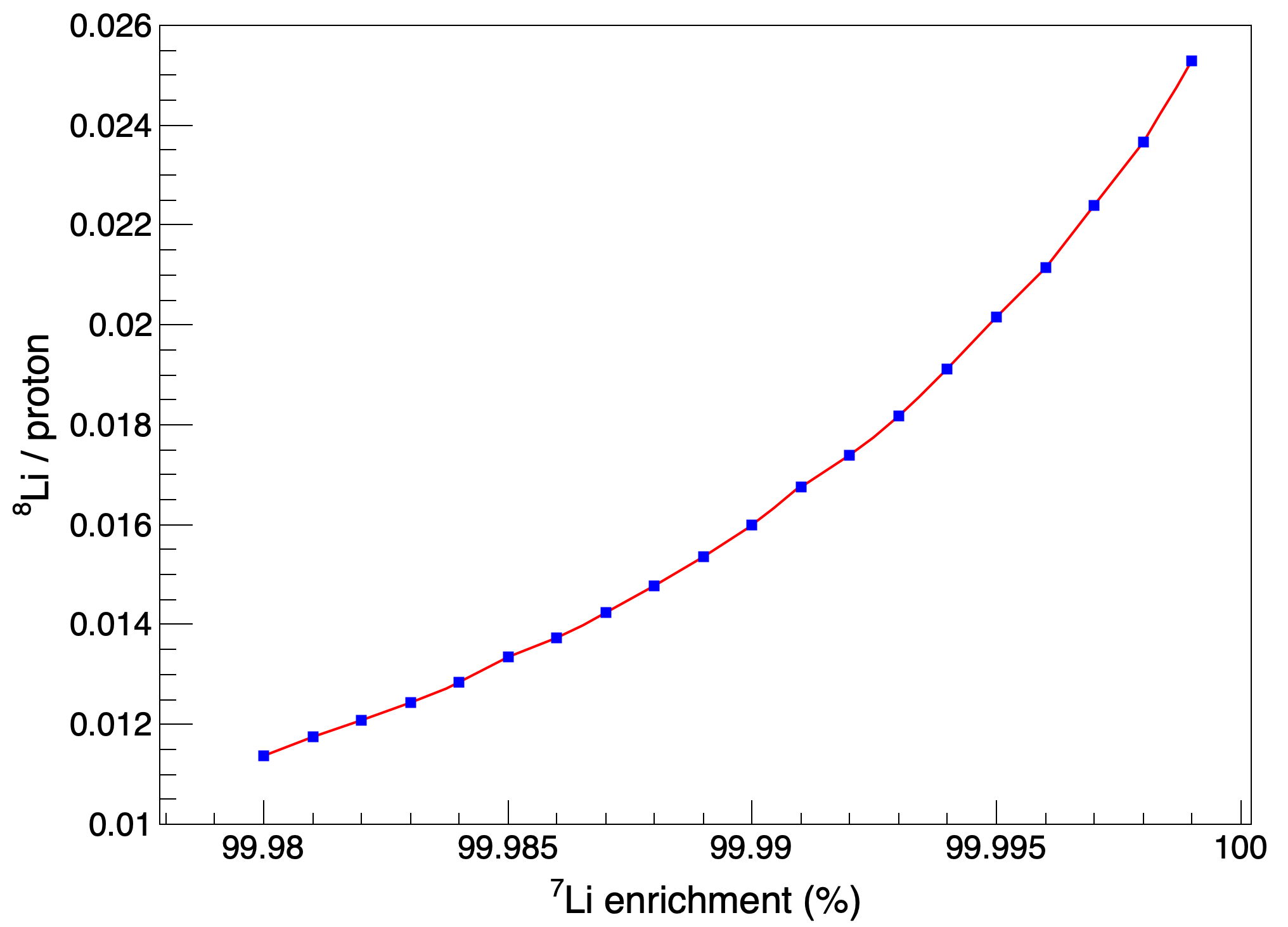} 
\vspace{0.3cm}
\caption{\label{fig:enrichment-nogrid} $^8$Li production as a function of $^7$Li  enrichment~\cite{bungau_optimizing_2019}. }
\end{centering}
\end{figure}

\begin{figure}
\begin{centering}
\includegraphics[width=1.0\textwidth ]{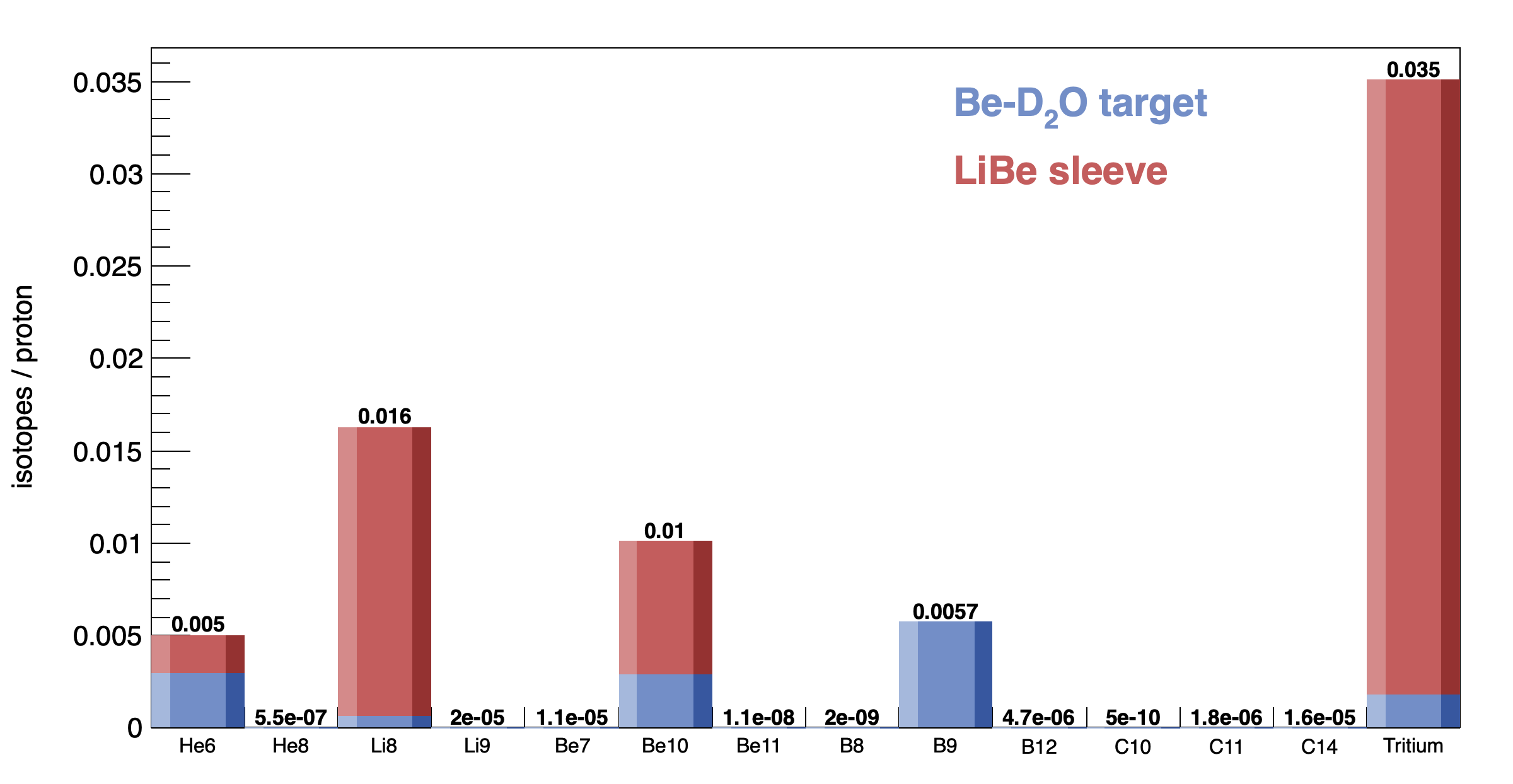}  
\vspace{0.3cm}
\caption{\label{fig:isotopes} Overall isotope production including $^8$Li in both the target and sleeve using the current design. }
\end{centering}
\end{figure}

\begin{figure}
\begin{centering}
\includegraphics[width=1.0\textwidth ]{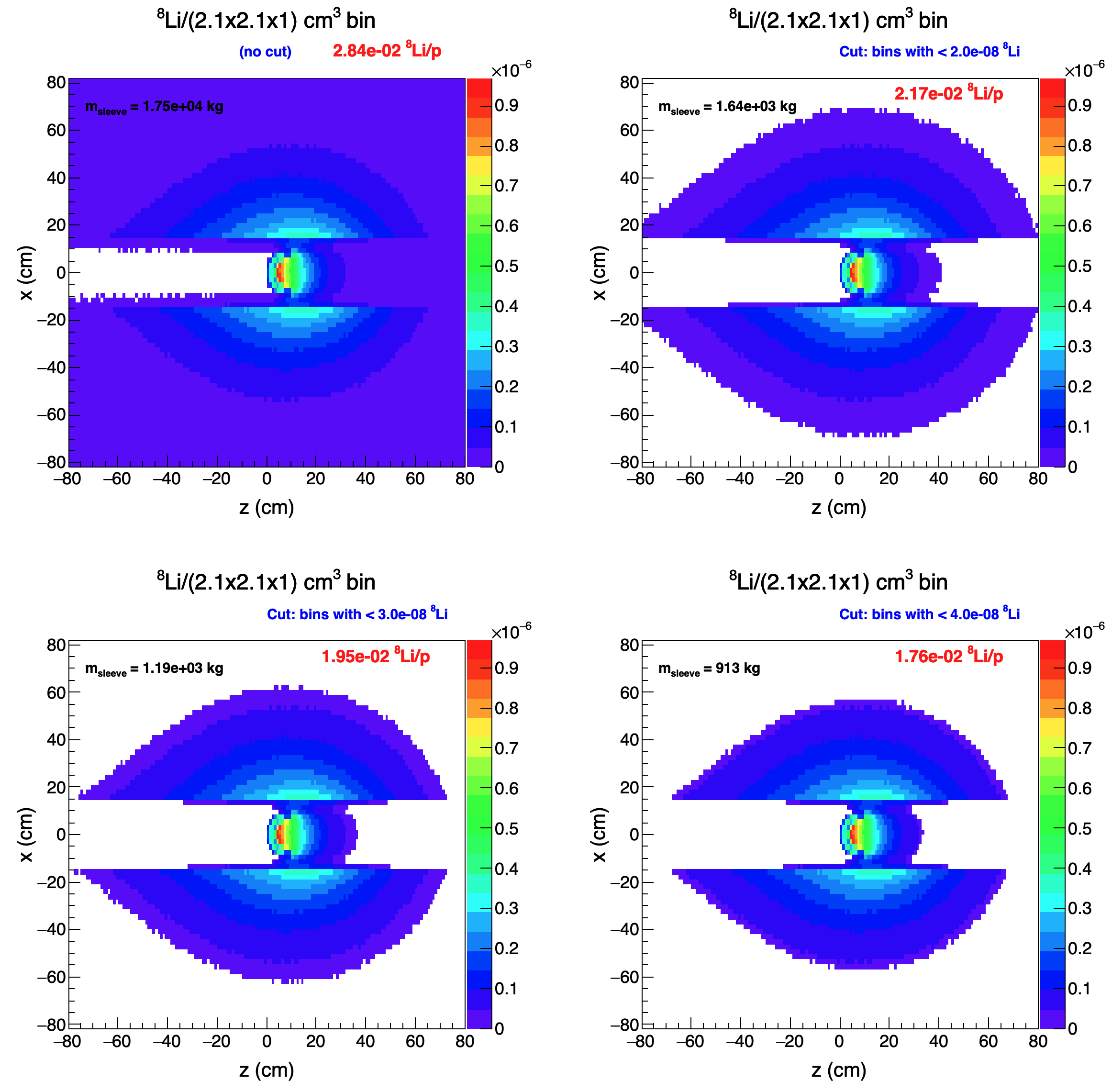}   
\vspace{0.3cm}
\caption{\label{fig:contour1} Optimization of the size and shape of the sleeve. The sleeve is made of a homogeneous mixture of lithium and beryllium (beryllium fraction mass is 75\%). The parameters of interest are the mass of the sleeve and the corresponding $^8$Li yield - part 1. }
\end{centering}
\end{figure}

\begin{figure}
\begin{centering}
\includegraphics[width=1.0\textwidth ]{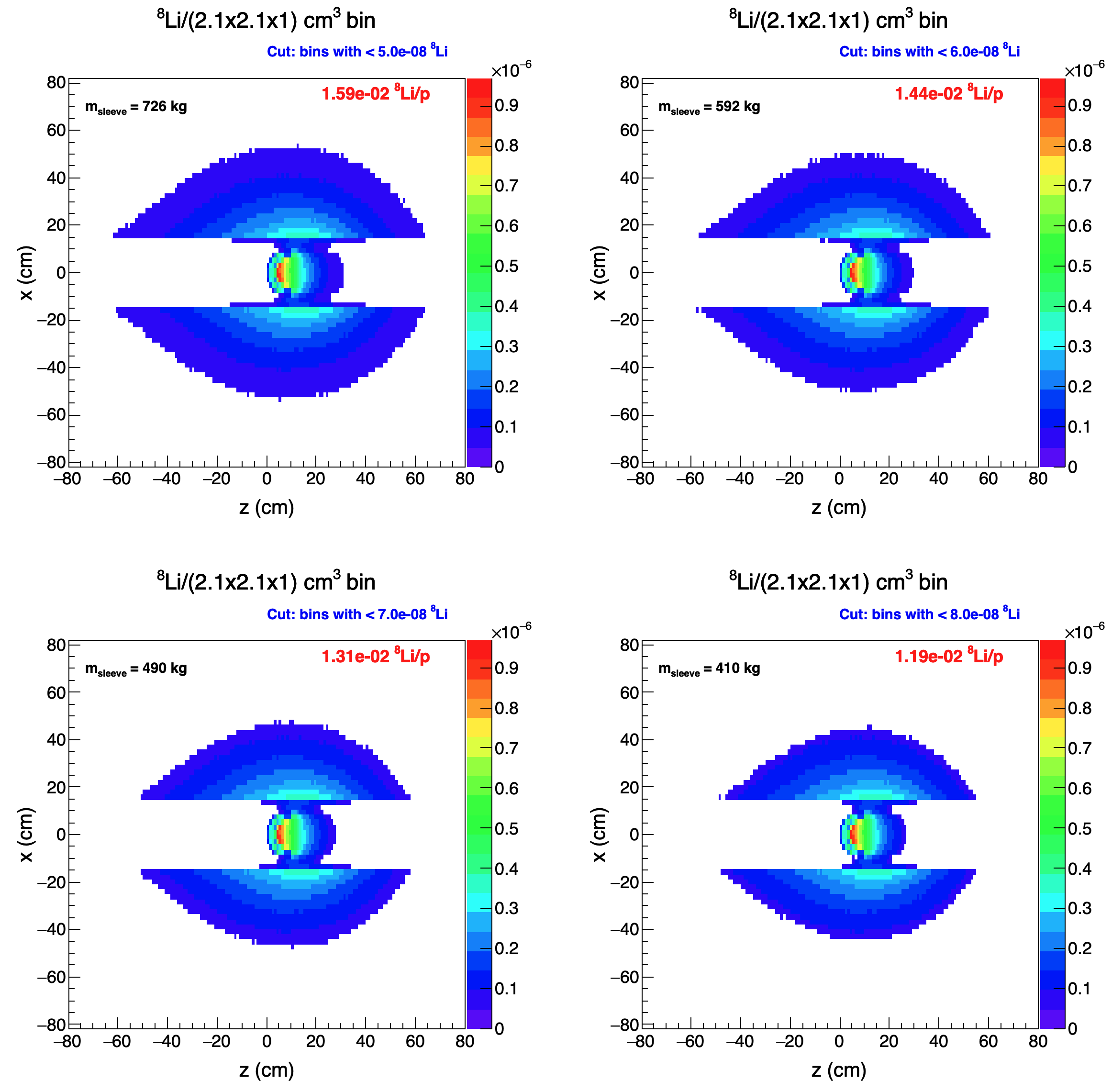}
\vspace{0.3cm}
\caption{\label{fig:contour2}Optimization of the size and shape of the sleeve. The sleeve is made of a homogeneous mixture of $^7$Li and beryllium (beryllium fraction mass is 75\%). The parameters of interest are the mass of the sleeve and the corresponding $^8$Li yield - part 2. }
\end{centering}
\end{figure}

To maximize antineutrino production, the shape and dimensions of the $^7$Li-beryllium sleeve were optimized and a detailed analysis was presented in Ref.~\cite{bungau_optimizing_2019}. The cylindrical sleeve of the previous design was replaced by a roughly spherical section to remove the least productive volumes. The sleeve is an ellipsoid, although in some simulations it was considered a sphere with a radius of 70 cm for simplicity. 

 Apart from neutron elastic scattering, the neutron interactions inside the sleeve are (1) inelastic processes in which other secondary particles are produced and (2) neutron capture processes which lead to isotope production. The $^8$Li is produced via neutron capture on  $^7$Li, or by neutron inelastic interactions with beryllium in which a neutron and a proton are produced. Previously, we considered a FLiBe mixture for the sleeve~\cite{Bungau:IPAC2014-THPRI083}; however, the presence of F was found to drastically reduce the number of neutron interactions with lithium and beryllium that can produce $^8$Li. Therefore, the material we consider for the sleeve is a mixture of $^7$Li-beryllium, and we found that the optimal sleeve material is a mixture of 75\% beryllium and 25\% enriched lithium ($^7$Li) fraction mass. The results of the parametric study for variable beryllium fraction (by weight) are shown in
Fig.~\ref{fig:Be-fraction-mass}. In a previous design, 
this material composition achieved a production rate of 0.018 $^8$Li per incident proton.

The figure of merit in these studies is the production of $^8$Li per incident proton. Natural lithium is a mixture of $^6$Li and $^7$Li (92.5\% pure $^7$Li), and even though the thermal neutron capture cross-sections on these two isotopes are similar in value, small amounts of $^6$Li will have significant effect on the physics experiment due to the very large inelastic neutron cross section on $^6$Li. The inelastic neutron cross section for $^6$Li(n,$\alpha$)t is a factor of $10^4$ higher. The experiment needs 0.016 $^8$Li per incident proton and Fig.~\ref{fig:enrichment-nogrid} shows the production of $^8$Li as a function of the enrichment of $^7$Li. A purity of 99.99\% will provide 0.016 $^8$Li per proton, and that number is taken as our baseline due to cost and availability, noting that even higher fractions are better.

The overall isotope production in both target and sleeve are shown in Fig.~\ref{fig:isotopes}. The current design produces about 0.016 $^8$Li per incident proton and the dominant source of $^8$Li is the sleeve. Another isotope of interest is tritium, produced in both the heavy water coolant and the sleeve (by interactions of neutrons with Li). The tritium produced in the sleeve will not pose a problem as it will remain trapped inside, however the tritium in the water coolant will represent a radiation hazard because of the possibility of a leak. Simulations have shown that the tritium produced in the heavy water with the Yemilab design is almost four times less than for the previous KamLAND-based study~\cite{alonso_isodarkamland_2017}.

\begin{figure}
\begin{centering}
\includegraphics[width=1.0\textwidth ]{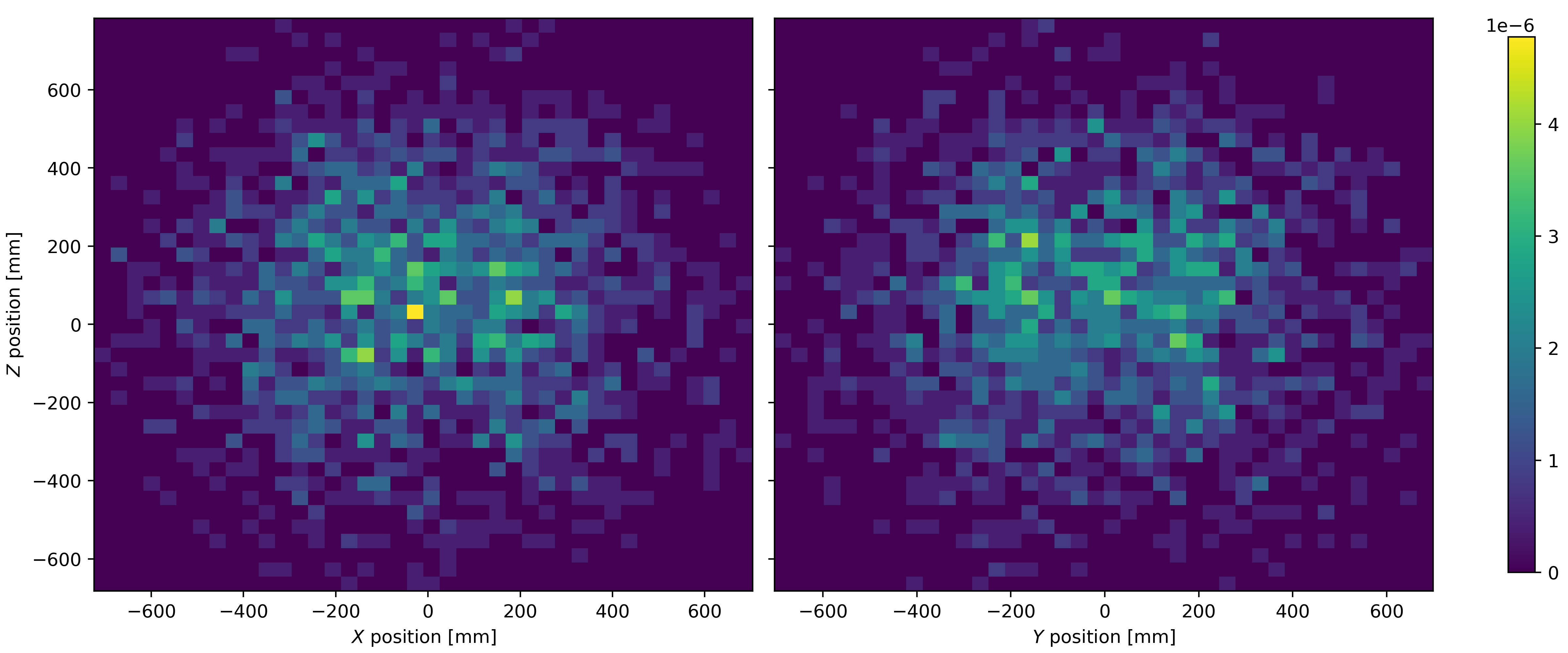}
\vspace{0.3cm}
\caption{\label{fig:li8_spatial_distribution}Spatial distributions of produced $^8$Li nuclei from 100,000 simulated 60 MeV proton injections. Protons are injected towards the $+\hat{z}$ direction.}
\end{centering}
\end{figure}

Due to the high cost of producing the enriched $^7$Li, the sleeve has required an optimization study of the actual size and shape.  The density as well as the beryllium and lithium mass fractions were maintained at constant values throughout this study. The study started with a larger sleeve in order to determine the regions where the $^8$Li yield is below certain threshold values. Once these regions were removed the total $^8$Li production was re-evaluated.
Figs.~\ref{fig:contour1} and ~\ref{fig:contour2} show a section plane cut through the sleeve, showing the shape of the sleeve while the $^8$Li production yield threshold was varied. As the volume and mass of the sleeve were reduced, so was the overall $^8$Li production rate. Keeping the axes ranges fixed, one can see the continuous drop in the sleeve volume. The optimum configuration is obtained when the $^8$Li production is 0.016 $^8$Li per incident proton, corresponding to a total sleeve mass of 726 kg. The corresponding spatial distribution of created $^8$Li nuclei from such a sleeve design is shown in Fig.~\ref{fig:li8_spatial_distribution}.

\subsection{The Lithium Injector for the Target Sleeve}

\begin{figure}
\begin{centering}
\includegraphics[width=0.8\textwidth]{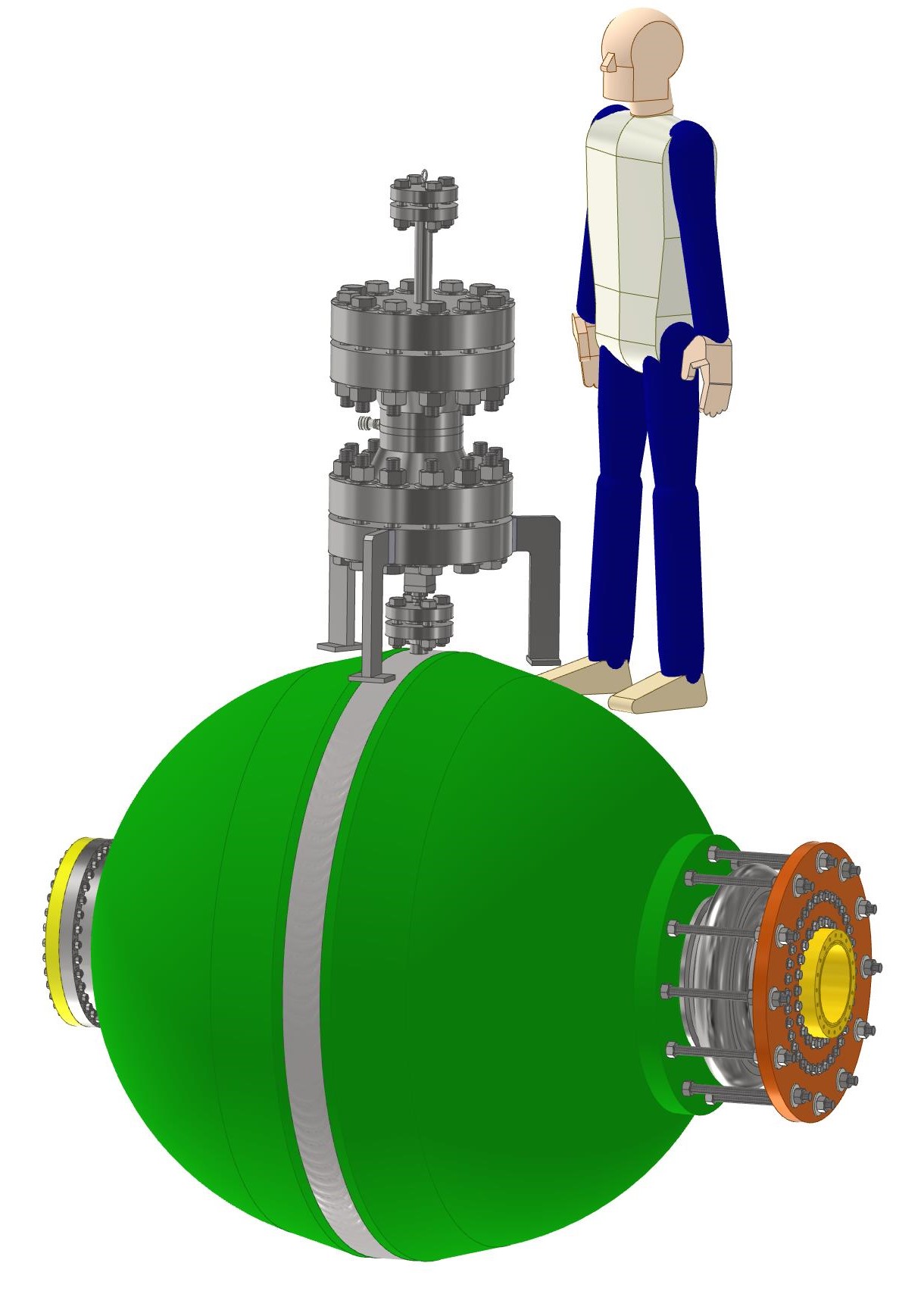}    \caption{\label{fig:Casteronsleeve-01} Overall view of the sleeve with an early version of the $^7$Li injector attached.  An actuator system was added to allow the bellows to be extended after it was collapsed in an injection.  The support connection between the injector and sleeve is not shown, nor is the argon glove box needed to surround the entire structure.}
\end{centering}
\end{figure}

\begin{figure}
\begin{centering}
\includegraphics[width=0.8\textwidth]{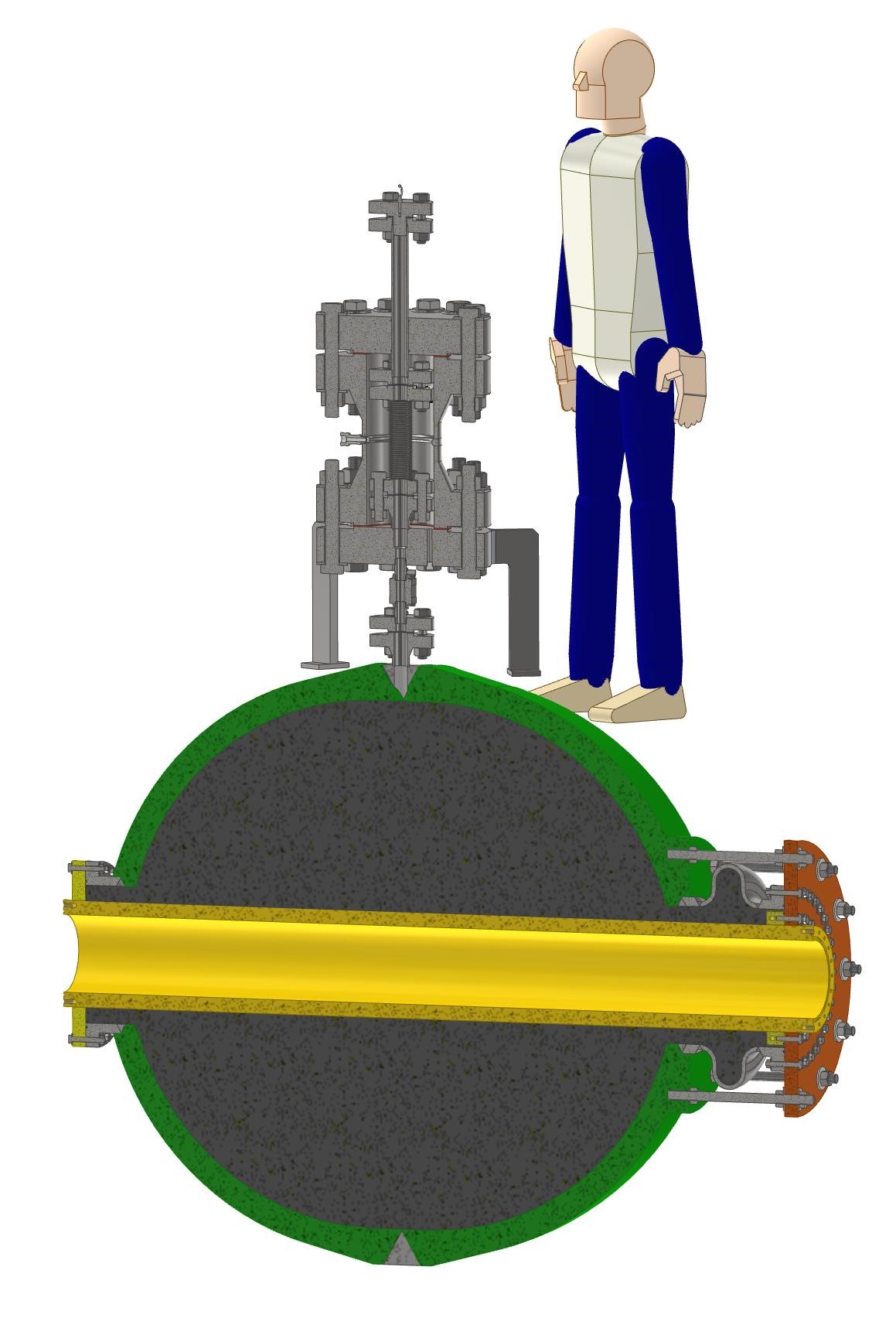}    \caption{\label{fig:Casteronsleeve-02} Section view through the sleeve with the $^7$Li injector attached at the top.}
\end{centering}
\end{figure}

\begin{figure}
\begin{centering}
\includegraphics[width=0.8\textwidth]{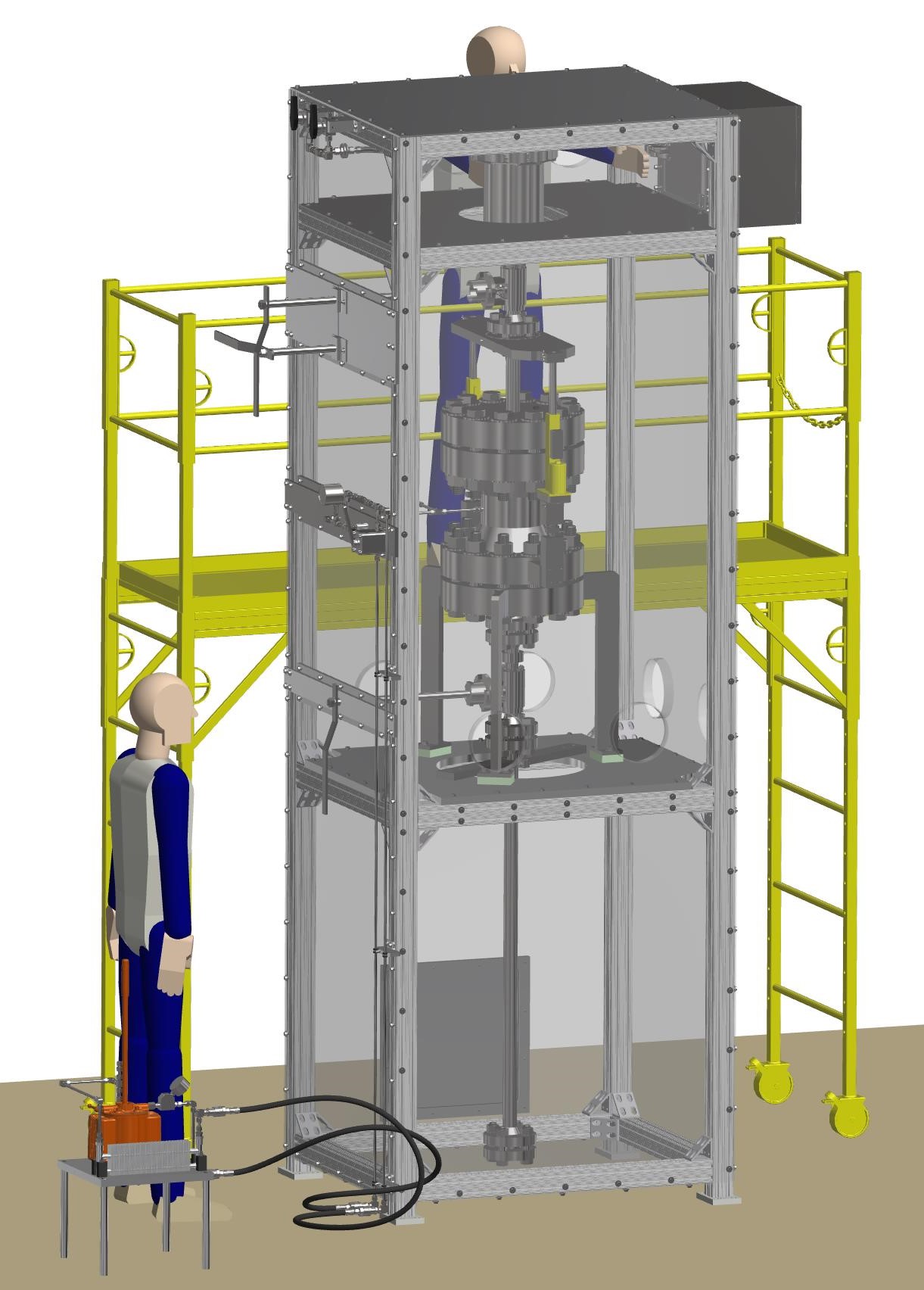}    \caption{\label{fig:licaster-01} Overall view of the $^7$Li Injector development facility at the University of Michigan Ann Arbor.}
\end{centering}
\end{figure}

\begin{figure}
\begin{centering}
\includegraphics[width=0.8\textwidth]{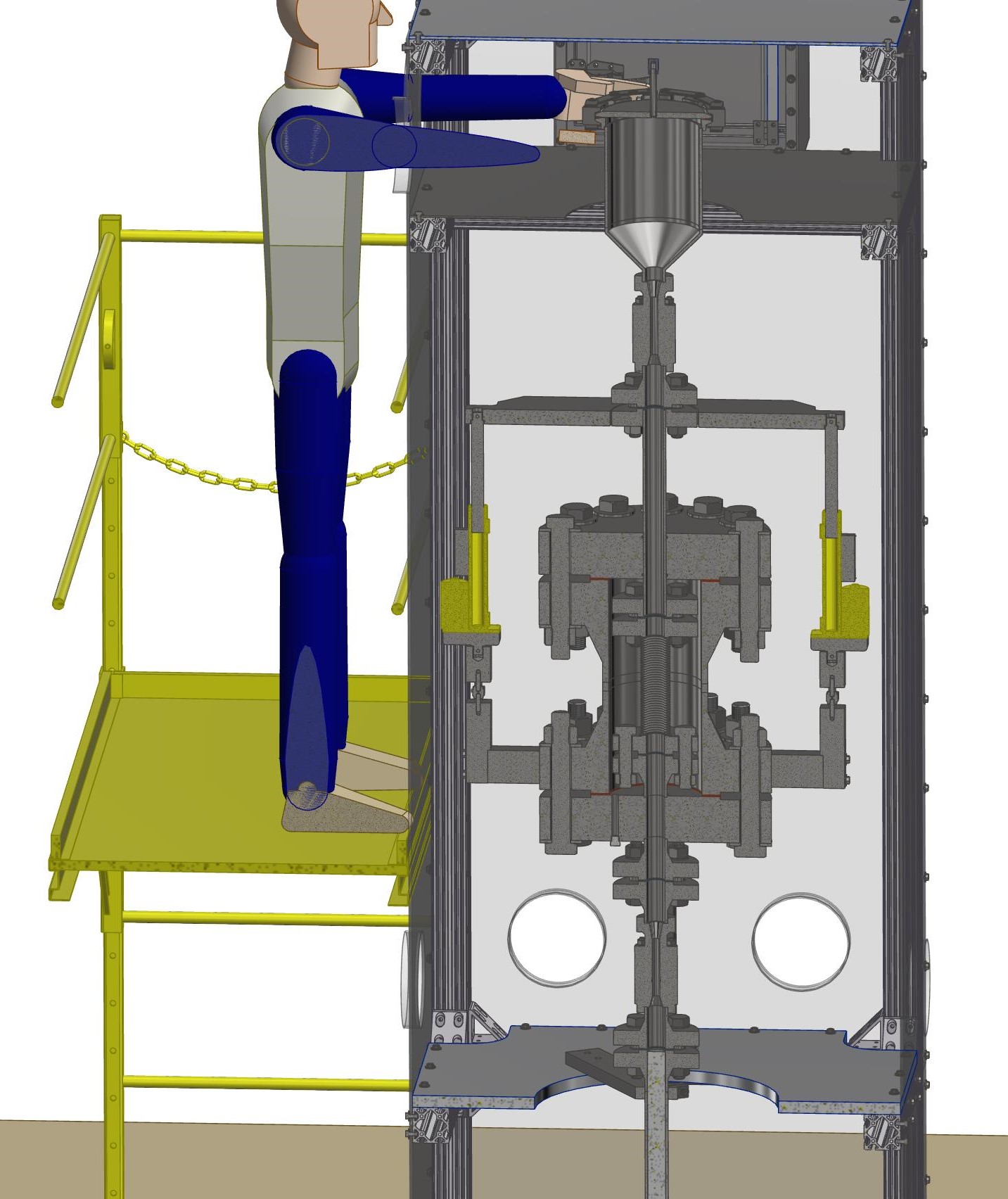}    \caption{\label{fig:licaster-03} Section view close-up of the pressure chamber and the bellows that injects the molten $^7$Li.  The conical vessel at the top is the melting chamber where solid $^7$Li is first melted.  The yellow linear jacks are shown to each side that extend the bellows after an injection.}
\end{centering}
\end{figure}

Towards creating the Li-Be mixture sleeve, we need to mechanically inject molten $^7$Li into beryllium powder. A machine has been designed to accomplish this, modeled after the process used at Fermilab to fill lithium lenses~\cite{1288558}.  The main difference between the Fermilab process and this injection process is that a lithium lens can be filled in a single batch.  The 479 kg of $^7$Li in the IsoDAR sleeve cannot be injected in a single batch, so the design principle of the Fermilab device had to be adapted to a multi-batch process.  The basic idea of the injector is that $^7$Li is first melted in a pot at the top of the injector, then allowed to flow through a valve into a bellows inside a pressure chamber.  Hydraulic fluid surrounds the bellows in the pressure chamber and the pressure of the molten $^7$Li inside the bellows is equalized with the pressure in the oil outside the bellows through the flexibility of the welded bellows.  Once the bellows is full of molten $^7$Li, the top valve is closed and the valve at the bottom of the pressure vessel is opened.  Hydraulic pressure is increased as necessary to cause the bellows to collapse, forcing the molten $^7$Li into the powder in the sleeve.  The entire sleeve assembly will be enclosed in an argon glove box to prevent molten $^7$Li from spilling into the environment and reacting with air in the event of an accident or seal failure.  The argon glove box for the full-size sleeve is not shown in the figures as it has not been designed yet.

The heating of the injector and sleeve pressure vessels is done by wrapping both vessels and their connecting plumbing with electrical heating tapes that can be programmed to heat to different temperatures in different zones.  The heating tapes are then surrounded by insulation to reduce the heat loss from the vessel. 
The $^7$Li must be solidified from the bottom of the sleeve up to the top to avoid large voids in the sleeve, so the bottom heat tapes are reduced in temperature first to begin the freezing of the molten $^7$Li.  None of the heat tapes or their surrounding insulation are shown in the rendered images.

\subsubsection{Hydraulic System}
Figure~\ref{fig:Casteronsleeve-01} shows the sleeve with an early version of the $^7$Li injector connected to it.  Support structures and the argon glove box needed around the injector and sleeve are not shown.  An actuator system was developed later to allow the collapsed bellows in the pressure vessel to be extended for the next fill of molten $^7$Li.  Figure~\ref{fig:Casteronsleeve-02} shows a section view through the hydraulic pressure vessel, showing the stainless welded bellows in the middle.  The upper flange moves up and down with the extension and collapse of the bellows during injection.  The melting pot at the top was also added later, as shown in the figures of the test stand at the University of Michigan Ann Arbor.

Typical hydraulic injection and pressing applications operate at fairly low temperatures, from ambient to approximately 50°C. The $^7$Li injector hydraulic system requires a fluid that can withstand much higher temperatures without significantly impacting performance.  The fluid selected to operate the bellows in the injector is Duratherm-600 from Duratherm Heat Transfer Fluids. Rated to maintain stable density and compressibility to 316°C (600°F), the Duratherm-600 is an environmentally friendly, non-toxic thermal fluid (heat transfer fluid) capable of providing precise temperature control.  The thermal expansion coefficient for this fluid is 0.10\% per degree Celsius which makes it a good selection for this application.

The pressure vessel of the injector shown in Figs.~\ref{fig:licaster-01} and \ref{fig:licaster-03} is constructed from two ASME B16.5 class 1500 8" pipe flanges welded together.  This is a stamped ASME pressure vessel rated for the temperature and pressure of the injection.

\subsubsection{Actuator System}

The actuator system is responsible for dynamically controlling the vertical position and orientation of the $^7$Li injector to ensure proper alignment with the sleeve and maintain optimal filling conditions. It is also responsible for lifting the bellows and resetting the injector for subsequent runs after each injection cycle. The system consists of two linear actuators mounted on either side of the injector, operating in coordination based on feedback from a tilt sensor (MPU6050) and strain sensors (Omega Electronics LC703-150).

Control logic is implemented on an Arduino Uno micro-controller, which receives sensor data and adjusts actuator positions via a Sabertooth dual motor driver. The system supports multiple operating modes, including manual override, lift mode (used to lift the bellows after an injection), leveling mode (stationary with automatic tilt correction), and follow mode (dynamic vertical positioning based on internal pressure changes during injection). Safety features include a force shutdown mode and configurable motion tolerances to prevent mechanical damage. Furthermore, the system calibrates the initial pressure and tilt states at startup, allowing additional components to be mounted or adjusted prior to injection without requiring manual recalibration.

\subsubsection{Thermal Control System}

Manipulation of $^7$Li in the molten state requires careful temperature control through the desired processing.  A multi-zone thermal control system has been developed to control viscosity and solid/liquid state changes as $^7$Li moves through the injection process.

\begin{figure}
\begin{centering}
\includegraphics[width=\textwidth]{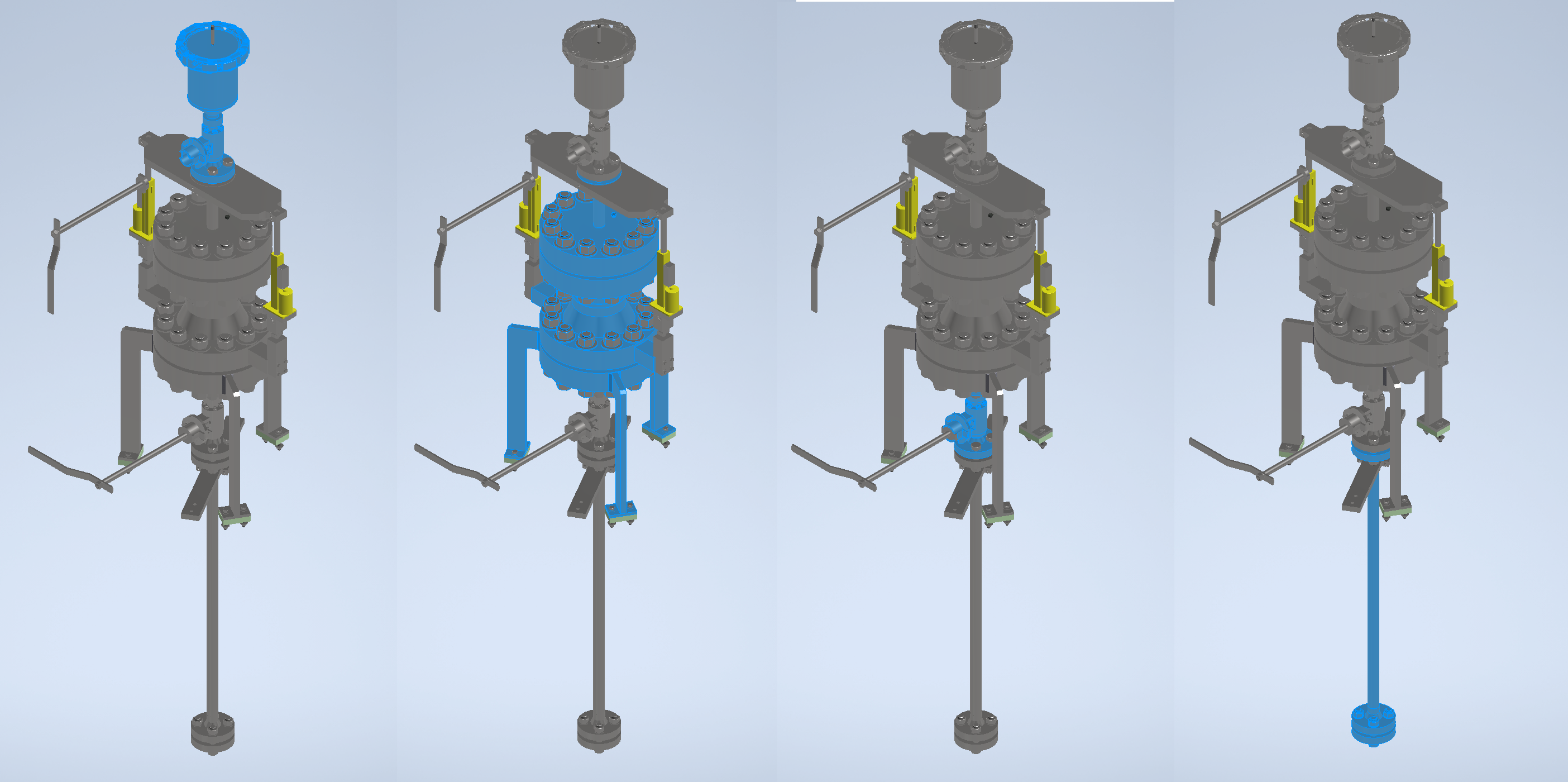}  
\caption{\label{temp_control} Regions within the injector have their own temperature control.}
\end{centering}
\end{figure}

\paragraph{Multi-zone Design}
The IsoDAR injector uses an electronic temperature control system to maintain molten $^7$Li during the injection process.   Tailoring the injector machine across multiple stages to an ideal temperature allows rapid heating to melt solid $^7$Li early in the process and permit controlled cooling to produce the most homogeneous possible $^7$Li-beryllium sleeve cast.  Zones in the heating system include an initial stage that can move vertically as the injector compresses the bellows, a large middle section and the target tube, which needs two temperature zones to ensure cooling from the bottom first. The various temperature control regions can be seen in Fig.~\ref{temp_control}.

\paragraph{Heating Control Hardware}

\begin{figure}
\begin{centering}
\includegraphics[width=\textwidth]{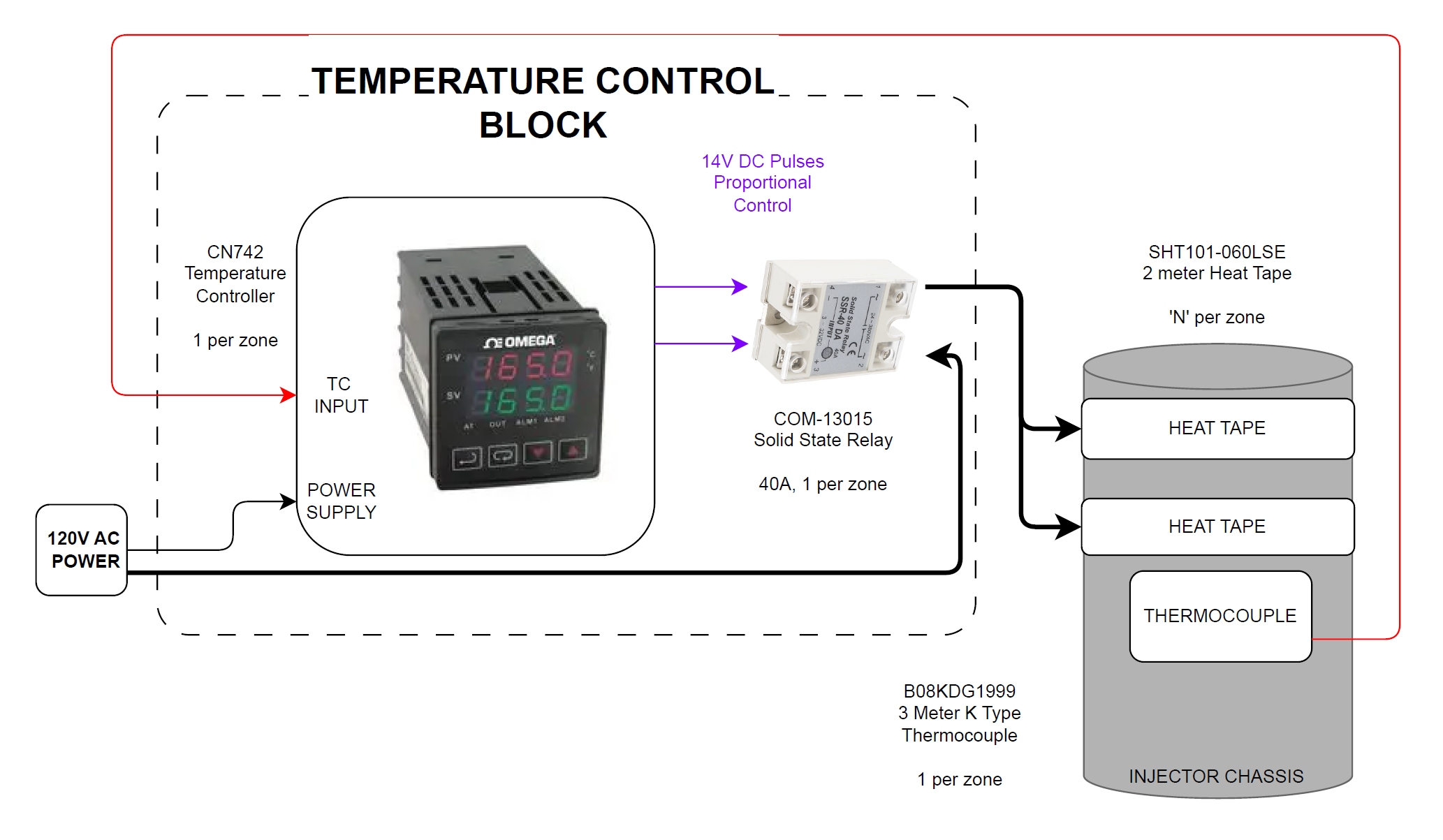}   
\caption{\label{fig:thermal-02} Simplified block diagram for each temperature control zone.  Heat tape and thermocouples provide the action and feedback for this closed loop system.}
\end{centering}
\end{figure}

The thermal control system is composed of several identical components, each serving the same function while maintaining individual temperatures for the different zones of the injector. A schematic of the system is shown in Fig.~\ref{fig:thermal-02}.

\textbf{Temperature controller}
The CN742 from Omega Engineering serves as an ideal proportional integral derivative controller. It can receive input from a wide range of temperature sensing hardware and can activate downstream hardware using direct relay switching, 14V DC pulses or proportional current. The control increases or decreases the output proportionately to the process temperature’s deviation from the set point. The integral time eliminates undershoot and overshoot of the set point by adjusting the proportioning control based on the amount of deviation from the set point during steady state operation. This control scheme maintains smooth transitions and eliminates overshoot and undershoot as the temperature approaches the target.

\textbf{Relay}
The power demand for raising steel to over 200°C exceeds the capability of the temperature controller.  To reduce the energy demanded of the temperature controller directly, an electrical relay is employed to power and control the heat energy added to the system. Operating the heat tape at 208VAC reduces current draw for the same power added to the system permitting smaller diameter wires and cables.

\textbf{Heating tape}
All electrical heating tape uses resistive heating to convert electrical energy into thermal energy. The heating tape used in the IsoDAR injector machine is made from fine gauge stranded resistance wires that are double insulated with braided Samox and knitted into flat tapes for maximum flexibility. The steel surface of the injector machine is completely covered with Omega STH-102 heat tape in various lengths.

\textbf{Thermocouple}
Sensing the temperature of the $^7$Li within the injector is the desired condition. The high pressure injection process and the lithium’s general incompatibility to ceramic and other silica makes direct measurement difficult.   A compromise solution is to measure the temperature of the objects in contact with the $^7$Li.   This means reading the temperature of the steel hopper, the lower valve and the outer surface of the target tube.
A better observation of the temperature of the $^7$Li contained in the pressure vessel is to observe the temperature of the Duratherm-600 hydraulic fluid in the injector.    The injector was designed with threaded ports to permit a slender thermocouple probe to contact the fluid in the hydraulic vessel.
\paragraph{Thermal Control Box}
To raise the temperature of the 1800 lb steel injector to over 200°C and maintain this temperature through the injection requires substantial power. Summing the power capability of all heat tapes, operating at close to 100\% duty cycle, (as is the case starting from ambient temperature) results in a power draw of 22 kW. 

\begin{figure}
\begin{centering}
\includegraphics[width=1.0\textwidth]{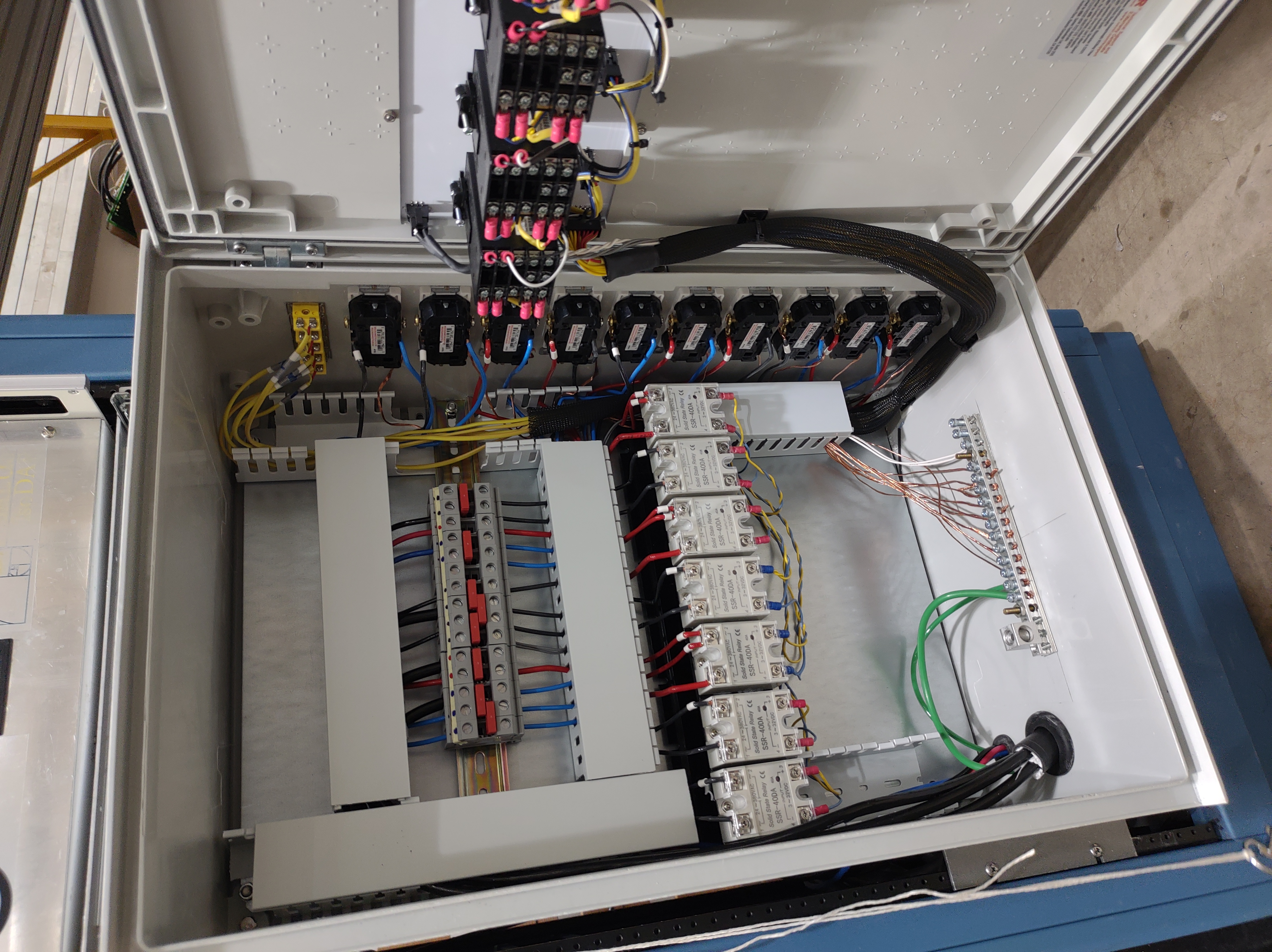}   
\caption{\label{fig:thermal-03} Mounting, wiring and power distribution are done within a fiberglass control box.}
\end{centering}
\end{figure}

To provide sufficient power and control, as well as reliable and robust connectivity, a central control panel box has been fabricated. A picture of the box can be seen in Fig.~\ref{fig:thermal-03}.
The thermal control box provides robust connectivity and firm mounting for all the components of the control blocks.   Heat tapes are connected to the system using commercial off-the-shelf plug and receptacle connections and route from the control box to the injector patch panel near the floor. The process controllers are mounted to a custom designed printed circuit board on the front panel. This circuit board interfaces each of the zone controllers with AC power, thermocouple connectivity and connections between the control and relays. The front panel design allows observation of the whole thermal control system, displaying both the process temperature and the target temperature for each of the five zones. The user controls are also located at the front panel, simplifying the transition from heating and injection to cooling and shutdown.

The Michigan sleeve injector test stand is poised to begin operations in Summer 2025. A small-scale sleeve cast will be created and analyzed with CT techniques in order to determine its homogeneity, among other quality metrics. In case we are satisfied with its properties, and with an eye towards creating the full-scale sleeve with the same device and technique, we may create further small-scale casts under different pressure and temperature (in terms of gradient and temporal) conditions. We expect that a publication describing this work and the results will be produced shortly after the tests. 

\subsection{Risks and Mitigation}

\noindent \textbf {Risk: availability of $^7$Li}  

\textit{Mitigation: Highly enriched $^7Li$ is an essential component in the nuclear industry.  Present-day light-water reactors utilize lithium hydroxide as a buffering agent, typically enriched to 99.99\%.  As discussed previously, a minimal isotopic fraction of $^6Li$ is required for optimizing antineutrino production.  The emerging generation of molten salt reactors utilize FLiBe as a primary agent, requiring also high enrichments.  We are aware of several independent projects for enrichment of $^7Li$, and are in possession of a cost-quote from one company to provide the amount and enrichment level we require.  Should this vendor not come through, we are certain there will be others.  The factor remains that we require metallic $^7$Li, not FLiBe nor LiOH.  We have verbal assurances that this will not be a problem, though it needs to be carefully specified in our contracting documents for the material. }

\clearpage
\section{Shielding Design}

\subsection{Shielding Requirements}  

Responsibility for development and deployment of the IsoDAR experiment rests jointly with the IsoDAR Collaboration and the South Korean Institute for Basic Science (IBS), who will build the $\nu$EYE detector and operate the Yemilab facility. 
While the IsoDAR Collaboration will provide the cyclotron, transport line, and target complex, it is expected that IBS will provide the basic infrastructure for such items as electrical power, water cooling, air handling.  
Radiation protection will be a joint responsibility:  IsoDAR will do everything possible to reduce the source of radiation by minimizing beam loss, and provide local shielding for the cyclotron, collimator areas, and target.  
Standards and Korean regulatory requirements will be determined by IBS.
These include allowed limits for personnel access, and residual activation of the rock in the cavern walls following decommissioning of the experiment. This is discussed below. 

An important component is the shielding between the target and $\nu$EYE; the basic requirement is that the interaction rate of fast neutrons in the detector originating from the target that are above 3~MeV must be kept at or below the natural rate of these particles, at most a few per year.  This is discussed more below.

\subsection{Target Shielding}

The full beam from the IsoDAR accelerator is required to be dumped into the target.  Extensive Geant4 studies of the neutrons produced in the target have been performed to optimize the design of the  sleeve surrounding the target, maximizing $^8$Li production, i.e. our desired outcome.  However, the majority of neutrons produced will not be absorbed by $^7$Li or $^9$Be and will need to be contained as best as possible to minimize adverse effects should they escape and activate surrounding material.

 \begin{figure}[b!]
    \centering
\includegraphics[width=1.0\textwidth]{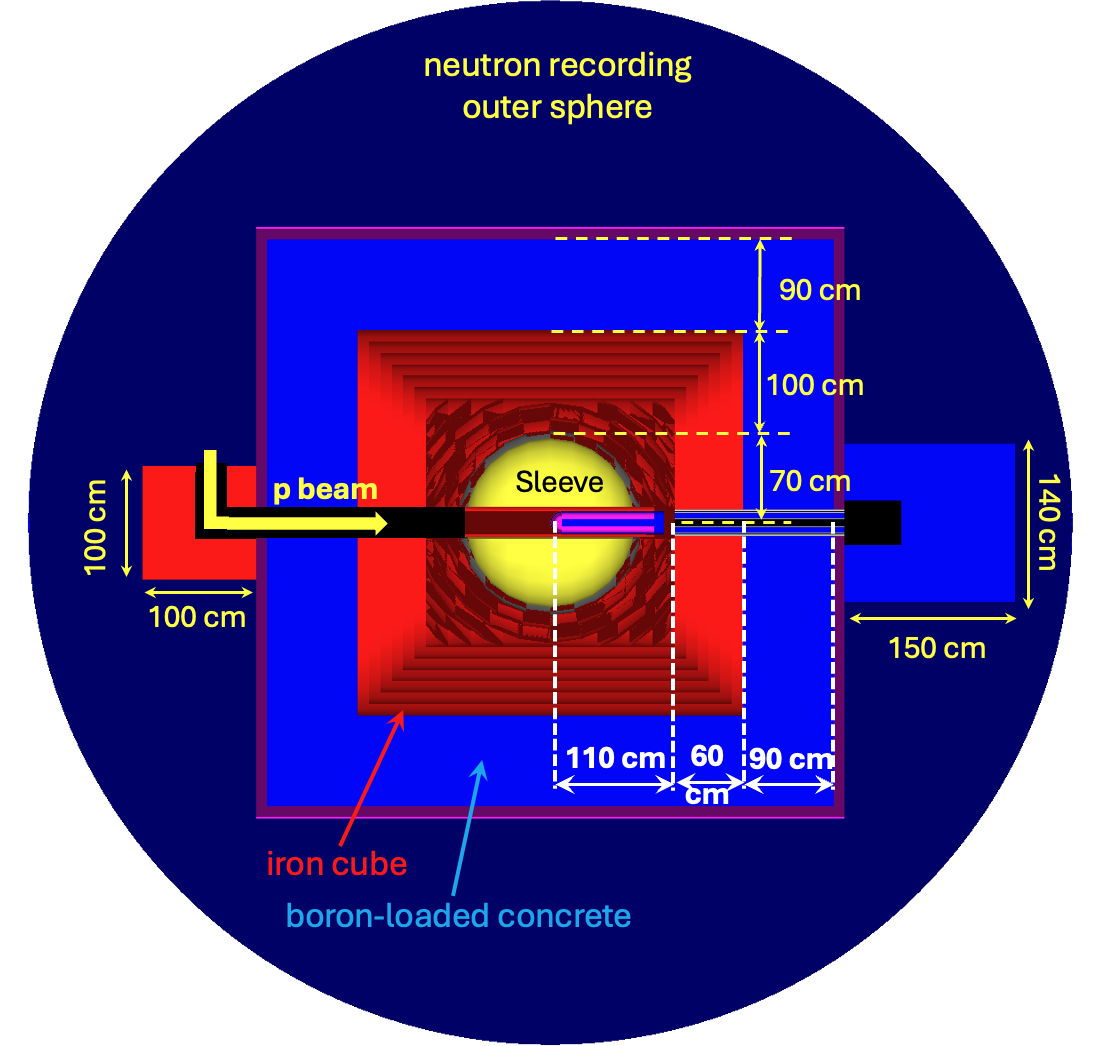}
    \caption{Concept of neutron shielding around the target: a block of steel surrounded by boron-loaded  concrete with thicknesses optimized for environmental shielding.  The $\nu$EYE detector is to the left.  The blue outer circle represents the 4-meter radius recording sphere relevant to Geant4 calculations.}
    \label{fig:TargetShieldingBlock}
\end{figure}

Figure~\ref{fig:TargetShieldingBlock} shows the design that has evolved for shielding around the target and sleeve.  It is comprised of iron and borated concrete.  The dimensions were originally established by requirements for rock activation, discussed in the next section, in the environment of KamLAND~\cite{KamLAND:2008dgz}, the original site planned for IsoDAR.  
It is an ultra-conservative design for the Yemilab environment, but as it will fit within the space available, it serves as a satisfactory preliminary design.  
Thicknesses can be reduced in a later design stage, as part of cost reduction and value engineering.

 \begin{figure}[t!]
    \centering
\includegraphics[width=0.75\textwidth]{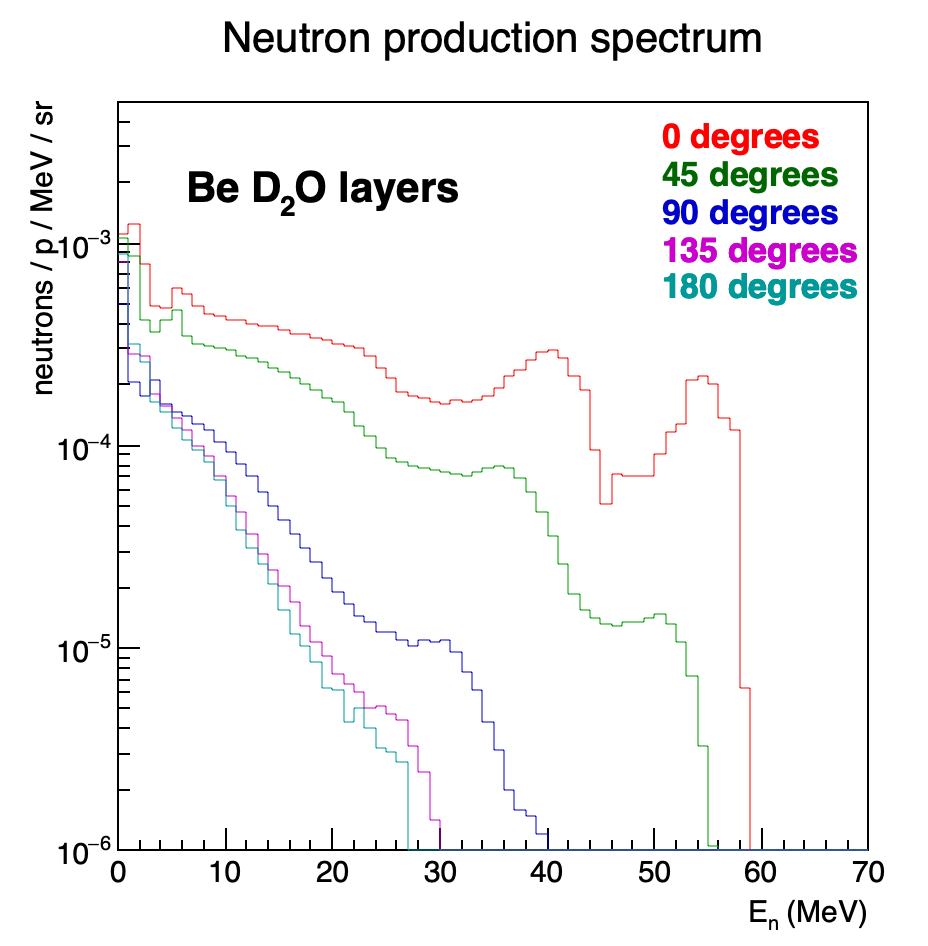}
    \caption{Neutron spectra from the hemispherical target with beryllium shells interspersed with cooling D$_2$O channels.  The highest-energy peak represents neutrons produced in the outer hemisphere, with the next peak at about 40 MeV representing neutrons produced in the second hemisphere, as the protons have lost energy before reaching this hemisphere~\cite{bungau2024neutrino}.}
    \label{fig:NeutronSpectra}
\end{figure}

As discussed in Sec.~\ref{BeamOnTarget}, the chosen orientation of the beam striking the target is the direction opposite the $\nu$EYE detector.  The rationale is presented in Fig.~\ref{fig:NeutronSpectra}, where neutrons emitted in the backward direction are lower in energy (peak energy is below 30 MeV vs. 60 MeV in the forward direction), and the total neutron flux is also vastly reduced. The drawback of this design, however, is that there is a ``hole'' in the shielding where the beam pipe brings the beam to the target.  Though the opening angle subtended by this pipe is very small, the flux of neutrons exiting the shielding is still substantially larger than from any other angle.

\begin{figure}[t!]
    \centering
\includegraphics[width=0.75\textwidth]{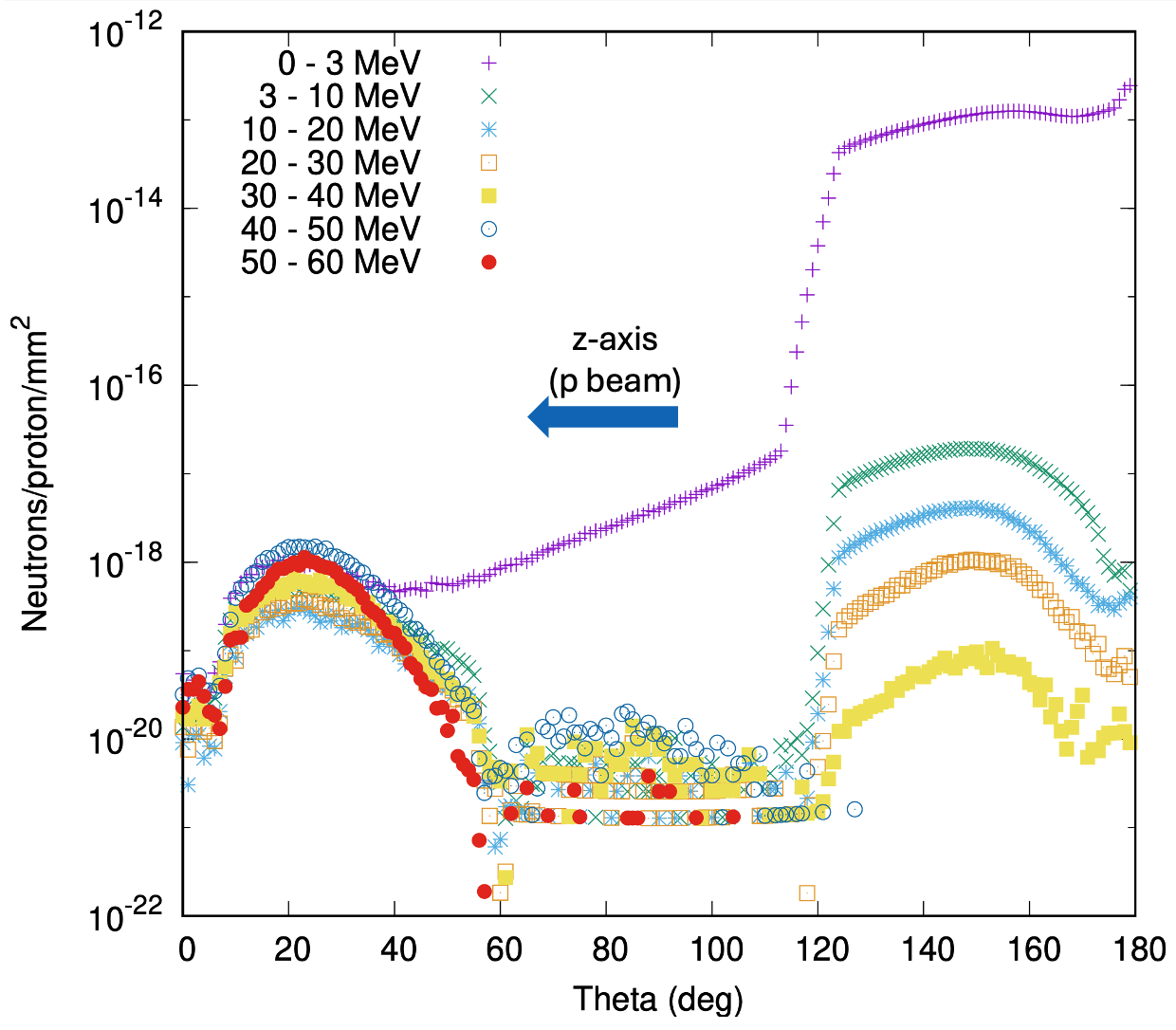}
    \caption{Neutron flux seen on the recording sphere with radius 4 meters centered on the target.  Attenuation factors of $10^{-10}$ or higher are seen for neutrons penetrating the bulk of the shielding, but more neutrons escape from the ends, in particular neutrons escaping through the beampipe are scattered in all directions from the last bending magnet steel, accounting for the steep rise just beyond the shadow from the edge of the concrete~\cite{bungau2024neutrino}.}
    \label{fig:NeutronsThruShield}
\end{figure}

A Geant4 study of the neutrons seen outside the shielding block for the geometry of Fig.~\ref{fig:TargetShieldingBlock} is shown in Fig.~\ref{fig:NeutronsThruShield}.  The vertical axis represents the neutrons (per incident proton on target) passing through one mm$^2$ on the surface of the 4-meter radius reference sphere versus azimuthal angle along the horizontal axis.
As there are about $10^7$ square millimeters on the reference sphere, one might view the neutrons absorbed in the shielding, for high-energy neutrons normal to the beam direction to be about a factor of $10^{13}$. Note, Geant4 only records the neutron passing through the sphere, not its direction vector.  The sharp rise in the backward direction at $120^\circ$ represents neutrons emitted in the backward direction emerging from the beam pipe and scattering off of the steel in the last bending magnet.  The steep rise represents the shadow of the corner of the concrete edge of the target shielding block.  These neutrons pass through the recording sphere at an angle slightly less than $90^\circ$.
The lower attenuation for neutrons below 3 MeV is not important for shielding for $\nu$EYE, but may lead to more activation in the local environment.

\subsection{Short Term Radiation Effects: Personnel Access Considerations}

While the accelerator RF is on, beam is being produced in the cyclotron and delivered to the target. During these times, radiation levels inside the IsoDAR areas will be exceedingly high.  This is due to x-rays from the cyclotron RF system, and  prompt radiation from nuclear reactions between the beam and material in the transport line, and ultimately from the target where the beam is intentionally stopped. During these times personnel cannot enter the areas so marked for personnel exclusion.

In addition, the above-mentioned nuclear reactions will almost always lead to radioactive products that decay with finite half-lives. After the beam and RF systems are interrupted, personnel may still be excluded until the radiation levels, measured by area monitors inside the caverns, drop to levels determined by local regulations to be acceptable for personnel access. Then, a radiation technician will enter with a radiation meter to verify the background levels.  This person will also survey and rope off areas likely to exhibit higher radiation fields, usually around beam line elements where beam is lost.

While the design of the experiment is to have all of the beam strike the well-shielded target, transport can never be perfect, and some beam will be lost between the cyclotron and the target. We can examine how beam can be lost, and describe tools for estimating the effects of this loss.

The first area of potential loss is at the stripping section.  Some neutrons will be produced in the stripping foil, though this is sufficiently thin that nuclear reactions will be rare.  The largest source of loss will be the collimator along the path of the protons from the analyzing magnet.  This collimator must be adequately cooled, and shielded to minimize radiation fields outside the immediate area.   
The shield will be a combination of borated polyethylene to slow down and absorb neutrons, and high Z material to shield gammas.

The MEBT is roughly 60 meters long, and losses can occur along this length.  Losses can occur by beam interactions with the residual gas in the beam pipe, which will affect the whole beam, or with the walls of the vacuum enclosure for particles whose orbits stray sufficiently from the central axis.  
Extrapolations from calculations of \htp losses inside the cyclotron due to gas stripping \cite{calvo_analyzing_2023} give confidence that by transporting protons instead of \htp and maintaining a pressure in the beamline better than $10^{-6}$ torr, we will see essentially no interactions with residual gas.  

With a vacuum pipe of 10 cm diameter, and a strong FODO lattice transport design, the edges of the beam will be far from the walls of the pipe.  Beam halo still must be evaluated.    
Studies indicate~\cite{allen2002beam} that halo can result from mismatching the front of the transport lattice with the incident beam parameters at our current-level\footnote{At 60~MeV, the 10 mA beam of IsoDAR (while a high cw-current) does not exhibit strong space-charge effects. Space-charge effects are most pronounced with high peak beam currents or at low energies. The OPAL simulations discussed in Section~\ref{TransportLine} confirm that space charge has a very low effect on the beam transport.}. Should detailed calculations point to difficulty in achieving proper matching, leading to the likelihood of halo, shielded scrapers can be inserted in the beam line to generate ``controlled'' loss points.  
Protons lost on the scraper will undergo nuclear reactions, but placing the scrapers inside a well-designed shielding block should contain reaction products and significantly attenuate neutrons.
In the end, if we can identify the location and amount of beam loss, methodology employed in a study of activation of beamline elements for the European Spallation Source~\cite{bungau2014induced} can be employed to evaluate radiation fields around the cyclotron and MEBT magnets.  The massive shielding around the target is discussed in the next section.

\subsection{Long Term Issues:  Rock Activation}

The underground environment of the Yemilab area is inherently one where radiation levels are very low. 
This will not be the case in the IsoDAR areas when the experiment is running.  
The neutrons in the environment will activate components of the experimental equipment:  magnets, pumps, instrumentation,  electronics, and the rock of the caverns.  The equipment brought in for the experiment  will be removed at the end of the run.  The rock in the surrounding walls, however, cannot be that easily removed.  
A legal requirement, then, is that when the experiment is over, and fully decommissioned, the remaining activation in the rock must meet a regulatory standard; For South Korea, this is less than 10 Bq/g. 
This is interpreted as follows:  one does a survey of the rock in the cavern, finds the hottest area, chips off a small sample, weighs it, and measures the activity level.  
This level must be less than 10 disintegrations per second for each gram of the material.

We  are at liberty to choose when the experiment is officially over and the time when the assay is conducted.  
We expect to be able to wait about two years for the cyclotron and other activated components to cool to the point where they can be safely removed.  
The decommissioning and removal will take another year or two, so the date for conducting the assay can be four or five years after the final ``Beam Off''.
We need to be concerned, then, only with isotopes with half-lives over about 2 years; isotopes with shorter half-lives will have mostly decayed away.  Table~\ref{tab:LongLivedIsotopes} lists the long-lived isotopes we need to follow, with their half-lives and production channel.

\begin{table}[h]
\centering
\caption{\label{tab:LongLivedIsotopes} Long-lived isotopes produced.}
\begin{tabular}{c|c|c|l}
Isotope & Half-life & Parent in rock & Production channel \\
\hline
$^{22}$Na & 2.6 years & $^{23}$Na  & (n,2n) fast 11 MeV threshold \\
$^{60}$Co & 5.3 years & $^{59}$Co  & (n,$\gamma$) slow \\
$^{152}$Eu & 13.5 years & $^{151}$Eu & (n,$\gamma$) slow \\
$^{135}$Eu & 8.6 years & $^{153}$Eu & (n,$\gamma$) slow \\
\end{tabular}
\end{table}

To quantitatively study this, we need to know the composition of the rock, and the concentration of each of the parent isotopes, then we need to evaluate the neutron flux striking the wall and how it activates the relevant isotopes.
These studies were first undertaken in 2020 for the original site at KamLAND~\cite{bungau_shielding_2020}. 
At this site, the Japanese regulatory limit was 0.1 Bq/g of total activity. 

\begin{figure}[t!]
    \centering
\includegraphics[width=\textwidth]{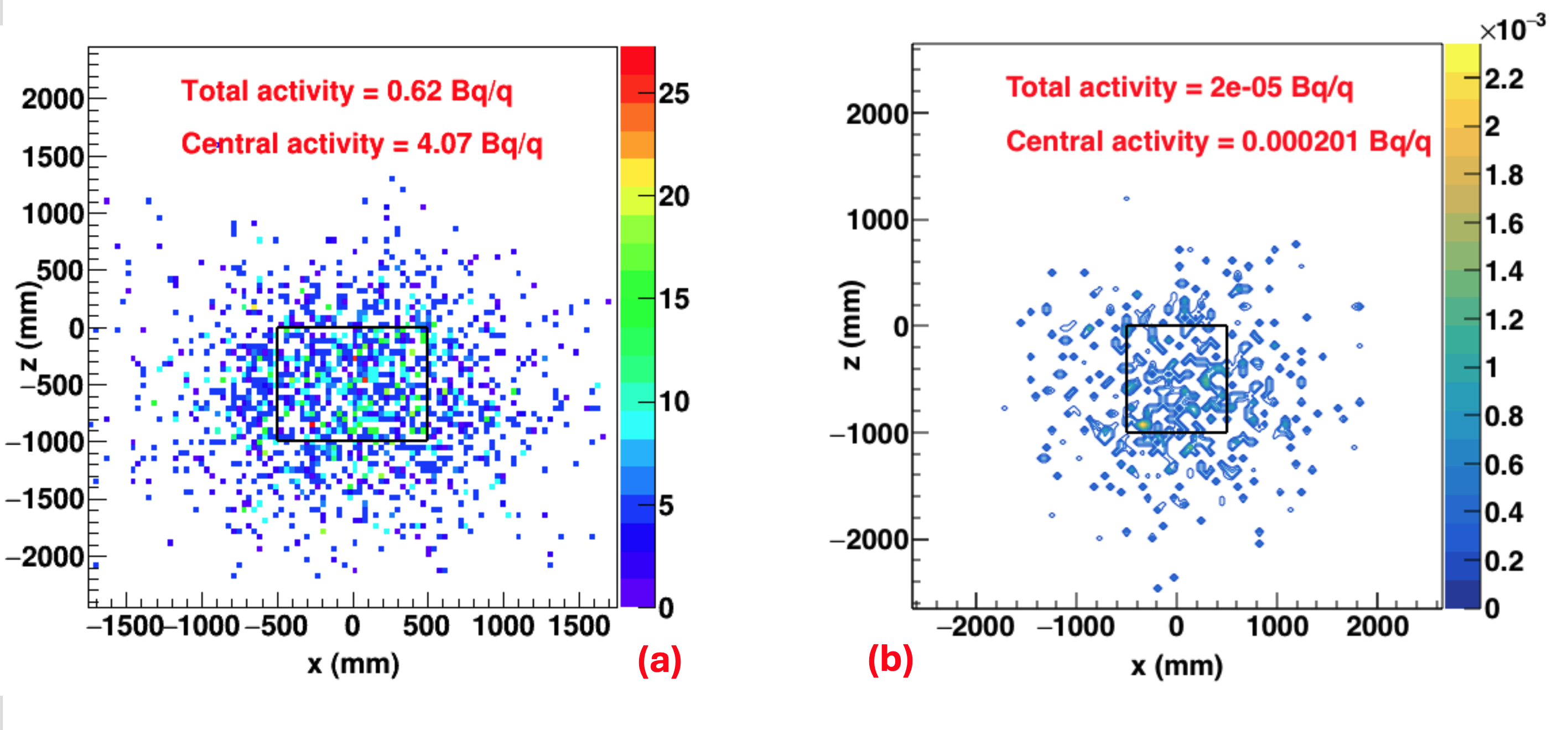}
    \caption{Geant4 calculation of activation on the surface layer of cavern rock at the Kamioka Observatory after 5 years of running and 2 years of cooldown for shielding blocks of (a) 40 cm of steel and 80 cm of boron-loaded concrete, and (b) 1 meter of steel and 2 meters of concrete. The point (0,0) is directly over the target.  Notice the differences in the scales.}
    \label{fig:RockActivation}
\end{figure}

Geant4 studies were performed to calculate the activity level after 5 years of full-intensity running, and 2 years of cooling. 
This is perhaps an overestimate as we stated above that in all likelihood we could count on 4 years of cooldown.
The studies considered two different target shielding configurations from which neutron flux was evaluated that impinged on the rock walls.  
The rock was simulated by a total of 20  5-cm rock layers that were placed directly over the shielding block, and activity of the isotopes listed in Table~\ref{tab:LongLivedIsotopes} produced in each of these layers was evaluated, reaching a total depth of 1 meter into the rock.
As might be expected, the maximum activity was seen in the first two layers, and because the assay for regulatory compliance would be done by scraping material from the surface, only the first layer is of real interest.
Figure~\ref{fig:RockActivation} shows the results of these  studies,  (a) for shielding of 40 cm of steel and 80 cm of concrete, our original concept, for which the total activity level is substantially above the 0.1 Bq/g requirement, and (b) for doubling the shielding to 1 meter of steel and 2 meters of concrete.   The second, essentially what is shown in Fig.~\ref{fig:ShieldPlan}, is our current baseline design.  

As stated earlier, the Korean standard is 10 Bq/g, a factor of 100 over the Japanese one, so even the original shielding thickness would be satisfactory.  

There is another aspect too: the rock composition at Yemilab, essentially pure limestone, is substantially different from that at Kamioka.  
Table~\ref{tab:RockComparison} compares the rock assays of the two sites.  Most noteworthy is that sodium content is greatly reduced, which significantly affects the production of $^{22}$Na, the primary activation product at Kamioka. 

\begin{table}[h]
\centering 
\caption{\label{tab:RockComparison} Comparison of rock assay at Yemilab and Kamioka.}
\begin{tabular}{c|c|l}
Element & Kamioka & Yemilab  \\
\hline
Sodium & 6.4\% &  0.022\%   \\
Cobalt & 16 ppm &  6 ppm\\
Europium & 1 ppm & 0.2 ppm \\
\end{tabular}
\end{table}

We have not repeated the calculations, as we feel that rock activation will not be a dominant factor at Yemilab.  
The overall shielding requirement will probably be driven by short-term activation and the need to access the cavern shortly after beam is interrupted. 
Again, though, reducing the target shield will be a significant value-engineering benefit.

\subsection{Controlling Neutrons Above 3 MeV in the Detector}

\begin{figure}[t!]
    \centering
\includegraphics[width=0.9\textwidth]{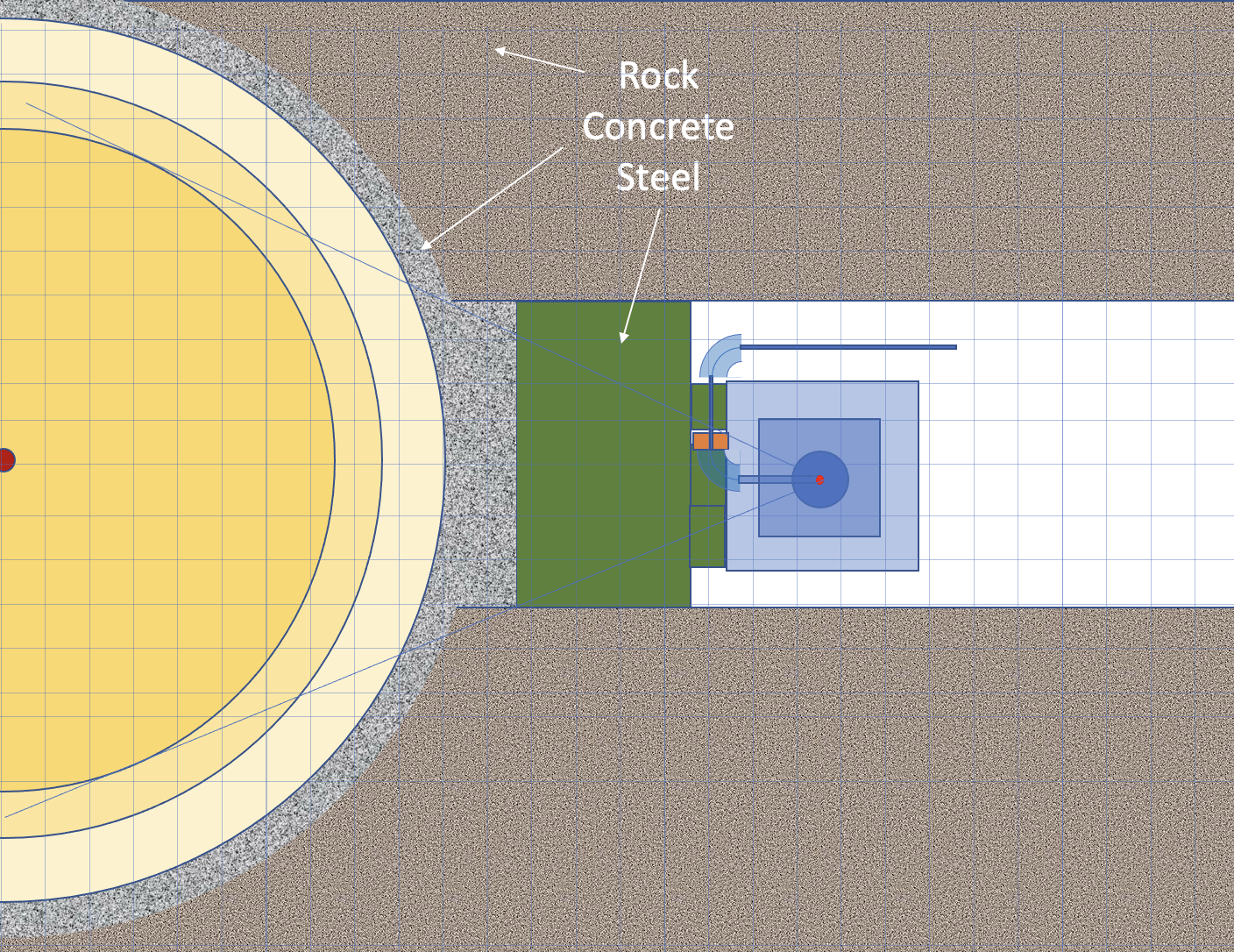}
    \caption{Concept of neutron shielding in IsoDAR target area (blue): steel and concrete block for environmental shielding, and thick steel block to keep neutrons above 3 MeV from the fiducial volume of the $\nu$EYE detector (yellow).}
    \label{fig:ShieldPlan}
\end{figure}

Events depositing more than 3 MeV of energy in the $\nu$EYE detector are flagged as potentially due to signal antineutrinos. Backgrounds below this energy from natural sources inside or outside the detector are relatively high compared with expected data rates~\cite{bungau2024neutrino}, leading to our decision to use a cutoff at this energy.  For neutrons above the 3~MeV threshold, we set our limit equal to the estimated neutrons
produced by the penetration of high-energy muons at the Yemilab depth, which is 3 neutrons per year. Our shielding calculations were performed with this goal in mind. To accomplish this we must place as much iron shielding as necessary, filling the space between the beam line and target and the aperture to the $\nu$EYE cavern.  
This is shown schematically in Fig.~\ref{fig:ShieldPlan}, with the  block of steel (shown in green) being about 3~m thick.  

Geant4 studies established the necessary steel thickness from an estimated attenuation requirement, derived as follows.
For 10 mA of beam on target, the number of protons per year is $2 
\cdot 10^{24}$.  
We can estimate the neutrons emitted towards the detector that must be absorbed using this, and the information in Fig.~\ref{fig:NeutronsThruShield}.  We take as the source term for the Geant4 studies those neutrons crossing the recording sphere at angles greater than $120\degree$, that on this graph show an attenuation factor of at least $10^{-17}$ (neutrons per proton per mm$^2$).  We can take this as a conservative requirement (higher energy neutrons are attenuated one or two orders of magnitude more).  As stated earlier, there are about $10^7$ mm$^2$ on the recording sphere, so in this region the effective attenuation is of the order of $10^{-10}$ for the total neutron number.  The solid angle subtended by the $120\degree$ cone represents about 25\% of the total, the total number of neutrons we need to absorb is around $5 \cdot 10^{13}$.
 
\begin{figure}[t!]
    \centering
\includegraphics[width=0.9\textwidth]{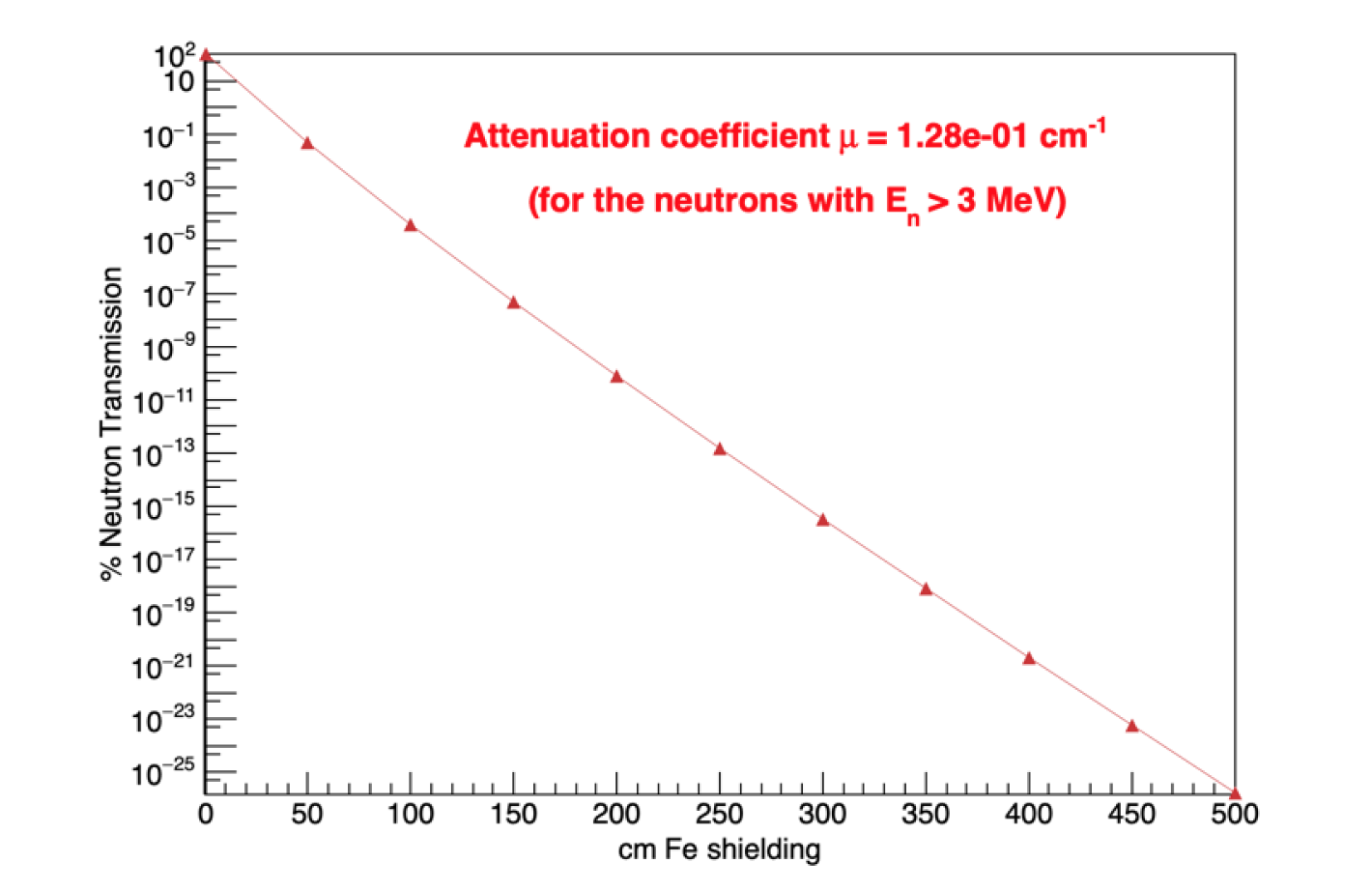}
    \caption{Attenuation factor for neutrons above 3 MeV penetrating through iron~\cite{bungau2024neutrino}.}
    \label{fig:AttenuationInIron}
\end{figure}

Evaluating this level of attenuation with a Monte Carlo code is difficult, but can be accomplished in a straightforward manner by segmenting the problem.  
One starts with a 50~cm block of iron and establishes the spectrum and distribution of neutrons emerging from its face.  This is used as the starting source for the next 50~cm block of iron, and is repeated as needed.
Figure~\ref{fig:AttenuationInIron} shows the total transmission of neutrons through an increasing thickness of iron blocks.  
To achieve our required attenuation will require about 2.5 to 3.0~m of iron.  

Some comments about the assembly of this thick shield are in order. First, the shape of the opening that must be filled is irregular, with a domed profile:  base is 7~m across, walls are 3.5~m high, and top has a semicircular profile with a 3.5~m radius.  
For a total thickness of 3~m, the volume of iron is 131~m$^3$, the weight is about $10^3$ metric tons.
A significant concern is the support of this weight by the rock located at the ledge of the larger $\nu$EYE detector cavern.  The floor of the IsoDAR cavern is about 10~m above the base of the detector cavern, and this weight located right at the lip is likely to cause an unacceptably high compression load on the rock in this area.  
Support for this weight of iron will need to be carefully engineered, and installed prior to the assembly of the detector. Secondly, this level of neutron attenuation will require very careful design of the iron structure--it must fit tightly with the irregular surface of the shotcreted rock walls, and must also not have cracks that neutrons can propagate through.

\subsection{Risks and Mitigation}

\noindent \textbf {Risk:  Supporting shield wall.}  The 3-meter thick iron wall filling the opening between the target and $\nu$EYE weighs about 1000 metric tons.  It will be installed close to the lip of the target hall as it opens into the detector cavern.  It is approximately 10~m above the floor of the detector cavern.  If this floor is not properly reinforced, there is a risk that it may crumble, dropping the iron block into the detector cavern.

\textit{Mitigation:  A thorough engineering study should be performed, and an engineering design developed to ensure adequate load-handling capability of this floor.  It may consist of steel columns installed directly under the iron block that extend down to the floor of the detector cavern.}

\bigskip
\noindent \textbf {Risk:  Neutron counts in $\nu$EYE are higher than anticipated.}  When the beam is turned on, the background of neutron events with energies higher than 3~MeV is higher than anticipated.

\textit{Migitation:  (1) Ensure that Geant4 calculations are thorough and valid. Calculations must include all paths for neutrons to reach the fiducial volume, including scattering through the rock and avoiding the iron wall.  The current attenuation studies do not cover this.  (2)  Ensure that the mechanical design of the iron wall contains no paths for neutrons to slip through, and that installation has not introduced any such paths.  Critical areas will be cracks between steel blocks, and space between the concrete walls and iron pieces.  }

\clearpage
\section{Monitoring and Instrumentation}
\label{monitoring_chapter}

In this section we address beam safety, describe beam line instrumentation and area monitoring, the data acquisition and control systems, as well as tuning and commissioning strategies.

\subsection{Beam Safety and Tuning}

Full-intensity beam extracted from the cyclotron will be 5 mA of \htp ions, or after stripping 10 mA of protons.  This is 600 kW of beam power.  
This level of beam can be extremely hazardous if not properly controlled.  Providing a means of containing the full beam if it deviates from its intended path is challenging. 
In fact the actively-cooled target itself is the only component capable of absorbing the full beam power. 
We must rely on quick and sensitive instrumentation to detect imminent failures that might lead to uncontrolled beam losses, and a safety system that responds as quickly as possible to cut beam off before it does excessive damage.

The best strategy is to interrupt the beam before it reaches the cyclotron.  
This will be primarily by the Low Energy Beam Transport (LEBT) chopper, which should be activated in less than a microsecond, followed by clamping the high voltage supplies in the ion source and the spiral inflector.  The main RF system powering the RFQ and cyclotron dees can be shut off, but because of the high stored energy in these systems they need to be handled more carefully.  While the LEBT chopper provides fast cutoff of the beam, it is not fail-safe as it must be energized to prevent beam passage.  Including high-voltage supplies or RF systems that must be on for beam to pass through satisfies the overall safety requirement for the cutoff system.

The LEBT chopper plays a truly important role in that beam must be brought up slowly to full power, and the only way of limiting beam power is by reducing the duty factor.  
The vortex motion effect requires bunches within the cyclotron to contain the designed full amount of charge, which means that reducing overall beam power can only be done by reducing the number of bunches.
In this regard it should be noted that the chopper risetime should be as quick as possible.  Ideally one would want to cut the beam off within the time for a single bunch, about 30 nanoseconds.  If the risetime is slower then partially-filled bunches will be accelerated while the chopper voltage is rising and falling, possibly leading to higher beam losses in the cyclotron.

During commissioning and tuning, the efficiency of beam transport to the target will be low, so keeping the duty factor low is imperative to prevent these beam losses from causing damage or excessive activation to components.  
In fact, the progress from initial startup to full power will be dictated by keeping beam losses below a fixed amount, this amount being determined during commissioning by characterizing the overall system performance, and establishing what the ultimate efficiency of the entire system can be.  As efficiency improves and beam losses decrease, the duty factor can be increased to increase beam power to the next commissioning level.

The trip point for the safety system needs to start out being very low, as access to the vaults will undoubtedly be frequent and activation levels must remain low during this period for rapid access.

Instruments that need to be included in this safety system will include at least:
\begin{itemize}
    \item The dump at the \htp port of the analyzing magnet just following the stripper, which monitors the health of the stripper foil;
    \item Area gamma and neutron monitors, that detect an unexpected rise in the overall radiation level in the vicinity of the transport line;
    \item The optical sensor monitoring target temperature for excessive heat;
    \item Instrumented curtains and collimators in the vicinity of the target to ensure centering of the beam and minimizing beam striking the edges of the target.
\end{itemize}

An extensive literature exists for safety systems in high-power accelerators, and the design of the IsoDAR system should be based on this experience.  We summarize here some salient features.

The time between detection of a fault and cessation of beam is not instantaneous.  Establishing this time delay will determine the requirements for handling beam power that must be absorbed before the beam itself is interrupted.

\begin{itemize}
    \item Latency in the detecting instrument will depend on the instrument type, but should be in the microsecond range for most devices.
    \item Triggering of the safety system cutoff electronics will also contribute microseconds, possibly as high as a millisecond.
    \item Triggering the fast chopper in the LEBT before the RFQ entrance. The risetime of this chopper will be on the order of a microsecond or less. 
    \item  Ions circulating in the cyclotron will continue being accelerated until the RF system can be safely shut off. From injection to extraction, ions reside in the cyclotron approximately 10 $\mu$s.
\end{itemize}
 
In all likelihood the maximum time delay could be on the order of a millisecond.  Beam power that must be absorbed would then be on the order of a kilojoule.  

It is anticipated that a large number of trips will occur due to failure of the stripper foil, in which case the signal will come from the unstripped ion beam dump.  This must be designed with sufficient cooling to accept the expected level of beam energy, which will depend on the actual evolution of the foil failure.  If it is gradual, then the rise of current in this dump will not be fast, but a trigger level should be set at a suitably low value to prevent overheating of this dump. 
For example, setting the threshold at 0.1\%, or 10 $\mu$A would require cooling capability for about 1 kW, easy to provide.

Another important consideration is operational recovery from a trip.  Experience at PSI has been that many trips can occur during normal operation, not all related to actual emergency conditions.
It is important that operators be given relevant information as to the cause of the trip immediately, and that they also have the ability to quickly resume normal operation if the trip has not been due to an actual failure.

\subsection{MEBT Beam Monitoring Instrumentation}
A number of essential beam monitoring instrumentation devices are underlined and discussed below.

Any instrumentation requiring intercepting even a very small portion of the beam must be used only with vastly-attenuated beams.  
Such would be \underline{harps or wire-grid beam profile monitors} that are used for initial tuning at very low beam current levels.  
These measure the {x/y} beam size and position, centering of the beam in quadrupole and dipole magnets for controlling focusing and steering of the beam.
These devices must be removed from the beam line when the current is brought up to operating levels.

During normal operation \underline{inductive and capacitive} (non-intercepting) instruments will be used liberally along the beam line.  These measure beam current and centroid at their location, but cannot measure beam size or shape.  

\underline{Collimators and scrapers} can be wired for detecting current, and are designed for intercepting only very small portions of the beam.  Nevertheless these should be carefully designed and adequately cooled.

\underline{Target shield.}  A water-cooled collimator protecting the outer edges of the target torpedo should be installed in the vacuum beam pipe just upstream of the target, consisting of a ring fitting on the inside of the beam pipe and extending about 2 cm into the aperture.  
This will ensure beam does not hit the target in areas where water flow is not high.  
It would be instrumented for current readout, probably in quadrants, and so also serving as a beam-steering diagnostic.

\underline{Current transformers} will be needed at the cyclotron extraction point, and after the stripper analysis magnet in the proton line.  These are simplest and work best if the beam still contains microstructure, which will most probably be true at the extraction point, but the beam may have debunched sufficiently after the stripper, so measuring the proton current may need to be done with a DC transformer.  Bergoz Instrumentation has excellent devices for both these applications~\cite{bergoz}.

\underline{Faraday cups} located at the exit of the secondary ports in the stripper analysis magnet.  Note, none will survive full beam current.  The most important is monitoring for unstripped \htp, as it signals failure of the stripper foil.

\underline{Stripper sensor}:  If possible, it would be good to have an IR camera to image the stripper; catching irregularities might help to detect imminent foil failure and initiate the process of placing a new stripper in the line.

\underline{Target face monitoring}:  This is extremely important for target health.  
Target face monitoring should be done by IR cameras, with light focused on a remote camera by means of mirrors aimed at the target.  Neutron fluxes will destroy any camera, and will cloud any refractive optical elements, requiring these cameras to be very well shielded from the target neutrons, and that mirrors be highly radiation resistant, such as shaped and polished copper surfaces.  FRIB has a system like this that has been shown to operate well~\cite{FRIB1,FRIB2,FRIB3}.

Note:  Water (D$_2$O) temperature in heat exchangers will also be a measure of the efficiency of beam coupling, a careful energy summing should be done to verify all beam energy is accounted for, and that there are not unacceptable beam losses, for instance in the beam pipe just upstream of the target.

\subsection{Area Radiation and Neutron Monitors}

\underline{Area Radiation monitors.} Versatile and reliable radiation-monitoring instrumentation is important for safety and operations.  The dynamic range of these instruments is also important--they cannot be rendered useless by saturation when radiation levels become exceedingly high.  But, at the same time they should accurately register the background levels when personnel access is required to the accelerator and target areas. Fast response is also required when radiation levels jump radically during operation, to trigger immediate cessation of beam because of an abnormal beam event.  

In all likelihood no single instrument can cover the entire dynamic range required. Normal Geiger-Mueller monitors can be used for personnel access monitoring, as they will saturate at high radiation levels; while a type of ionization detector could be employed for the safety-system fast-cutoff detector needs.

There is ample experience with this type of device at the larger National Laboratory accelerator facilities; these will be consulted to plan for the instrumentation needed for our applications.

\underline{Scintillator Neutron Monitors.}
To complement the charged-particle and gamma monitoring, we will deploy scintillator-based neutron monitors around the beam line and shielding perimeter. The neutron monitors will consist of modular neutron-sensitive scintillators like boron- or lithium-loaded plastic scintillators (e.g., EJ-426 from Eljen Technology) surrounded by a polyethylene moderator. The scintillators will be coupled to a small array of silicon photomultipliers (SiPMs), with sufficient photocathode coverage to enable pulse height discrimination with backgrounds. Each neutron monitor will operate independently, and communicate asynchronously to the DAQ computer, recording time stamp and pulse amplitude. A Geant4 simulation will be developed to study this. Positioned around the beam shielding, the neutron monitoring system will provide detailed insights into neutron rate variations and detect potential breaches in containment. This setup enables continuous monitoring of secondary particle production and shielding effectiveness, contributing to the overall safety framework of the IsoDAR project.

\subsection{Data Acquisition and Control System}

The control system will serve as the central platform for collecting and processing data from all systems of IsoDAR, providing the interface with operators, and managing the safety system.   The widely-used open source EPICS platform~\cite{EPICS} will be utilized. This system will standardize data formats, ensuring compatibility with analysis tools and consistency across data streams.  EPICS  supports secure remote access via SSH, allowing authorized personnel to monitor and interact with the system from any location. Data acquisition will occur in real-time, with a software interface to display beam conditions, monitor neutron rates, and track interlock status. The system will be designed for scalability, enabling future integration with additional monitoring devices or expansion of the IsoDAR experimental setup. With a well-defined data format, the DAQ system will streamline data handling, allowing for immediate visualization, monitoring, and archiving of critical parameters for safety and research purposes.

\subsection{Risks and Mitigation}

\noindent \textbf {Risk:  Instrument failure due to radiation damage.}  Aside from scrapers, all instrumentation will be non-intercepting:  inductive or capacitive pickups.  Beam loss, to some level, is inevitable, and this will produce a radiation field that can affect electronics.  
Though the monitors and probes themselves are unlikely to suffer radiation damage, most of these have front-end electronics that will be affected by radiation fields that might cause deterioration of performance.

\textit{Mitigation:  Regular maintenance and re-calibration schedules can identify effected electronics, and an adequate supply of spare components should be available to change out damaged units.}

\bigskip
\noindent \textbf {Risk:  Critical target-monitoring infrared camera detector may become inaccurate or may fail.}  The infrared camera itself is very radiation sensitive, it may not perform reliably.  The optical elements in the path between target and camera may lose efficiency.

\textit{Mitigation:  (1) The location of the camera must be adequately shielded.  It will most likely still need to be in the Target Hall, so its separation from sources of radiation may not be optimal. It should reside inside a radiation-resistant enclosure. Maintenance and recalibration protocols, and monitoring of the local radiation environment, should be adequate to ensure proper performance of the camera.  (2) The optical path from the target to the camera should be frequently recalibrated, optical surfaces of mirrors checked for radiation damage.  A radio-resistant IR optical source should be mounted in the vicinity of the target torpedo to provide a calibration source to check the optics and camera response.  There should in addition be a protocol for periodically calibrating this IR source to ensure it is not being degraded by radiation.}

\clearpage
\section{Installation}

\subsection{Overview}

Adding to the discussions of installation issues in Volume I, the elements of the MEBT, target, sleeve, and most importantly the shielding components must be brought to the experimental site from the surface, installed, and commissioned.  

While the stated goal of this PDR volume was to be a site-independent analysis; the specifics of installation, commissioning and operations must of necessity be linked to an example location.  The same issues must be addressed for any site, and the details presented here can be modified according to those specifics.

For us, the example site is Yemilab.
This new underground laboratory, excavated about 1~km under Mount Yemi, is about 200~km east-southeast from Seoul.  It is adjacent to the Handuk iron mine, whose system of ramps and shafts provides access to the lab.  
The lab is managed by the South Korean Institute for Basic Sciences (IBS), and specifically its Center for Underground Physics (CUP). 
This Center is also responsible for the construction of the 
$\nu$EYE detector.
The deployment of the IsoDAR experiment will be a joint effort of the IsoDAR Collaboration and the IBS CUP.
Those elements of the deployment that relate to infrastructure, such as electrical power, cooling water, ventilation, operations management, radiation safety, and probably much of the actual installation, will most likely be shouldered by CUP, while the IsoDAR Collaboration will provide the physical equipment:  cyclotron, beam transport, target, sleeve, and its shielding.

This section then will focus on an installation overview, including specifications, and a few suggestions related to the installation process.

 \begin{figure}[b!]
    \centering
\includegraphics[width=\linewidth]{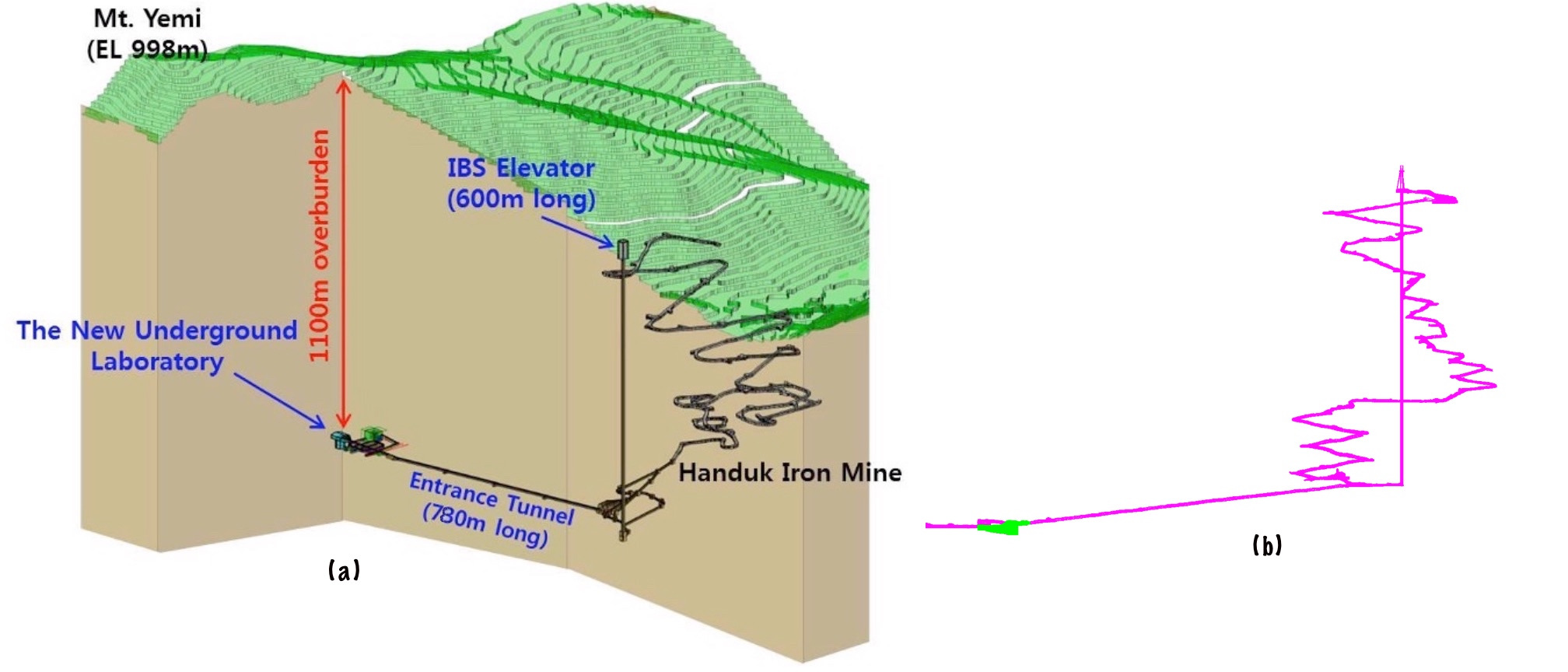}
    \caption{(a) Schematic of Handuk/Yemilab area, (b) Low resolution laser scan of Handuk mine and Yemilab access ramps, green area denotes high resolution scans of IsoDAR deployment area, shown in more detail in the next figure. From Ref.~\cite{kimYemilabNewUnderground2024}.}
    \label{Yemilab environment}
\end{figure}

Figure~\ref{Yemilab environment} shows the geographical area of the Laboratory, including the access points through the Handuk mine.  These include the 600 meter vertical shaft with a personnel carrier which, unfortunately, cannot be used for transporting heavy or oversized pieces.  These must be brought down through the 6 km mine ramp.  This ramp has a nominal 5 meter x 5 meter clearance, and can be navigated by the large mine trucks used for hauling ore to the surface.  

There is little doubt that all of the elements of the MEBT, target, sleeve and shielding can be broken into components that will fit in the bed of these mine trucks, whose dimensions are roughly 3 meters wide, 3 meters high and 7 meters long.

The mine trucks can transport material from the surface to the base of the vertical shaft, but beyond that point, which is the  access gateway to the lab, the area is clean, and diesel trucks will not be allowed.  There is ample space to transload material to electric conveyances for the remainder of the transport distance,

\begin{figure}[tb]
    \centering
    \includegraphics[width=0.6\linewidth]{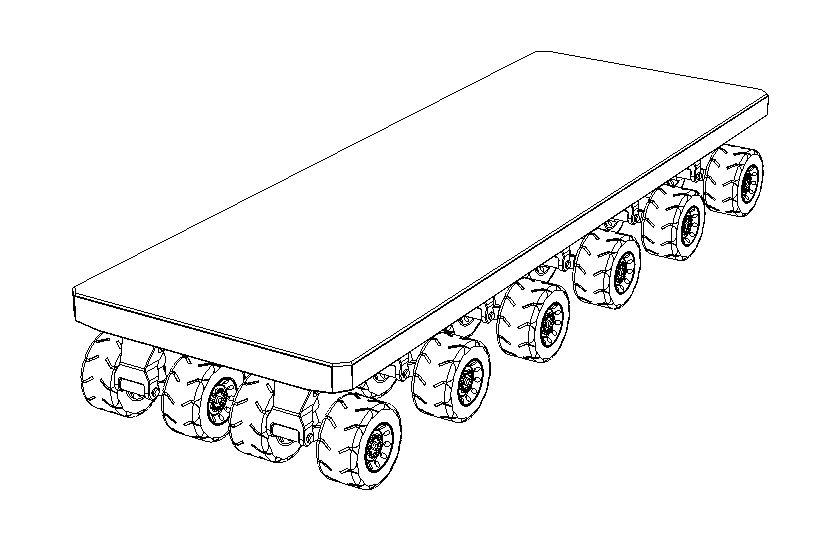}
    \caption{A typical SPMT (Self-Propelled Motorized Transport) vehicle, a versatile way to transport heavy loads over irregular terrain.}
    \label{fig:SPMT vehicle}
\end{figure}

Alternately, in Volume I we presented a flexible motorized (battery-driven) platform that could be customized to navigate the twists and turns of the mine ramp, that could be used for transporting the MEBT, target and shielding components to their installation site (see Fig.~\ref{fig:SPMT vehicle}). It is also clear that all of the elements beyond the cyclotron can be broken into pieces that will meet the load and size limits of these conveyance devices.

\subsection{Cavern Preparation}

\begin{figure}[hbt!]
    \centering
    \includegraphics[width=0.9\linewidth]{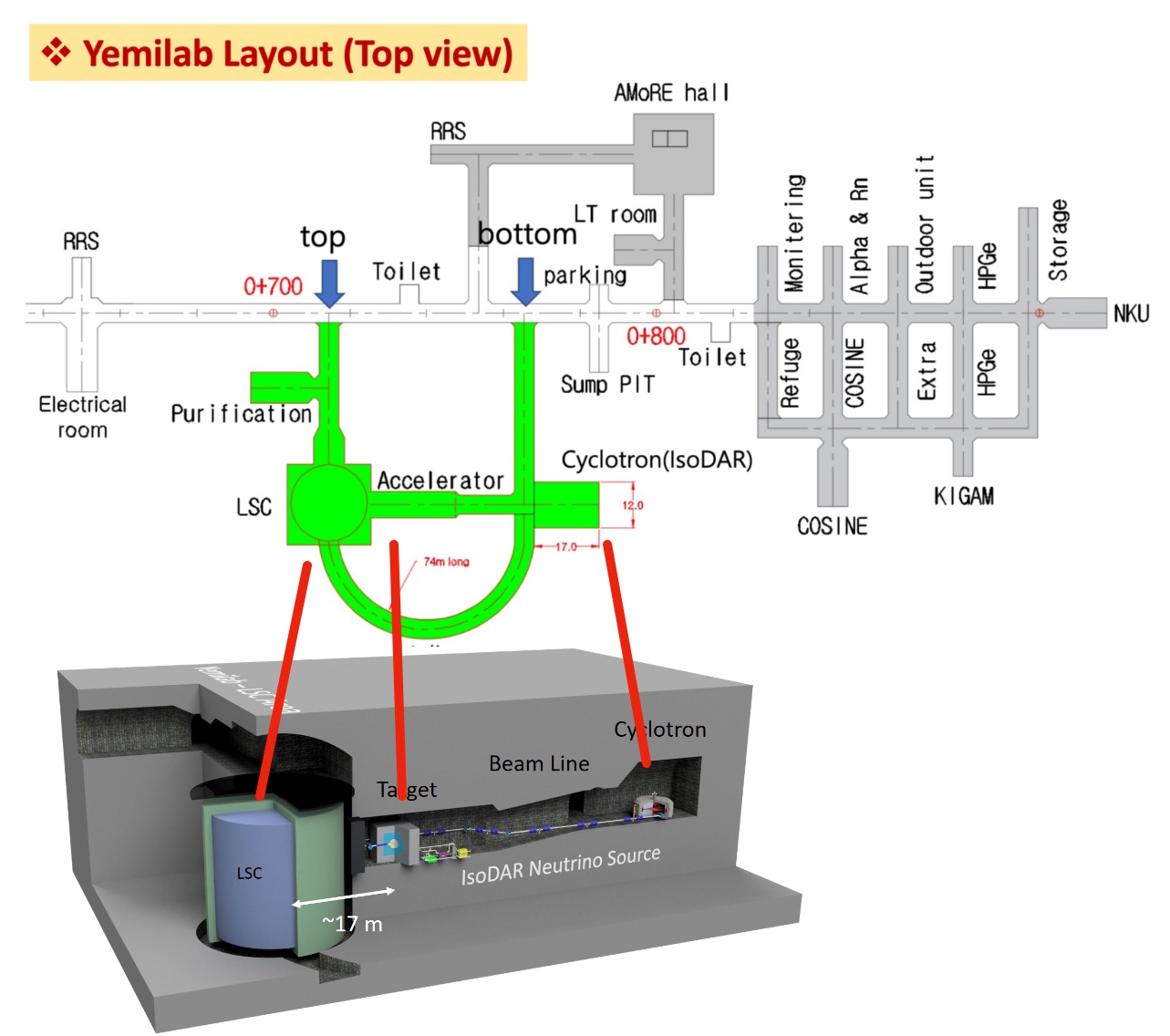}
    \caption{Layout of Yemilab, showing (in green) the area where IsoDAR will 
             be installed. The cyclotron will be assembled in the dedicated 
             cyclotron vault as shown.}
    \label{Yemilab Layout}
\end{figure}

Space for all the IsoDAR components has already been provided by the Yemilab facility.  Figure~\ref{fig:CavePhotos} shows photographs taken during a tour of the IsoDAR/$\nu$EYE area, while Fig.~\ref{Yemilab Layout} shows the deployment plan for IsoDAR at the Yemilab site.  Volume I Chapter 4 described the transport, assembly, and installation of the cyclotron in its dedicated vault; here we will address the installation of the MEBT, target, and shielding components in the hall labeled ``Accelerator''  which we are renaming as the ``Target Hall''.

\begin{figure}[hbt!]
    \centering
    \includegraphics[width=0.9\linewidth]{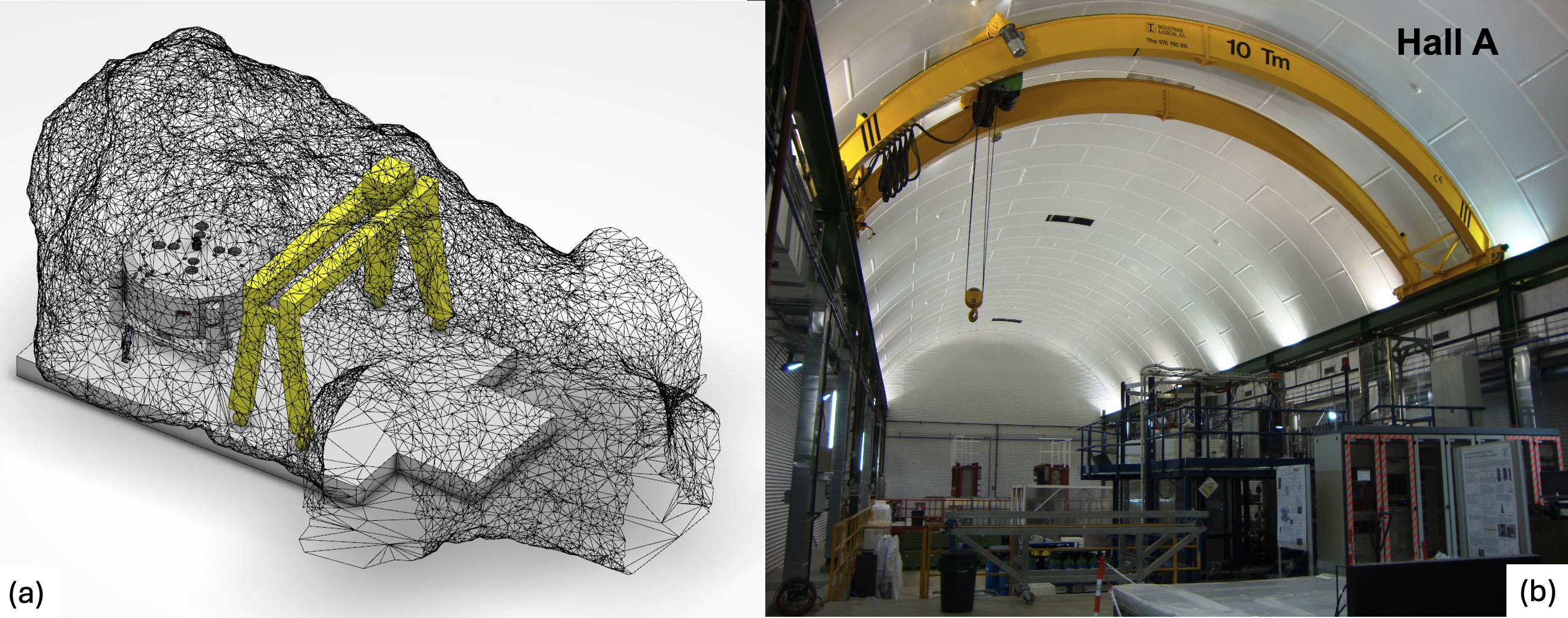}
    \caption{(a) Cyclotron vault showing a bridge crane.  Once the cyclotron is assembled the bridge crane will be dismantled, and reconfigured for installation in the Target Hall.  (b) Possible reconfiguration of the crane to adapt to the domed ceiling, photo is the crane in the Canfranc Laboratory hall in the Spanish Pyrenees~\cite{Canfranc2016}. }
    \label{fig:cyclotron vault}
\end{figure}

Assembly of the cyclotron will be facilitated by means of a bridge crane mounted on rails on the floor of the cyclotron vault, shown in Fig.~\ref{fig:cyclotron vault}. The same crane can be dismantled and reassembled in the target hall. The height and span of this crane are completely adjustable, as it must be brought down in pieces as well, and pieces can be changed to fit the span and height of the target hall.  

Consideration should be given to the crane style used at the Canfranc Laboratory in the Pyrenees, the arc shape allows access to more vertical space within the cavern. This crane will remain in the target hall, and can be used for maintenance of the target such as replacing the ``torpedoes.''  Specialized bore holes will be provided in the rock to house spent, and new targets, so this target changing can be accomplished with maximum efficiency and minimum personnel exposure (see Fig.~\ref{fig:target-change}).

While having such a fixed crane would provide convenience and efficiency, it is not an absolute necessity, and flexible fork lifts and portable cranes may be a more cost-effective choice.  
One consideration is that radiation exposure to personnel would undoubtedly be lower for an installed bridge crane during maintenance operations such as changing the target torpedo.

\subsection{Iron Wall Installation}

Figure~\ref{fig:ShieldPlan} shows a concept of the principal shielding elements, which constitute the heaviest pieces to be installed in the target cavern.  The iron shield would be the first to be installed, as clear access to the entire area is required.  It will be beyond the reach of the crane, so specialized rigging techniques will be required. The design of the wall must take into account the capacity of available rigging equipment.
In all probability the lining of the $\nu$EYE cavern will have been completed, so the cavern end will show a solid concrete face.  

Note that supporting the weight of the iron wall, approximately 1 kiloton, will undoubtedly require reinforcement of the floor area adjacent to the $\nu$EYE cavern.  This work must be completed before the detector lining is installed.

Iron pieces will be assembled against this lining wall, the design needs to be carefully done to ensure control of cracks between iron pieces.  (Slow) neutrons will migrate through these cracks, so pieces should be interlocked and fitted to minimize such cracks running along the direction pointing to the detector.  Care needs to be taken too of the top dome shape of the cavern, and the irregular rock surfaces along the edges. These areas must be filled with specially fitted pieces, or tightly-packed bags of steel shot.  

\subsection{Target Shielding}

The target shielding, shown in Fig.~\ref{fig:TargetShieldingBlock}, consists of an inner layer of iron that contains the target torpedo and $\sim$1.4 meter diameter sleeve, and an outer layer of boron-loaded concrete.  This assembly should be fully under the coverage of a bridge crane, which would greatly facilitate its assembly.  Pieces will be brought down the ramp with a fork lift or SPMT, and picked up by the crane for positioning in their proper places.

The lower portions of the shield, bottom concrete and steel pieces, will be installed first, up to the level of the beam height.  This, along with steel filling the space between the steel wall and the target shield, will form a base for the installation of the beam line and sleeve components, which can be rigged into place with the bridge crane.

Once beam line elements are installed, the upper portions of the shielding assemblies can be constructed.  While the target block will be sealed, there are no parts whose access requires removal of any parts of this block, the channel between this block and the steel shield wall where beam line elements -- two 90° dipoles and possibly other magnetic elements -- and diagnostic instrumentation are located must be accessible for maintenance.  This will require stacking of shielding components that can be easily removed by the overhead crane.

\subsection{MEBT and Target Support Equipment}

Support stands for the MEBT components:  magnets, vacuum equipment, and instrumentation, will be rigged with fork lifts for the areas not under crane coverage, and inside the crane rails in the target hall.  Heavy beamline components are installed next, and are wired and aligned.  Beam pipes, vacuum connections and pumps, and instrumentation come next.

The equipment cart for servicing the target is installed behind the target block, with connections made to the end of the torpedo. This is located in an easily accessed and serviced location, and has connections through heat exchangers to the main laboratory water system. A low-conductivity water system to cool the cyclotron, and all magnets and components in the transport line must be provided, again providing heat exchangers to transfer heat to the main laboratory water system. Note that an assessment needs to be made to determine if  second sets of heat exchangers are needed located outside the IsoDAR controlled area, as any chances of activating the main Yemilab water cooling system must be avoided.

\subsection{Target ``Torpedo'' Handling}

\begin{figure}[hbt!]
    \centering
    \includegraphics[width=0.9\linewidth]{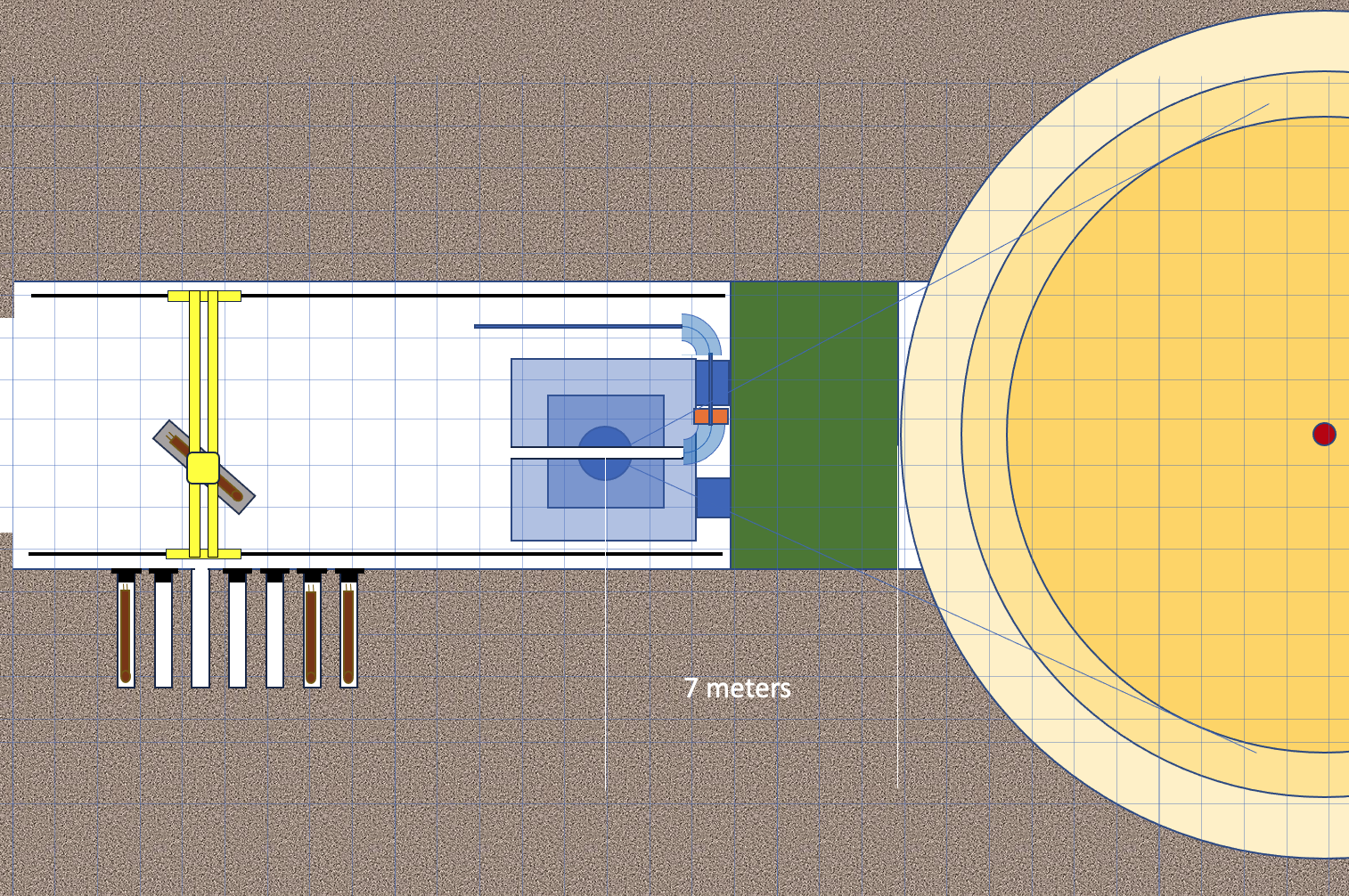}
    \caption{Schematic showing spent target torpedo in lead-lined steel casket being transferred to a shielded storage borehole, one of several drilled into the target cavern walls. This storage area also serves for new replacement targets.  The torpedo is slid out from the beam pipe, which is shown up to air. Note that, for clarity, the MEBT is not shown.}
    \label{fig:target-change}
\end{figure}

It is anticipated that the target ``torpedo'' will need to be changed several times during the lifetime of the experiment.  The presence of the bridge crane will greatly facilitate this process.  A lead-lined steel casket will be attached to the end of the beam pipe, and the old torpedo drawn into the casket.  The casket will be moved via the overhead crane and positioned at the port of one of several storage bore holes, and the spent target inserted into this bore hole.  These bore holes also serve as storage for new targets, and the process can be reversed for bringing a new target into the beam pipe inside the target shield.  The process is shown schematically in Fig.~\ref{fig:target-change}.  

The spent targets will remain in these storage boreholes, where they will cool for the duration of the lifetime of the experiment, and will be removed as part of decommissioning of the entire experiment.

\subsection{Support Equipment and Control Area}

Areas need to be identified and prepared for locating power supplies, interlock and control equipment, vacuum pumping infrastructure, and cooling water systems.  
These will need proper connections to the laboratory infrastructure. In addition, a space needs to be identified and provided which contains instrumentation and control equipment for monitoring and operation of all the active components in the experiment.  
Ideally this should be in relatively close proximity to the operating equipment so that delay to respond to a service need is not excessive. There will be significant advantages to having this control center connected to, and ideally in close proximity to the control areas allocated for $\nu$EYE.

\subsection{Radiation Protection}

Part of the radiation protection infrastructure must be air handling, that maintains the IsoDAR cyclotron and target halls at a slight negative pressure, to prevent the migration of any air activation products, which are all short-lived, from reaching other areas of the laboratory.  Radiation protection gates and access-control interlocks must be provided, it is expected, by IBS, that are compliant with Korean regulations.

\subsection{Risks and Mitigation}

\textbf{      Risk:  Handuk ramp roadbed too rough for planned transport equipment.}  The very flexible material transport handler may have difficulty navigating the 6 km winding ramp of the Handuk mine if the roadbed is not smooth enough.  This ramp can be navigated by the mine ore trucks, specifically designed for rough terrain with very large tires, but this does not guarantee that the SPMT we acquire will.  

\textit{Mitigation:  All of the components for the target and shielding can be sized to fit in the bed of the Handuk mine-trucks, and be brought underground by these trucks.  There will be a requirement then for transloading the material onto the SPMT at the interface between the end of the mine ramp and the access drift to the lab.  
Mine trucks cannot pass this point because only battery-operated equipment can be used in the laboratory area.  This will require proper rigging equipment to transfer items from the truck to the SPMT.}

\bigskip
\textbf{Risk:  Bridge crane cannot be provided in the target hall.}

\textit{Mitigation:  For the most part, the overhead crane is a convenience, not a necessity.  Installation of the iron shield wall will no doubt require a forklift as the wall will be beyond the reach of the crane rails.  This heavy-duty forklift can also serve for intalling the target and shielding, as well as the MEBT magnets and systems.  The only issue will be rigging of highly radioactive components for servicing or replacement.  For this, either remote control of the forklift, so the operator does not need to be in close proximity to the activated component, or provision of adequate shielding  on the forklift for the operator.}

\clearpage
\section{Conclusion}

This preliminary design report document has addressed the elements of the IsoDAR experiment following the extraction of the beam from the cyclotron, up to the experiment's interface with the $\nu$EYE detector.  
The deployment of the experiment will be a partnership effort with the Korean Institute for Basic Science (IBS), and while the regulations and requirements of the local authorities and the site need to be adhered to, we present observations of how the experiment and its requirements would interface into the Yemilab environment.

The roles and responsibilities of each partner need still to be established in detail, but identifying all the facets of the experiment can provide a basis on which to work out these details. We trust that this document can provide the basis for the work that lies ahead for the IsoDAR Collaboration and our IBS partners towards making the IsoDAR experiment coupled to the 
$\nu$EYE detector at Yemilab a reality.

\clearpage
\backmatter





\bmhead{Acknowledgements}
This work was primarily supported by NSF grant PHY-2012897. Preliminary work leading up to this was supported by NSF grants PHY-1912764 and PHY-1626069. Work for Volume I was supported by DOE grants DE-SC0024138 and DE-SC0024914 and NSF grant PHY-2411745.

\section*{Declarations}


\begin{itemize}
\item Data availability: Data is available from the authors upon reasonable request.
\end{itemize}

\bibliography{PDR_vol2}

\begin{thebibliography}{43}
\bibitem[\protect\citeauthoryear{Engebretson}{}]{Engebretson}
\begin{botherref}
\oauthor{\bsnm{Engebretson}, \binits{S.}}:
The data were provided by {S}am {E}ngebretson
\end{botherref}
\endbibitem
\end{thebibliography}

\end{document}